\documentclass[aps,amsmath,amssymb,showpacs,showkeys]{revtex4}
\usepackage[dvips]{graphicx,color}
\usepackage[utf8]{inputenc}
\usepackage{multirow}
\usepackage{times}
\usepackage{xcolor}
\usepackage[%
  colorlinks=true,
  urlcolor=blue,
  linkcolor=red,
  citecolor=blue
]{hyperref}

\begin{document}
\title{Quasinormal Modes of 
Black Holes with Non-Linear-Electrodynamic sources in Rastall Gravity}

\author{Dhruba Jyoti Gogoi}
\email[Email: ]{moloydhruba@yahoo.in}

\affiliation{Department of Physics, Dibrugarh University,
Dibrugarh 786004, Assam, India}

\author{Umananda Dev Goswami} 
\email[Email: ]{umananda2@gmail.com}

\affiliation{Department of Physics, Dibrugarh University,
Dibrugarh 786004, Assam, India}

\begin{abstract}
One of the notable modifications of General Relativity (GR) is the Rastall 
gravity. We have studied the quasinormal modes of black holes in Rastall gravity in 
presence of non-linear electrodynamic sources. Here the impacts of cosmological
field, dust field, phantom field, quintessence field and radiation field on 
the quasinormal modes in presence of electrodynamic sources have been 
investigated. Apart from this, we have also checked the dependency of 
quasinormal modes with the Rastall parameter $\lambda$, black hole structural 
constant $N_s$ and charge of the black hole $Q$. The study shows that the 
quasinormal modes corresponding to the black hole with non-linear 
electrodynamic sources show significant deviations from a general charged 
black hole in Rastall gravity under certain conditions. Further, the behaviour
of black holes and hence the quasinormal modes depend on the type of 
surrounding field considered.
\end{abstract}


\maketitle
\section{Introduction}
The existence of black holes and Gravitational Waves (GWs) are two most 
significant predictions of General Relativity (GR). The binary black hole 
systems predicted by the ground based Laser Interferometer Gravitational Wave 
Observatory (LIGO), and Variability of Solar Irradiance and Gravity Oscillations
(Virgo) detector systems stand as a valid evidence for the viability of GR. 
The direct detection of GWs by LIGO and Virgo provided us a new way to test 
different theories of gravity including GR. GR had already passed experimental 
tests in the weak field regimes and moderately relativistic regimes such as in 
the solar system tests \cite{will2014} and in binary pulsars \cite{hulse1975, 
damour1992}. The recent detections of GWs by the LIGO and Virgo have also
established the viability of GR in the highly relativistic strong gravity 
regime connected to binary black holes. Nonetheless GR has two main issues
or drawbacks. Firstly, in the high energy regime the theory is not 
re-normalizable \cite{stelle} and secondly, in the infrared regime the theory 
deviates largely from the experimental observations, especially the 
observations of accelerated expansion and hidden matter content of the 
universe are outside of the predictions of GR \cite{Riess1998, Perlmutter1999, Bahcall1999}. Due to these 
issues, extensions of GR known as Modified Theories of Gravity (MTGs) are 
introduced. Basically, the extensions of GR are of two types, in one type
the spacetime geometry is modified and in the other type, the modifications are 
done in the matter-energy sector. The simplest extension of GR in the 
matter-sector is the $\Lambda$CDM model, but with it one has to bear with the 
problems of dark matter and dark energy together with the fine tuning issue 
\cite{Bull}. Some other extensions of GR are - $f(R)$ gravity, Rastall gravity, 
$f(R, T)$ gravity etc. In different MTGs the properties of GWs may change. 
For example, in f(R) gravity metric formalism the polarization modes of GWs 
increase to three, where the first two are GR modes or tensor plus and cross 
modes, and the third is scalar polarization mode which is a mixture of 
massless breathing mode and massive longitudinal mode 
\cite{Liang_2017, gogoi1, gogoi2}. The tensor plus 
and cross modes are transverse, traceless and massless in nature. They 
propagate with the speed of light through the spacetime. The massless 
breathing mode is transverse but not traceless. These new properties of GWs in 
different MTGs have increased the need to study the properties of black holes 
and compact stars in the regime of MTGs. It is because the black holes, 
although which are considered as the cleanest objects in the universe, and 
the neutron stars are the potential candidates for the generation of GWs. 
Realizing this necessity, a significant amount of studies have already been 
pursued to understand their behaviours and properties in MTGs.

Apart from the f(R) gravity, another widely studied extension of GR is the 
Rastall gravity \cite{Rastall72}. This MTG was originally introduced by P.\ 
Rastall in 1972, 
but it did not get sufficient attention at that time. In the last few years, 
Rastall gravity has drawn the attention of so many researchers due to some of 
its unique behaviours including the violation of normal conservation law in 
presence of non-vanishing background curvature \cite{Saleem2021, Bronnikov2021, Ghosh2021, Sakti2021, Maurya2020, Visser, Darabi, Moraes, Shabani,Rawaf01,Rawaf02, Fabris, Rahman, Rahman2, Moradpour}. In Rastall gravity, 
the original conservation law is modified by making the covariant divergence of 
 stress-energy tensor proportional to the covariant divergence of the Ricci 
curvature scalar. However, it is to be noticed that the usual conservation law 
can be easily recovered by setting the background curvature zero. In other 
words, the Rastall gravity is equivalent to GR in absence of any matter source.
Thus at a sufficient distance away from the actual source of GWs, we get 
only two tensor polarization modes of GWs in Rastall gravity.
However, it is seen that in a recent study it was claimed that Rastall gravity 
 is equivalent to GR \cite{Visser}. But it can be clearly seen from the field 
equations of Rastall gravity that this theory is different from GR in many 
aspects \cite{Darabi}. The field equations, as mentioned in Ref. \cite{Visser},
can be rearranged to give a GR like form by defining an effective energy 
momentum tensor say $T^{eff}_{\mu\nu}$ as a sum of the actual energy momentum 
tensor and the rest curvature dependent part (curvature fluid energy momentum 
tensor). But this is true for other metric theories also. For example, one can 
consider $f(R)$ gravity in which it is very easy to express the field 
equations in GR like form with the help of an effective energy momentum tensor 
say $T^{eff}_{\mu\nu}$ \cite{Darabi}. But this does not make $f(R)$ gravity 
equivalent to GR. These points are discussed in Ref. \cite{Darabi} in details. 
Thus for any matter source, the field equations of Rastall gravity are always 
different from the field equations in GR. Apart from these, Lagrangian 
formulation for Rastall gravity has been introduced recently 
\cite{Moraes, Shabani}, which shows that Rastall gravity can be treated as a 
special case of $f(R, T)$ gravity as well as $f(R, \mathcal{L}_m)$. 
In recent times the Rastall gravity gains significant importance due to 
some good behaviours of it. Rastall gravity may provide an alternative 
description for the matter dominated era with respect to the 
GR \cite{Rawaf01}. The theory also seems to be in significantly good agreement 
with observational data on the age of the Universe and on the Hubble parameter 
\cite{Rawaf02}. Another positive side of Rastall gravity is that the theory 
does not suffer from the entropy and age problems of standard cosmology 
\cite{Fabris}. The theory is also consistent with the gravitational lensing 
phenomena \cite{Rahman, Rahman2}. Apart from these, traversable asymptotically
flat wormholes are also studied in Rastall gravity and it is seen that 
feasible traversable wormhole solutions are possible for Rastall gravity 
\cite{Moradpour}. Moreover, the study of a theory that deviates from GR only 
in presence of matter or any exotic fields can have several merits. Since the  
normal conservation law is measured in the flat space only \cite{Rastall72}, 
that leaves a possibility of violation of the conservation law in curved 
spacetime. This feature easily allows the Rastall gravity to pass solar system 
tests. As the theory tends to GR in the weak field regime, experimental results 
that support GR in this regime will also support Rastall gravity. Another 
important feature of the theory is that the field equations being directly 
obtained by modifying the normal conservation law, they do not depend on the 
metric or palatini formalism unlike $f(R)$ gravity.

In this work, we shall study the scalar quasinormal modes for black holes in 
Rastall gravity. The quasinormal modes are some complex numbers related to the 
emission of GWs from compact and massive perturbed objects in the universe \cite{Kokkotas}. The real part of the quasinormal modes is related to the emission frequency, while 
the imaginary part is connected to its damping. The idea of quasinormal modes 
was initially put forwarded by Vishveshwara in 
1970 \cite{Vishveshwara}. It was later confirmed by Press \cite{Press}. He 
found that an arbitrary initial perturbation decays as a pure frequency mode. 
After around 5 years of Vishveshwara's idea, Chandrasekhar and Detweiler 
calculated the quasinormal eigenfrequencies \cite{Chandrasekhar}. The GWs or 
the related quasinormal modes provide a strong possibility to test GR and 
other extended theories of gravity \cite{Berti2015, Dreyer2004}. 

The study of black holes and neutron stars in Rastall gravity is not very old. 
In 2015, Oliveira et al., studied the neutron stars in Rastall gravity 
\cite{Oliveira}. In 2017, Heydarzade et al., studied the black hole solutions 
in Rastall gravity \cite{Heydarzade}. In the same year, another work by 
Heydarzade and Darabi appears in which they have studied different black hole 
solutions surrounded by perfect fluid in Rastall gravity \cite{Heydarzade2}. 
The quasinormal modes for the black holes in GR regime surrounded by 
quintessence field have received significant interests and the modes were 
explicitely studied in \cite{Chen, Zhang, Zhang2, Ma, Zhang3}. In another 
work, the quasinormal modes for higher dimensional black holes surrounded by 
quintessence field in Rastall gravity were studied \cite{Graca}. In this work 
they have shown the shifts in the quasinormal modes in Rastall gravity from 
GR. Jun Liang studied the quasinormal modes of black holes surrounded by 
quintessence field in Rastall gravity \cite{Liang}. In his study, he 
investigated the variation of quasinormal modes with respect to the Rastall 
parameter in details and found that when $\kappa \lambda < 0$ the gravitational
field, electromagnetic field and massless scalar field damp more rapidly. In 
this case, the real frequencies of quasinormal modes were larger in comparison 
to GR. On the other hand, for $\kappa \lambda > 0$, he noticed that the 
gravitational field, electromagnetic field and massless scalar field damp more 
slowly and real frequencies of oscillations are comparatively smaller. From 
this study it was also clear that the variation patterns of the real and 
imaginary frequencies of quasinormal modes with respect to $\kappa \lambda$ are
similar for different values of $l$ and $n$. However, in this study only fixed 
type of the surrounding field and structural parameter $N_s$ were considered 
for the charge neutral black holes. Some other studies in this direction are 
\cite{Xu, Lin, Hu}. Another study was made on the regular black holes, where 
non-linear electrodynamic sources were used \cite{Balart}. However, this work 
was done in the GR regime. 

In this work, we introduce the non-linear electrodynamic sources in the Rastall
gravity black holes for five different surrounding fields. This work will shed 
some light on the possibility of formation of regular black holes with 
non-linear electrodynamic source in Rastall gravity for different surrounding 
fields. Apart from this, it will also focus on the dependency of real and 
imaginary quasinormal frequencies with charge of the black hole $Q$, the 
Rastall parameter $\lambda$ as well as the structural parameter $N_s$. The work
is organized as follows. In the next section \ref{section2}, we have included 
a very short review on the field equations of Rastall gravity. 
The surrounded black hole solutions are obtained in section 
\ref{section4}. Here we have focused on a special type of static black hole 
having non-linear electrodynamic sources and we show that depending the 
surrounding field, it may give rise to regular black hole solutions. We have 
obtained the quasinormal modes of black holes in section \ref{section5}. In 
this section, we have tried to show the dependency of the quasinormal modes 
on the surrounding fields of the black holes. We have studied five different 
possible surrounding fields and calculated the quasinormal modes of the black 
holes. Finally, we have summarized our results and findings in section 
\ref{conclusion}. Throughout the paper, we have used the geometric unit 
system $(G=c=1)$.

\section{A short review of Rastall Gravity} \label{section2}
In this interesting modification of GR, the covariant conservation 
condition $\nabla_\nu T^{\mu \nu} = 0$ was changed to a more generalized 
version as \cite{Rastall72}
\begin{equation}\label{r1}
\nabla_\nu T^{\mu\nu} = a^\mu.  
\end{equation}
To make the theory consistent with GR, one must have the right hand side of 
the above equation equal to zero when the scalar curvature or the background 
curvature vanishes. Thus as a convenient option, we can set the four vector 
$a^\mu$ as
\begin{equation}\label{r2}
 a^\mu = \lambda \nabla^\mu R,  
\end{equation}
here $\lambda$ is a free parameter known as Rastall's parameter. From 
Eq.s \eqref{r1} and \eqref{r2}, we can have the Rastall field equation as
\begin{eqnarray}
&& R_{\mu\nu}-\frac{1}{2}\left(\, 1 -2\kappa\lambda\,\right) g_{\mu\nu}R=\kappa T_{\mu\nu}\ .\label{E1}
\end{eqnarray}
The trace of the above equation gives,
\begin{equation}
R=\frac{\kappa}{\left(4\,\kappa\,\lambda-1\right)}\,T\ ,\quad \kappa\lambda\ne 
\frac{1}{4}. \label{E2}
\end{equation}
From Eq.s \eqref{E1} and \eqref{E2}, it is clear that GR is easily recovered 
when $\lambda \rightarrow 0$. Moreover, the theory deviates from GR only in 
the non-zero curvature regime, i.e. when $T$ is non-zero. As soon as we move 
away from any matter source, the Rastall gravity tends to GR. But since, the 
Rastall gravity provides significant deviations from GR in high curvature 
regime, we can expect prominent changes or deviations of the theory near 
massive objects like black holes and neutron stars. Rastall gravity was 
recently constrained by using the rotation curves of 16 LSB spiral galaxies. 
The lowest value of $\kappa\lambda$ was found to be $0.048$ for galaxy F579-v1 
and the maximum value of $\kappa\lambda$ was found to be $0.15$ for galaxy 
F583-1. In general, the value of $\kappa\lambda$ was found to be around 
$0.1$ \cite{Tang}. 

\section{Surrounded Black hole Solutions in Rastall Gravity} \label{section4}
In order to get the surrounded black hole solutions we consider first a 
spherically symmetric general spacetime metric in Schwarzschild coordinate as
given by
\begin{equation} \label{metric}
ds^2 = -f(r) dt^2 + \dfrac{dr^2}{f(r)} +r^2 d\Omega^2,
\end{equation}
where $f(r)$ is the metric function which depends on the radial coordinate 
$r$ and $d\Omega^2 = d\theta^2 + \sin^2\theta d\phi^2$. Now defining the 
Rastall tensor from the field Eq.~\eqref{E1} as $\Theta_{\mu\nu} = G_{\mu\nu} + \kappa \lambda g_{\mu\nu} R$, we can obtain the following non-vanishing 
components of the field equation:

\begin{eqnarray}\label{H}
&&{\Theta^{0}}_{0}={G^{0}}_{0}+\kappa\lambda R=-\frac{1}{f(r)}G_{00}+\kappa\lambda R=\frac{1}{r^2}\big[rf^{\prime}(r) + f(r) -1 \big]+\kappa\lambda R,\nonumber\\
&&{\Theta^{1}}_{1}={G^{1}}_{1}+\kappa\lambda R=f(r) G_{11}+\kappa\lambda R=\frac{1}{r^2}\big[rf^{\prime}(r) + f(r) - 1  \big]+\kappa\lambda R,\nonumber\\
&&{\Theta^{2}}_{2}={G^{2}}_{2}+\kappa\lambda R=\frac{1}{r^2}G_{22}+\kappa\lambda R=\frac{1}{r^2}\big[rf^{\prime}(r)+\frac{1}{2}r^2 f^{\prime\prime}(r)\big]+\kappa\lambda R,\nonumber\\
&&{\Theta^{3}}_{3}={G^{3}}_{3}+\kappa\lambda R=\frac{1}{r^2 \sin^2 \theta}G_{33}+\kappa\lambda R=\frac{1}{r^2}\big[rf^{\prime}(r)+\frac{1}{2}r^2 f^{\prime\prime}(r)\big]+\kappa\lambda R,
\end{eqnarray}
where the Ricci scalar reads as
\begin{equation}\label{R}
R=-\frac { 1
}{{r}^{2}}\big[{r}^{2}f^{\prime\prime}(r) + 4r f^{\prime}(r) + 2\,f(r) -2 \big].
\end{equation}
Here the prime denotes the derivative with respect to the radial coordinate 
$r$. It is seen that the mixed Rastall tensor components ${\Theta^{0}}_{0} = 
{\Theta^{1}}_{1}$ and ${\Theta^{2}}_{2} = {\Theta^{3}}_{3}$. This is the 
consequence of the spherical symmetric nature of the metric (\ref{metric}) in 
the mixed tensor form. In this work, we would like to consider a general total 
energy momentum tensor $T^\mu_\nu$ defined by
\begin{equation} \label{gen_T}
{{T}^{\mu}}_{\nu}={E^{\mu}}_{\nu}+{\mathcal{T}^{\mu}}_{\nu},
\end{equation}
where ${E^{\mu}}_{\nu}$ is the trace-free Maxwell tensor which is expressed by
\begin{equation}\label{E*}
E_{\mu\nu}={\frac{2}{\kappa}}\left(F_{\mu\alpha}{F_{\nu}}^{\alpha}-
\frac{1}{4}g_{\mu\nu}F^{\alpha\beta}F_{\alpha\beta}\right),
\end{equation}
and $F_{\mu\nu}$, the antisymmetric Faraday tensor that satisfies the 
following conditions:
\begin{eqnarray}\label{max}
&&{F^{\mu\nu}}_{;\mu}=0,\nonumber\\
&&\partial_{[\sigma} F_{\mu\nu]}=0.
\end{eqnarray}
These equations in presence of the spherical symmetry give,
\begin{equation}
F^{01}=\frac{Q}{r^2},
\label{addeq}
\end{equation}
where the parameter $Q$ plays the role of electrostatic charge in the theory. 
Hence, the Maxwell tensor can be expressed as
\begin{equation}\label{E**}
{E^{\mu}}_{\nu}={\frac{Q^2}{\kappa r^4}}~\begin{pmatrix}-1 & 0 & 0 & 0 \\
0& -1 & 0 & 0 \\
0 & 0 & 1 & 0 \\
0 & 0 & 0 & 1\\
\end{pmatrix}.
\end{equation}

The other term ${\mathcal{T}^{\mu}}_{\nu}$ in Eq.~\eqref{gen_T} describes the 
energy momentum tensor of the surrounding field expressed as \cite{Kiselev}
 \begin{eqnarray}\label{sur}
&&{\mathcal{T}^{0}}_{0}=-\rho_s(r),\nonumber\\
&&{\mathcal{T}^{i}}_{j}=-\rho_{s}(r)\alpha\left[-(1+3\beta)\frac{r_i r^j}{r_n r^n}+\beta{\delta^{i}}_{j}\right].
\end{eqnarray}
Here the parameters $\alpha$ and $\beta$ are related to the internal structure 
of the black hole surrounding field. $\rho_s(r)$ is the energy density of 
the field. The barotropic equations of state for the 
surrounding field are,
\begin{equation}
p_s=\omega_s \rho_s, ~~~\omega_s=\frac{1}{3}\alpha,
\end{equation}
where $p_s$ and $\omega_s$ are the pressure and equation of state parameter,
respectively. In this study, we shall use five different surrounding 
fields corresponding to five different values of the equation of state 
parameter $\omega_s$. For $\omega_s=-\dfrac{4}{3}$, the surrounding field is 
the phantom field. With increase in the value of $\omega_s$ towards 
$\omega_s=- 1$, the surrounding field is the cosmological constant field. 
Around $\omega_s=- \dfrac{2}{3}$, the surrounding field is quintessence type 
and when $\omega_s= 0$, we get the dust field. Finally for the radiation field 
we have $\omega_s=\dfrac{1}{3}$. Now following Ref.~\cite{Kiselev}, the free 
parameter $\beta$ can be given as
 \begin{equation}
\beta=-\frac{1+3\omega_s}{6\omega_s}.
\end{equation}
Consequently, the non-vanishing components of the ${\mathcal{T}^{\mu}}_{\nu}$ 
tensor are,
\begin{eqnarray}
&&{\mathcal{T}^{0}}_{0}={\mathcal{T}^{1}}_{1}=-\rho_s,\nonumber\\
&&{\mathcal{T}^{2}}_{2}={\mathcal{T}^{3}}_{3}=\frac{1}{2}(1+3\omega_s)\rho_s.
\end{eqnarray}

Thus the ${\Theta^{0}}_{0}={T^{0}}_{0}$ and ${\Theta^{1}}_{1}={T^{1}}_{1}$ 
components of Rastall field equations give,
 \begin{equation}\label{e00}
\frac{1}{r^2}\big(rf^{\prime} + f  -1 \big)-\frac {\kappa\lambda
}{{r}^{2}}\big({r}^{2}f^{\prime\prime} + 4r f^{\prime}+2\,f -2\big)
=-\kappa\rho_s-\frac{Q^2}{ r^4},
\end{equation} 
and ${\Theta^{2}}_{2}={T^{2}}_{2}$ and ${\Theta^{3}}_{3}={T^{3}}_{3}$ 
components can be read as
\begin{equation}\label{e22}
\frac{1}{r^2}\big(rf^{\prime}+\frac{1}{2}r^2 f^{\prime\prime}\big)-\frac {\kappa\lambda
}{{r}^{2}}\big({r}^{2}f^{\prime\prime}  +4r f^{\prime}+2\,f  -2 \big)
=\frac{1}{2}(1+3\omega_{s} )\kappa\rho_{s}+\frac{Q^2}{ r^4}.
\end{equation}
Solving Eq.s \eqref{e00} and \eqref{e22}, we get the general solution for the 
metric function,
\begin{equation}\label{f1}
f(r)=1-\frac{2M}{r}+\frac{Q^2}{r^2}
+\frac{N_s}{ r^{\frac{1+3\omega_{s}-6\kappa\lambda(1+\omega_s)}{1-3\kappa\lambda(1+\omega_s)} }},
\end{equation}
and the energy density,
\begin{equation}\label{rho}
\rho_s (r)=- \frac{3\mathcal{W}_s N_s  }{\kappa r^{\frac{3(1+\omega_s)
-12\kappa\lambda(1+\omega_s)}{1-3\kappa\lambda(1+\omega_s)}}},
\end{equation}
where the integration constants $M$ and $N_s$ represent the black
hole mass and the structural nature of the black hole surrounding field, 
respectively. The other term $\mathcal{W}_{s}$ is given by
\begin{equation}\label{Ws}
{\mathcal{W}_{s}=-\frac{\left(1-4\kappa\lambda\right)
\left[\kappa\lambda\left(1+\omega_s\right)-\omega_s\right]}{\left[1-3\kappa\lambda(1+\omega_s)
\right]^2},}
\end{equation}
which is a geometric constant. This geometric constant depends on the Rastall 
parameter $\lambda$ and on the black hole surrounding field. The geometric
constant $\mathcal{W}_{s}$ is basically related to the energy density 
distribution of a surrounding field depending on the value of the $\lambda$. 
In the GR limit it is equal to the equation of state parameter $\omega_s$ of
the concerned field. But it deviates from the $\omega_s$ of a field for any 
allowed nonzero value of $\lambda$, which is the implication of the change of 
geometry of that field. It also defines the feasible range of the structural 
parameter $N_s$ with the help of the weak energy condition. The weak energy 
condition gives,
\begin{equation} \label{weak_energy}
\mathcal{W}_{s} N_s \le 0.
\end{equation}
Thus the choice of $N_s$ also depends on the Rastall parameter $\lambda$ and 
the nature of the surrounding field. The weak energy condition for this 
case is derived from Eq.\ \eqref{rho} imposing the condition 
$\rho_s (r) \ge 0$. Eq.\ \eqref{weak_energy} restricts the structural 
parameter for a physical $\rho_s (r)$. However, a detail study of this 
condition and other implications from it might give more insights to the 
properties of black holes and we leave it as a future scope of the study.

\begin{figure}[htb]
\centerline{
   \includegraphics[scale = 0.3]{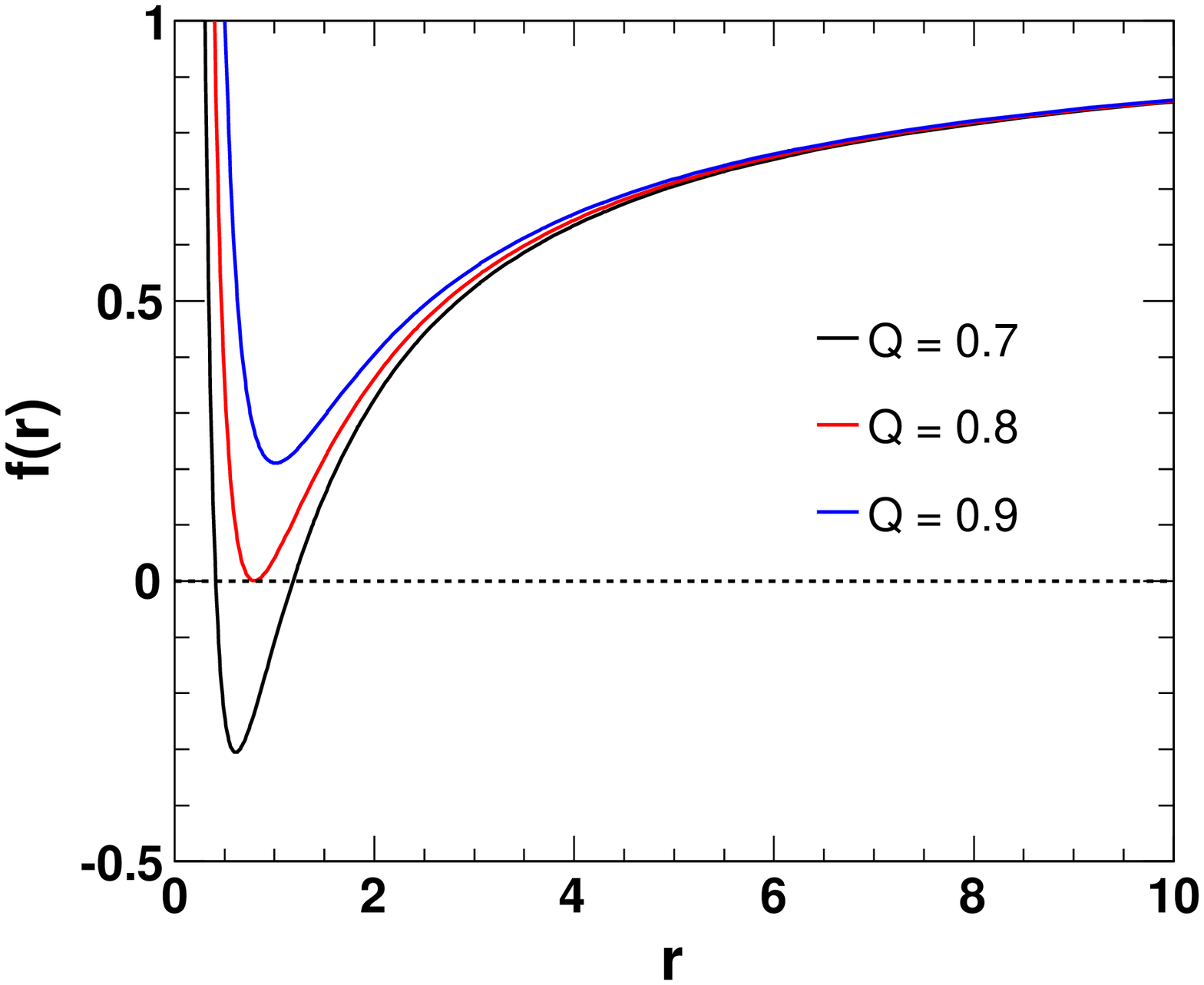}\hspace{0.5cm}
   \includegraphics[scale = 0.3]{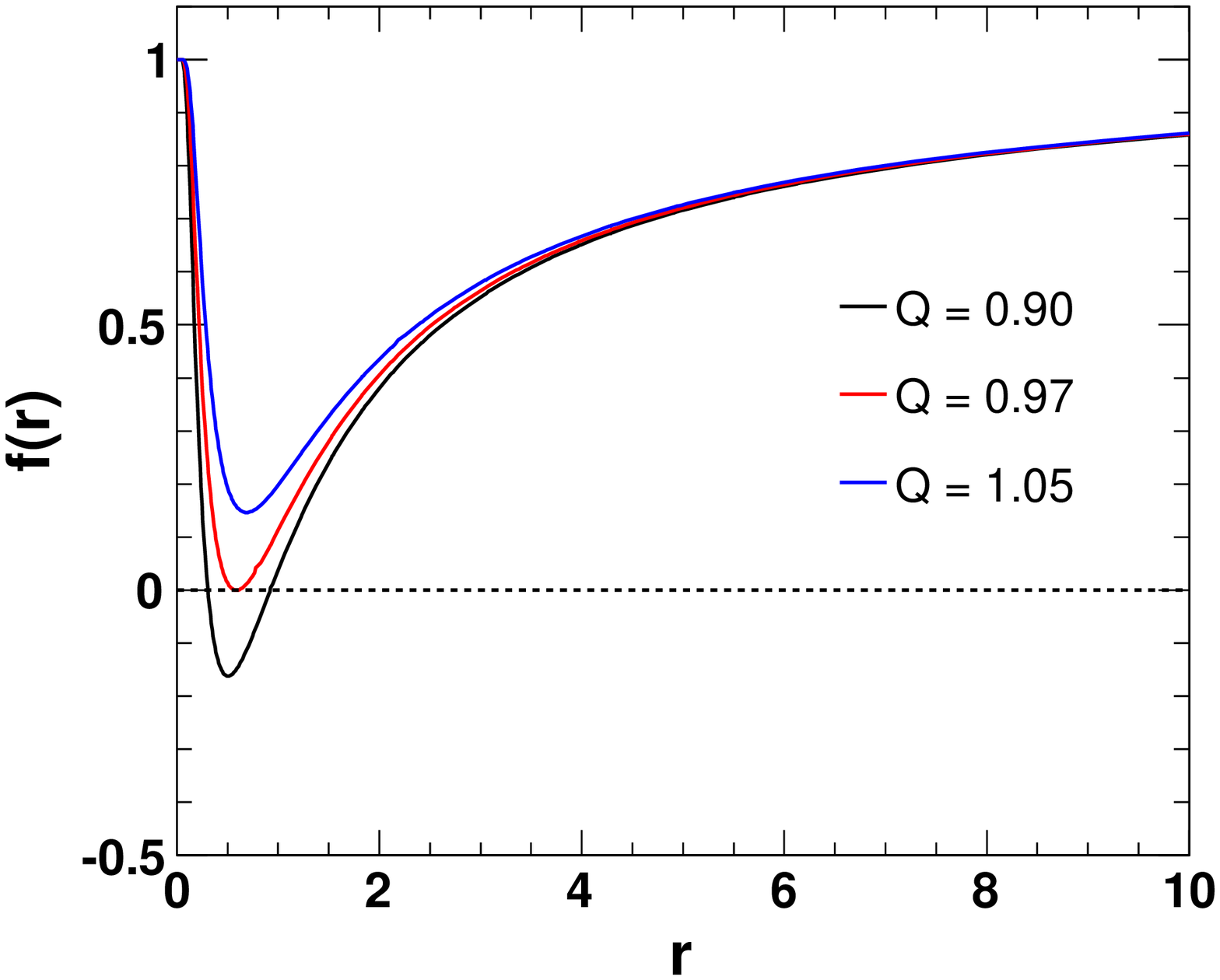}}
\centerline{
   \includegraphics[scale = 0.3]{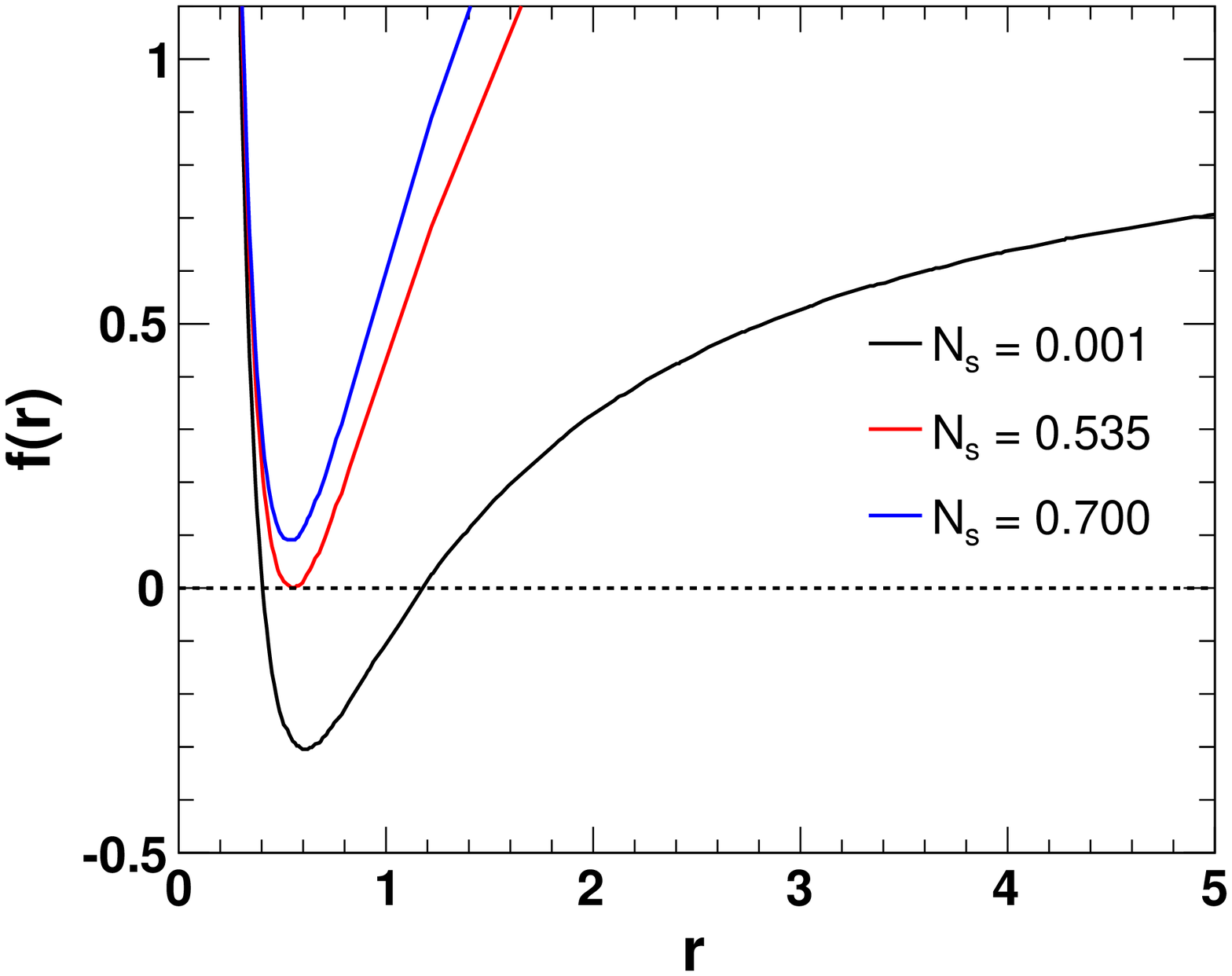}\hspace{0.5cm}
   \includegraphics[scale = 0.3]{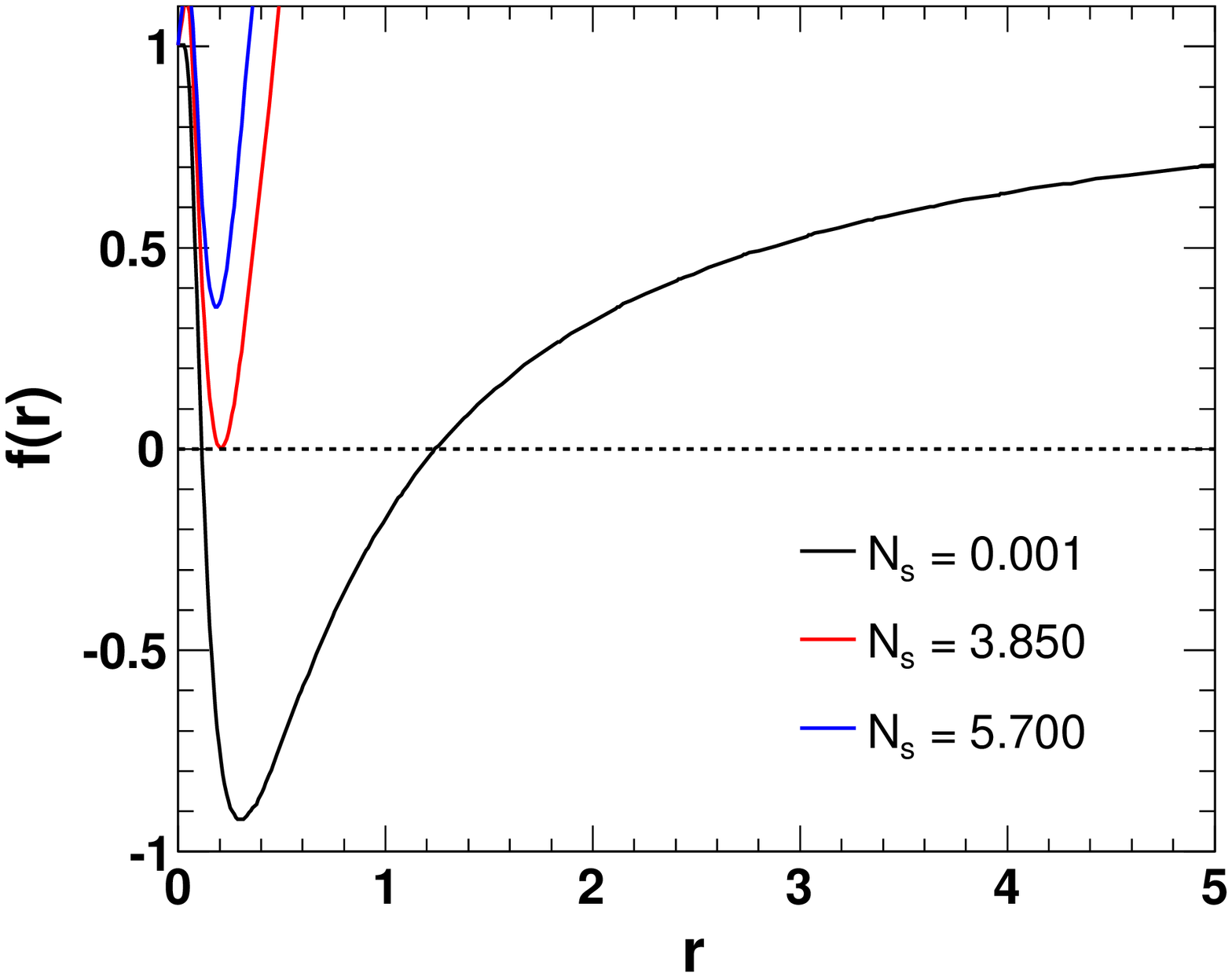}}
\vspace{-0.2cm}
\caption{Variation of the metric function for the black holes defined by the 
metric \eqref{metric01} (on left) and metric \eqref{metric03} (on right) 
surrounded by quintessence field with a viable set of common parameters 
$\lambda=0.01$, $\kappa = 1$ and $M = 0.8$. The top panel is for different 
values of $Q$ with $N_s = 0.001$ and the bottom panel is for different values 
of $N_s$ with $Q = 0.7$.}
\label{fig01}
\end{figure}

\begin{figure}[htb]
\centerline{
   \includegraphics[scale = 0.3]{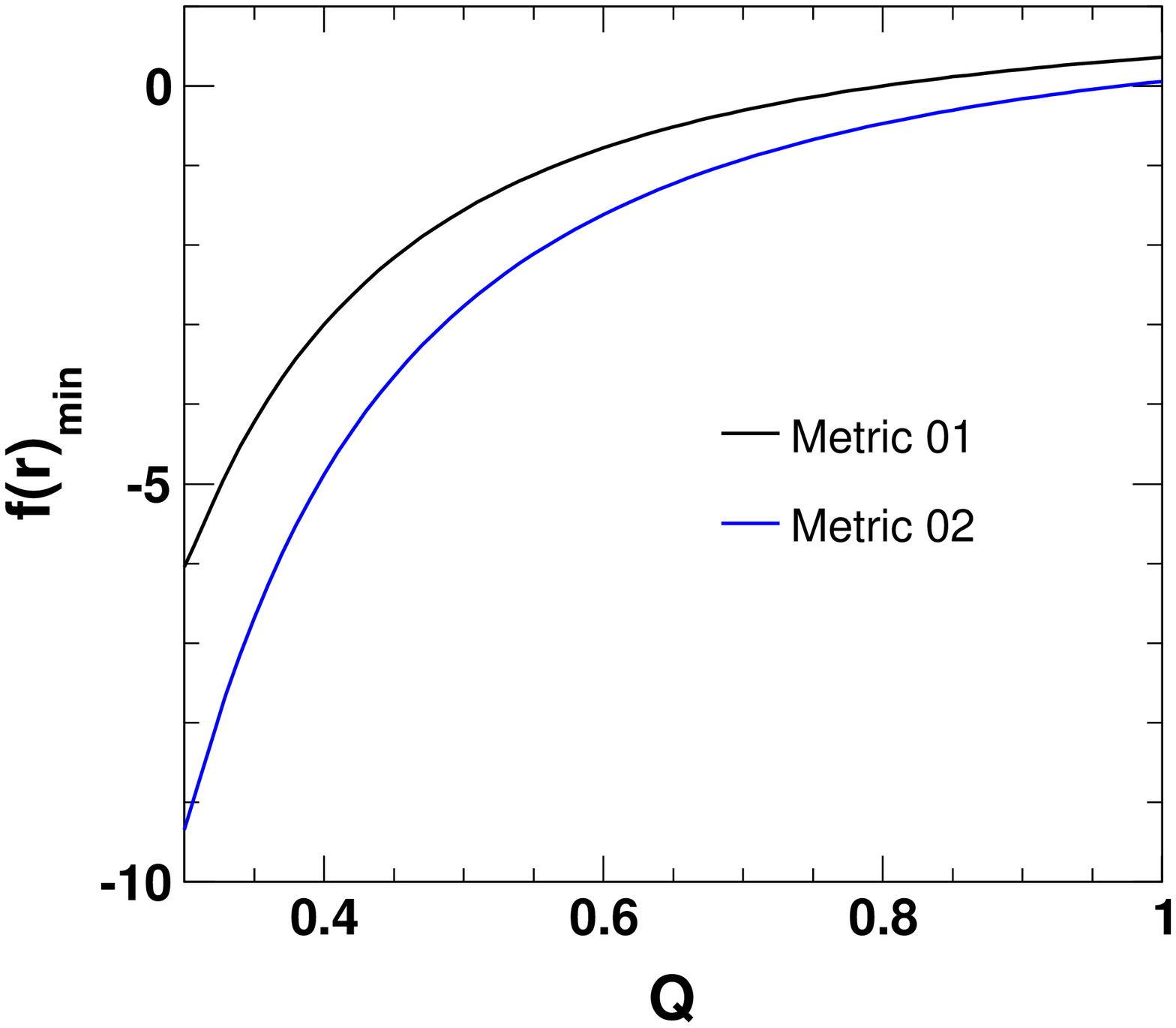}\hspace{0.5cm}
   \includegraphics[scale = 0.3]{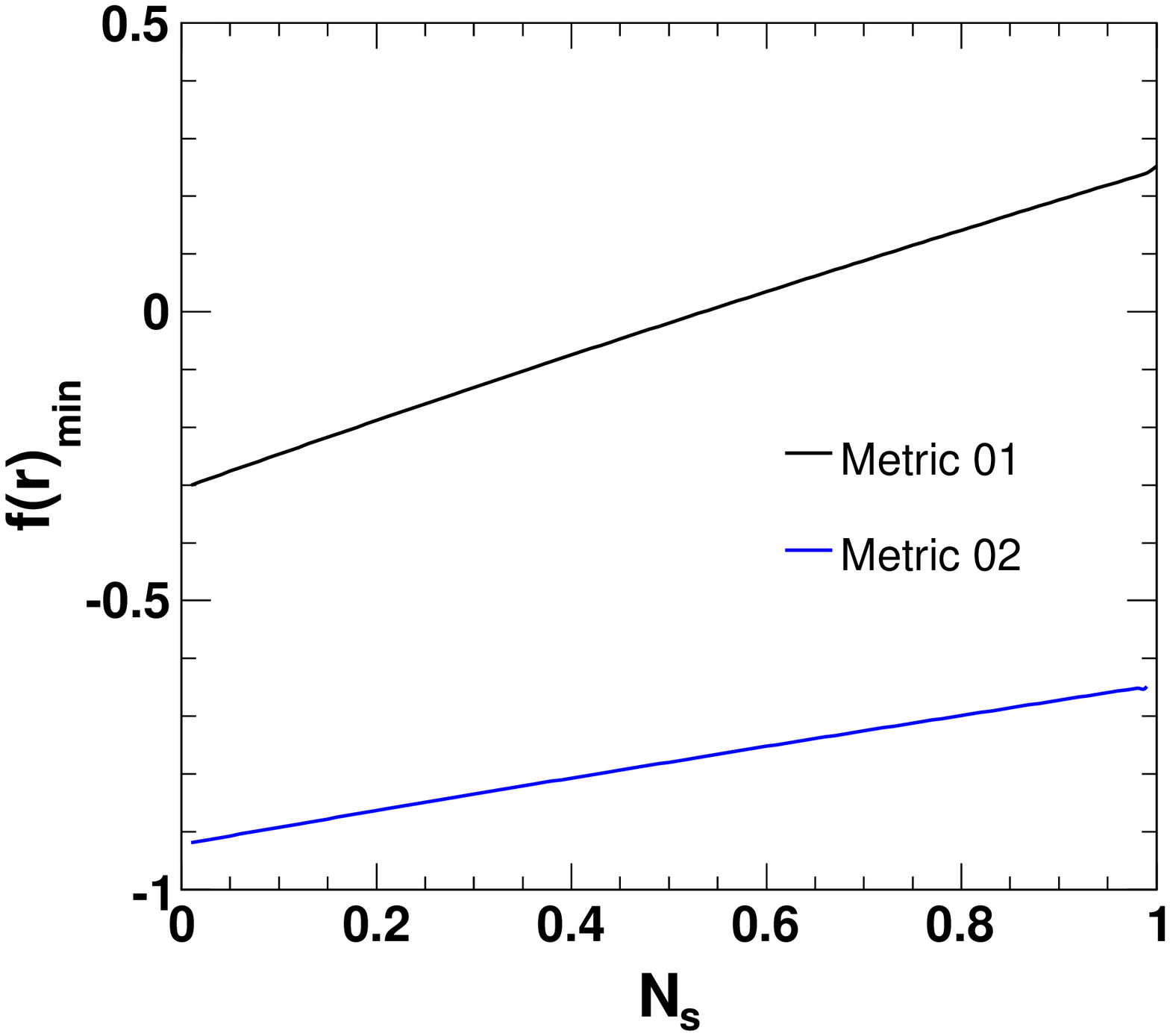}}
\vspace{-0.2cm}
\caption{Variation of the minimum value of the metric function with respect to
$Q$ and $N_s$ for the black holes surrounded by quintessence field with 
$\lambda=0.01$, $\kappa = 1$ and $M = 0.8$. Here the Metric 01 is defined by 
Eq.~\eqref{metric01} and the Metric 02 is defined by Eq.~\eqref{metric03}. The
left plot is for $N_s = 0.001$ and the right plot is for $Q = 0.7$.}
\label{fig02}
\end{figure}

Now for the function $f(r)$ given in Eq.~(\ref{f1}), the metric (\ref{metric}) 
takes the form: 
\begin{equation}\label{metric01}
ds^2=-\left(1-\frac{2M}{r}
+\frac{Q^2}{r^2}+\frac{N_s}{r^{\frac{1+3\omega_s -6\kappa\lambda(1+\omega_s)}{1-3\kappa\lambda(1+\omega_s)}}}\right)dt^2
+\frac{dr^2}{1-\frac{2M}{r}+\frac{Q^2}{r^2}
+\frac{N_s}{r^{\frac{1+3\omega_s -6\kappa\lambda(1+\omega_s)}{1-3\kappa\lambda(1+\omega_s)}}}}
+r^2 d\Omega^2.
\end{equation}
The Ricci scalar for the above metric is given by
\begin{equation} \label{ricci_m01}
R = \frac{3 N_s (3 \omega_s -1) \left[\kappa  \lambda  (\omega_s +1)-\omega_s \right] r^{-\frac{3 (\omega_s +1) (4 \kappa  \lambda -1)}{3 \kappa  \lambda  (\omega_s +1)-1}}}{\left[1-3 \kappa  \lambda  (\omega_s +1)\right]^2},
\end{equation}
the Ricci squared is given by
\begin{align} \label{riccisq_m01}\notag
R_{\mu\nu}R^{\mu\nu} &= \frac{9 N_s^2 \left[\omega_s -\kappa  \lambda  (\omega_s +1)\right]^2 \left[9 \left(8 \kappa ^2 \lambda ^2 (\omega_s +1)^2-4 \kappa  \lambda  (\omega_s +1)^2+\omega_s ^2\right)+6 \omega_s +5\right] r^{-\frac{6 (\omega_s +1) (4 \kappa  \lambda -1)}{3 \kappa  \lambda  (\omega_s +1)-1}}}{2 \left[1-3 \kappa  \lambda  (\omega_s +1)\right]^4}\\
&+\frac{18 N_s Q^2 (\omega_s +1) (4 \kappa  \lambda -1) \left[\kappa  \lambda  (\omega_s +1)-\omega_s \right] r^{\frac{-24 \kappa  \lambda  (\omega_s +1)+3 \omega_s +7}{3 \kappa  \lambda  (\omega_s +1)-1}}}{\left[1-3 \kappa  \lambda  (\omega_s +1)\right]^2}+\frac{4 Q^4}{r^8},
\end{align}
and the Kretschmann scalar is
\begin{align} \label{kret_m01}\notag
R_{\alpha\beta\mu\nu}R^{\alpha\beta\mu\nu} &= \frac{4 \left(-2 M r+N_s r^{\frac{3 \omega_s -1}{3 \kappa  \lambda  (\omega_s +1)-1}}+Q^2\right)^2}{r^8}\\ \notag
&+\left[-\frac{4 M}{r^3} +\frac{N_s \big(6 \kappa  \lambda  (\omega_s +1)-3 \omega_s -1\big) \big(9 \kappa  \lambda  (\omega_s +1)-3 \omega_s -2\big) r^{-\frac{3 (\omega_s +1) (4 \kappa  \lambda -1)}{3 \kappa  \lambda  (\omega_s +1)-1}}}{\big(1-3 \kappa  \lambda  (\omega_s +1)\big)^2}+\frac{6 Q^2}{r^4}\right]^2\\
&+\frac{4 \left[\frac{2 M}{r^2}+N_s \left(\frac{3 \omega_s -1}{3 \kappa  \lambda  (\omega_s +1)-1}-2\right) r^{\frac{3 \omega_s -1}{3 \kappa  \lambda  (\omega_s +1)-1}-3}-\frac{2 Q^2}{r^3}\right]^2}{r^2}.
\end{align}
From the expressions \eqref{ricci_m01}, \eqref{riccisq_m01} and 
\eqref{kret_m01}, it is seen that the black formed by this metric is singular 
for any value of the structural constant and the Rastall parameter. The 
Kretschmann scalar as well as the Ricci squared is a function of the 
electrodynamic source, while the Ricci scalar is independent of the 
electrodynamic source. In the limit of $\lambda\rightarrow0$ in 
the metric \eqref{metric01}, one can recover the Reissner-Nordstr\"om black 
hole surrounded by a field in GR \cite{Kiselev} as 
\begin{equation}\label{metric02}
ds^2=-\left(1-\frac{2M}{r}+\frac{Q^2}{r^2}+\frac{N_s}{ r^{3\omega_s +1}}  \right)dt^2
+\frac{dr^2}{
1-\frac{2M}{r}+\frac{Q^2}{r^2}
+\frac{N_s}{{r}^{3\omega_s +1}}
}+r^2 d\Omega^2.
\end{equation} 
Again a regular black hole can be formed from the above ansatz by introducing 
a distribution function \cite{Balart}. Hence, we modify the metric 
\eqref{metric01} by introducing a non-linear charge distribution function of 
the type, 
\begin{equation}\label{mass_distribution}
\sigma(r) = \exp\left(-\dfrac{Q^2}{2Mr}\right)
\end{equation}
to form a black hole metric in the following way:
\begin{equation}\label{metric03}
ds^2=-\left(1-\frac{2m(r)}{r}
+\frac{N_s}{r^{\frac{1+3\omega_s -6\kappa\lambda(1+\omega_s)}{1-3\kappa\lambda(1+\omega_s)}}}\right)dt^2
+\frac{dr^2}{1-\frac{2m(r)}{r}
+\frac{N_s}{r^{\frac{1+3\omega_s -6\kappa\lambda(1+\omega_s)}{1-3\kappa\lambda(1+\omega_s)}}}}
+r^2 d\Omega^2,
\end{equation}
where $$m(r) = \dfrac{\sigma(r)}{\sigma_\infty}M.$$
Note that the distribution function $\sigma(r)>0$ and $\sigma'(r)>0$ for 
$r \geq 0$. Moreover, $\sigma(r)/r \rightarrow 0$ as $r \rightarrow 0$. 
$\sigma_\infty$ is a normalisation constant and is equal to $\sigma(r)$ at 
$r \rightarrow \infty$. The metric \eqref{metric03} represents a black 
hole with non-linear charge distribution in Rastall gravity. Here the equation 
of state parameter $\omega_s$ defines the nature of the surrounding field as
mentioned earlier. This metric is obtained directly by modifying the metric
\eqref{metric01}, which is for the linear charge distribution in Rastall 
gravity. To obtain metric \eqref{metric03}, we have replaced the term 
$-\dfrac{2M}{r} + \dfrac{Q^2}{r^2}$ in the metric \eqref{metric01} by the term 
$m(r) = \dfrac{\sigma(r)}{\sigma_\infty}M$ as mentioned above, where 
$\sigma(r)$ is defined by Eq.\ \eqref{mass_distribution}. Later we shall show 
that this metric can be a solution of the field equations in Rastall gravity 
and we shall obtain an analogous expression to Eq.~\eqref{addeq}, which 
satisfies the non-linear charge distribution function. The metric 
\eqref{metric03} basically represents a black hole with an exponential 
non-linear charge distribution function in Rastall gravity. One can see that 
introduction of such a mass and charge coupled non-linear distribution 
function removes the singularity issues arising from the mass term and charge 
term in metric \eqref{metric01}. Thus, if the term containing the structural 
parameter $N_s$ does not create a singularity issue, we can have a regular 
black hole from the metric \eqref{metric03}. We shall show in the next section 
that a wise choice of the surrounding field may give a regular charged black 
hole with this metric. The metric \eqref{metric03} asymptotically approaches 
the metric \eqref{metric01} in Rastall gravity and \eqref{metric02} in GR 
limit. We have already mentioned that the equation of state parameter 
$ \omega_s = - \dfrac{2}{3}$ implies a quintessence field. Now, using 
 this value of $ \omega_s$ in the metrics \eqref{metric01} and 
\eqref{metric03} with linear charge distribution and non-linear charge 
distribution respectively, we can obtain the respective black hole metrics 
surrounded by the quintessence field. We have plotted these metric functions 
for linear charge distribution and non-linear charge distribution for the 
quintessence field in Fig.~\ref{fig01} for different values of $Q$ and $N_s$. 
For the plots with different $Q$ values, we have used the fixed viable value
of $N_s = 0.001$ and for the plots with different $N_s$ values, fixed
value of $Q = 0.7$ has been used.
It is seen that for the metric \eqref{metric01}, the charge $Q \le 0.7$ results 
two horizons of the black hole and the increasing charge reduces the horizons 
to one near $Q = 0.8$. Beyond this the black hole has no horizons. Similarly, 
for the metric \eqref{metric03}, at $Q \le 0.90$ the black hole has two 
horizons and increasing $Q$ to $0.97$ reduces the number of horizons to one, 
and beyond this the black hole has no horizons. In asymptotic regions, for 
small charges, the black holes have two horizons in general. 
It is also seen that the black hole defined by the metric \eqref{metric01} has 
single horizon for the $N_s$ around $0.535$. For $N_s> 0.535$, the black hole 
has no horizons and for $N_s< 0.535$, the black hole has two horizons. Similar 
results are observed for the black hole with non-linear electrodynamic source 
in Rastall gravity. In this case, $N_s$ around $3.850$ results a single horizon. For any $N_s$ values greater than this, the black hole has no horizons and 
for $N_s$ 
values less than this, one can have two horizons for the black hole. From
the pattern of the metric functions for both cases, it is distinct that the 
dependency of the metric functions on $N_s$ is different from that of $Q$. For 
small values of $r$, the metric function shows no significant deviation for 
different structural constant $N_s$, however, for large values of $r$, for 
different $N_s$ the metric function behaves differently. The slope of $f(r)$ 
at large $r$ is smaller for small values of the structural parameter or 
constant $N_s$. Thus, in both cases of metrics, the charge and the structural 
parameter play a very important role for the nature of black holes.  We'll see 
the impact of charges on the quasinormal modes of black holes in the next 
sections in details. Another observation from Fig.~\ref{fig01} is that the 
minimum value of the function $f(r)$ changes with the charge of the black 
hole and the structural parameter. To see these variations explicitly, we have 
plotted the \texttt{minimum}$(f(r))$, i.e. $f(r)_{min}$ with respect to the 
charge $Q$ and structural parameter $N_{s}$ in Fig.~\ref{fig02}. In both 
cases, we see that the magnitude of $f(r)_{min}$ is more for the metric 
\eqref{metric03}. $f(r)_{min}$ shows non-linear variation with the charge $Q$ 
and for the smaller values of charge $Q$ this variation is more significant. 
The variation tends to cease when the charge $Q$ approaches $1$. On the other 
hand, $f(r)_{min}$ shows linear variation with $N_{s}$ for both the metrics 
\eqref{metric01} and \eqref{metric03}. But due to the presence of the 
distribution function \eqref{mass_distribution} in the metric \eqref{metric03}, the 
value of $d f(r)_{min}/dN_s$ for the metric \eqref{metric03} is smaller 
than that for the metric \eqref{metric01}. We have obtained all these results
corresponding to a viable set of common parameters $\lambda=0.01$, 
$\kappa = 1$ and $M = 0.8$.

For a black hole solution having mass function given by 
\eqref{mass_distribution}, it is possible to get the electric field from the 
field Eq.~\eqref{E1}, if we define the energy momentum tensor $T_{\mu\nu}$ as
\begin{equation}
T_{\mu\nu} = L(F) g_{\mu\nu} - L_F F_{\mu \lambda} F^\lambda_\nu + {\mathcal{T}^{\mu}}_{\nu},
\end{equation}
where $L(F)$ is an arbitrary function of $F = \dfrac{1}{4} F^{\mu\nu}F_{\mu\nu}$ (an Lorentz invariant) and $L_F = \dfrac{dL}{dF}$. Restricting the electric 
field by considering 
$F_{\mu\nu} = E(r)(\delta^0_\mu \delta^1_\nu - \delta^1_\mu \delta^0_\nu)$, 
we can express the components of the field equation as
\begin{eqnarray}\label{Feq_02}
&&{\Theta^{0}}_{0}={G^{0}}_{0}+\kappa\lambda R=\frac{1}{r^2}\big[f^{\prime}(r)r + f(r) - 1 \big]+\kappa\lambda R = \kappa \big[ L(F) + E^2 L_F - \rho(r) \big],\nonumber\\
&&{\Theta^{1}}_{1}={G^{1}}_{1}+\kappa\lambda R=\frac{1}{r^2}\big[f^{\prime}(r)r + f(r) -1  \big]+\kappa\lambda R= \kappa \big[ L(F) + E^2 L_F - \rho(r) \big],\nonumber\\
&&{\Theta^{2}}_{2}={G^{2}}_{2}+\kappa\lambda R=\frac{1}{r^2}\big[rf^{\prime}(r)+\frac{1}{2}r^2 f^{\prime\prime}(r)\big]+\kappa\lambda R = \kappa \big[ L(F) + \dfrac{1}{2} \rho(r) (3 \omega_s + 1) \big],\nonumber\\
&&{\Theta^{3}}_{3}={G^{3}}_{3}+\kappa\lambda R=\frac{1}{r^2}\big[rf^{\prime}(r)+\frac{1}{2}r^2 f^{\prime\prime}(r)\big]+\kappa\lambda R = \kappa \big[ L(F) + \dfrac{1}{2} \rho(r) (3 \omega_s + 1) \big].
\end{eqnarray}
The above mentioned restriction on the electric field confirms that the 
electric field is spherically symmetric, i.e. electric field does not depend on
the $\theta$ and $\phi$ coordinates. This will be shown explicitly in the 
following part. One can also see that the restriction respects the asymmetric 
behaviour of $F_{\mu\nu}$. Further, the electromagnetic field equation 
$\nabla_\mu (F^{\mu \nu} L_F) = 0$ gives,
 \begin{equation}
 E(r) L_F = - \dfrac{Q}{4 \pi r^2}.
 \end{equation}
 Using the above result in Eq.s \eqref{Feq_02} and using the expression 
\eqref{rho} we obtain,
\begin{equation}
 E(r) = \frac{\pi  Q e^{-\frac{Q^2}{2 M r}} \left(8 M r-Q^2\right)}{\kappa  M r^3}.
 \end{equation}
It is seen that the electric field is regular everywhere and asymptotically it 
results,
 \begin{equation}
 E(r) \approx \dfrac{Q}{r^2}.
 \end{equation}
This agrees with the result \eqref{addeq}.
 We have calculated the Ricci scalar for the ansatz \eqref{metric03} as given 
by
 \begin{equation} \label{ricci_m02}
 R = \frac{Q^4 e^{-\frac{Q^2}{2 M r}}}{2 M r^5}+\frac{3 N_s (3 \omega_s -1) \left[\kappa  \lambda  (\omega_s +1)-\omega_s \right] r^{-\frac{3 (\omega_s +1) (4 \kappa  \lambda -1)}{3 \kappa  \lambda  (\omega_s +1)-1}}}{\left[1-3 \kappa  \lambda  (\omega_s +1)\right]^2},
 \end{equation}
 the Ricci squared is
 \begin{align} \label{riccisq_m02}
 R_{\mu\nu} R^{\mu\nu} = \frac{e^{-\frac{Q^2}{M r}} \left[4 A M Q^4 r^2+B+16 M^2 N_s \left(P^2-P-2\right) Q^2 r^{P+4} e^{\frac{Q^2}{2 M r}}-8 M Q^6 r+Q^8\right]}{8 M^2 r^{10}},
 \end{align}
 where $A=8 M-N_s P (P+1) r^{P+1} e^{\frac{Q^2}{2 M r}}$,
 $ B=4 M^2 N_s^2 (P+1)^2 \left(P^2+4\right) r^{2 P+6} e^{\frac{Q^2}{M r}}$ and $ P = \frac{3 \omega_s -1}{3 \kappa  \lambda  (\omega_s +1)-1}-2,$
and finally the Kretschmann scalar is found to be,
\begin{align} \label{kret_m02}
R_{\alpha\beta\mu\nu}R^{\alpha\beta\mu\nu} &= \\ \notag 
&\left[\frac{N_s \big(6 \kappa  \lambda  (\omega_s +1)-3 \omega_s -1\big) \big(9 \kappa  \lambda  (\omega_s +1)-3 \omega_s -2\big) r^{-\frac{3 (\omega_s +1) (4 \kappa  \lambda -1)}{3 \kappa  \lambda  (\omega_s +1)-1}}}{\left[1-3 \kappa  \lambda  (\omega_s +1)\right]^2}\right.\\ \notag 
&\left.-\frac{e^{-\frac{Q^2}{2 M r}} \left(8 M^2 r^2-8 M Q^2 r+Q^4\right)}{2 M r^5}\right]^2  
+\frac{4 \left[N_s r^{\frac{3 \omega_s -1}{3 \kappa  \lambda  (\omega_s +1)-1}-2}-\frac{2 M e^{-\frac{Q^2}{2 M r}}}{r}\right]^2}{r^4}\\
&+\frac{4 \left[\frac{e^{-\frac{Q^2}{2 M r}} \left(2 M r-Q^2\right)}{r^3}+\frac{N_s \big(-6 \kappa  \lambda  (\omega_s +1)+3 \omega_s +1\big) r^{\frac{-9 \kappa  \lambda  (\omega_s +1)+3 \omega_s +2}{3 \kappa  \lambda  (\omega_s +1)-1}}}{3 \kappa  \lambda  (\omega_s +1)-1}\right]^2}{r^2}.
\end{align}
From the expressions \eqref{ricci_m02}, \eqref{riccisq_m02} and 
\eqref{kret_m02} it is seen that for some selected values of $\omega_s$, 
$\lambda$ and $N_s$ it is possible to get a regular black hole from the metric 
\eqref{metric03}. Thus we have obtained regular black hole solutions with 
non-linear electrodynamic sources in Rastall gravity. In the next part of the 
work, we'll see the situations explicitly in which we can obtain regular 
black hole solutions for the ansatz and how the quasinormal modes of the 
corresponding black holes change. 

\section{Quasinormal Modes of Black holes in Rastall Gravity} \label{section5}
In this section, we would like to study the quasinormal modes of two types 
black holes viz., one defined by the metric \eqref{metric01} and the other, a 
black hole with non-linear electrodynamic sources, defined by another metric 
\eqref{metric03}. Here we consider 5 types of surrounding fields viz., dust 
field, radiation field, quintessence field, cosmological field and phantom 
field, and calculate the quasinormal modes for each case. At first, we perturb 
the black hole with some probe minimally coupled to a scalar field the having 
the equation of motion,
\begin{equation}\label{scalar_eq01}
\dfrac{1}{\sqrt{-g}} \partial_\alpha(\sqrt{-g} g^{\alpha\beta} \partial_\beta) \Phi^m_l = \mu^2 \Phi^m_l,
\end{equation}
here $\mu^2$ is the mass square of the scalar field. It is possible and 
convenient to express the scalar field $\Phi$ in the form of spherical waves
as
\begin{equation}\label{phi_expression}
\Phi^m_l(t, r, \theta, \phi) = e^{-i \omega t} \dfrac{\psi(r)}{r} Y^m_l(\theta, \phi),
\end{equation}
where $\psi(r)$ is the radial part, $Y^m_l(\theta, \phi)$ is the well known 
spherical harmonics and $\omega$ is the frequency of oscillations of time part
of the wave. This $\omega$ will correspond to the quasinormal mode frequencies 
of black hole solutions as will be clear from the below. Using 
Eq.~\eqref{phi_expression}, we can transform 
Eq.~\eqref{scalar_eq01} into a Schr\"{o}dinger like equation, given by
\begin{equation}\label{scalar_eq02}
\dfrac{d^2 \psi}{dx^2} + \left[\omega^2 - V_l(x)\right] \psi = 0,
\end{equation}
where $x$ is defined as
\begin{equation}
x = \int \dfrac{dr}{f(r)},
\end{equation}
which is known as tortoise coordinate. In Eq.~\eqref{scalar_eq02}, the 
effective potential is 
\begin{equation}\label{gen_pot}
V_l(r) = f(r) \left\lbrace \dfrac{f'(r)}{r} + \dfrac{l(l+1)}{r^2} + \mu^2 \right \rbrace.
\end{equation}
In this study, we'll be interested in the massless scalar field only. Thus 
throughout this whole study, we'll set $\mu = 0$. For physical consistency, 
appropriate boundary conditions should be applied to Eq.~\eqref{scalar_eq02} 
both at the horizon of the black hole and infinity. For asymptotically flat 
spacetimes, the quasinormal conditions are \cite{Ferrari, Vishveshwara},
\begin{equation}
\psi(x) \rightarrow \begin{cases} A e^{+i \omega x} \;\;\; \text{if} \;\; x \rightarrow -\infty\\
B e^{-i \omega x} \;\;\; \text{if} \;\; x \rightarrow +\infty \end{cases},
\end{equation}
where $A$ and $B$ are constants. The purely ingoing wave physically expresses 
that nothing can escape from the horizon of the black hole. On the other hand, 
the purely outgoing wave expresses that no radiation comes from the infinity. 
These two expressions precisely give the quasinormal requirement or condition 
which ensures the existence of infinite set of discrete complex numbers, known 
as quasinormal modes. For a normal Schwarzschild black hole, omitting 
electric charge, the quasinormal modes basically depend on the mass M only. 
This is in accordance with the no hair theorem \cite{Carter1971, Israel1967} 
and it is a significant outcome of GR. This theorem simply states that the 
black holes are simple objects and a few observable properties, such as mass,
electric charge and angular momentum can completely describe them. In other 
words, the quasinormal modes of a charge-less black hole in GR can be simply 
characterized by mass $M$ and overtone $l$ of the black hole \cite{Bustillo, 
Gurlebeck}. But for the black holes defined by the metric 
\eqref{metric01} and \eqref{metric03}, it is seen that, apart from mass $M$, 
the parameters $N_s$, $Q$ and $\lambda$ affect the quasinormal modes 
significantly. In this work we 
have used 5th order and 6th order WKB approximation \cite{Kono2003} to 
estimate the quasinormal frequencies. 

\begin{figure}[htb]
\centerline{
   \includegraphics[scale = 0.3]{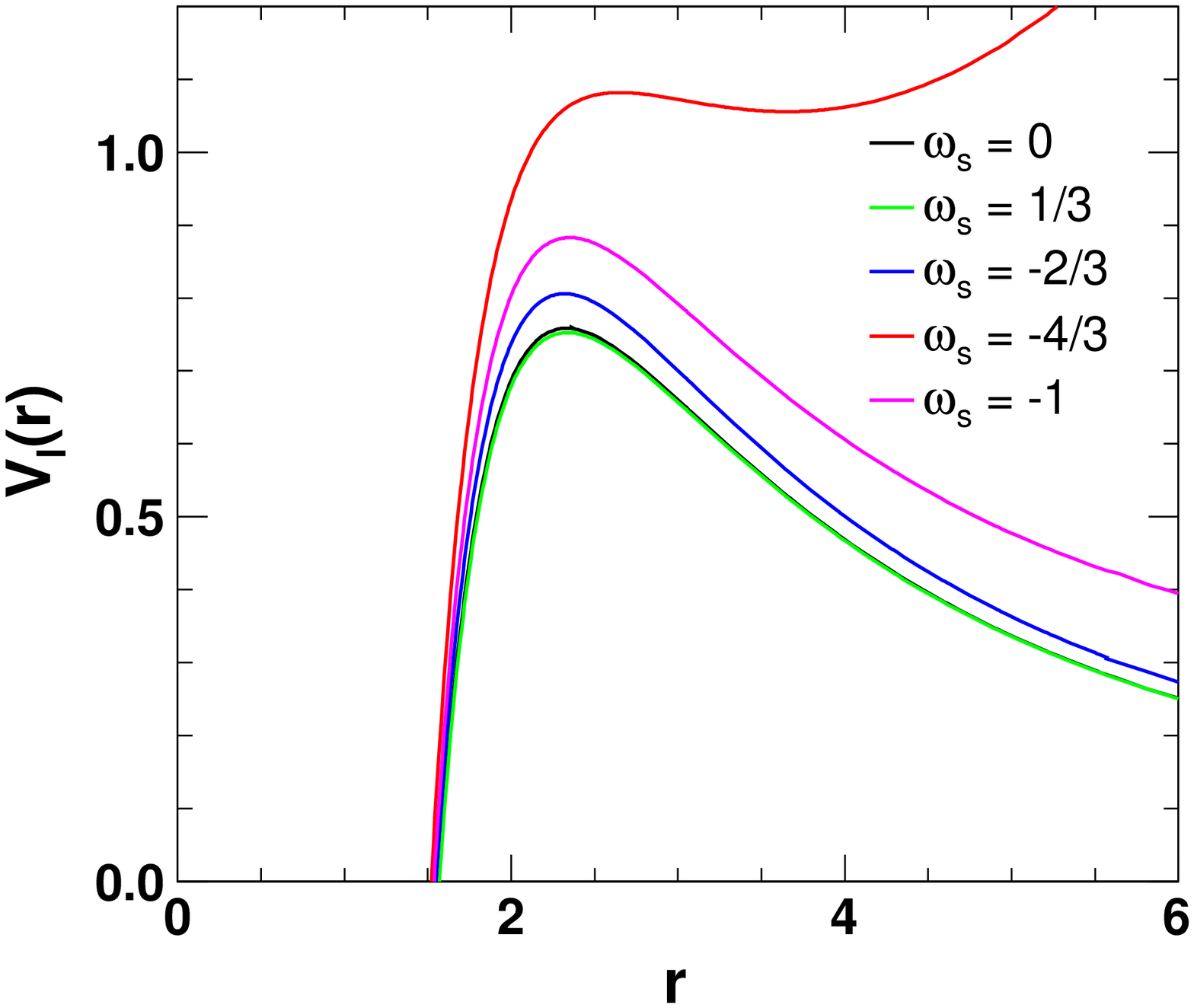}\hspace{0.5cm}
   \includegraphics[scale = 0.3]{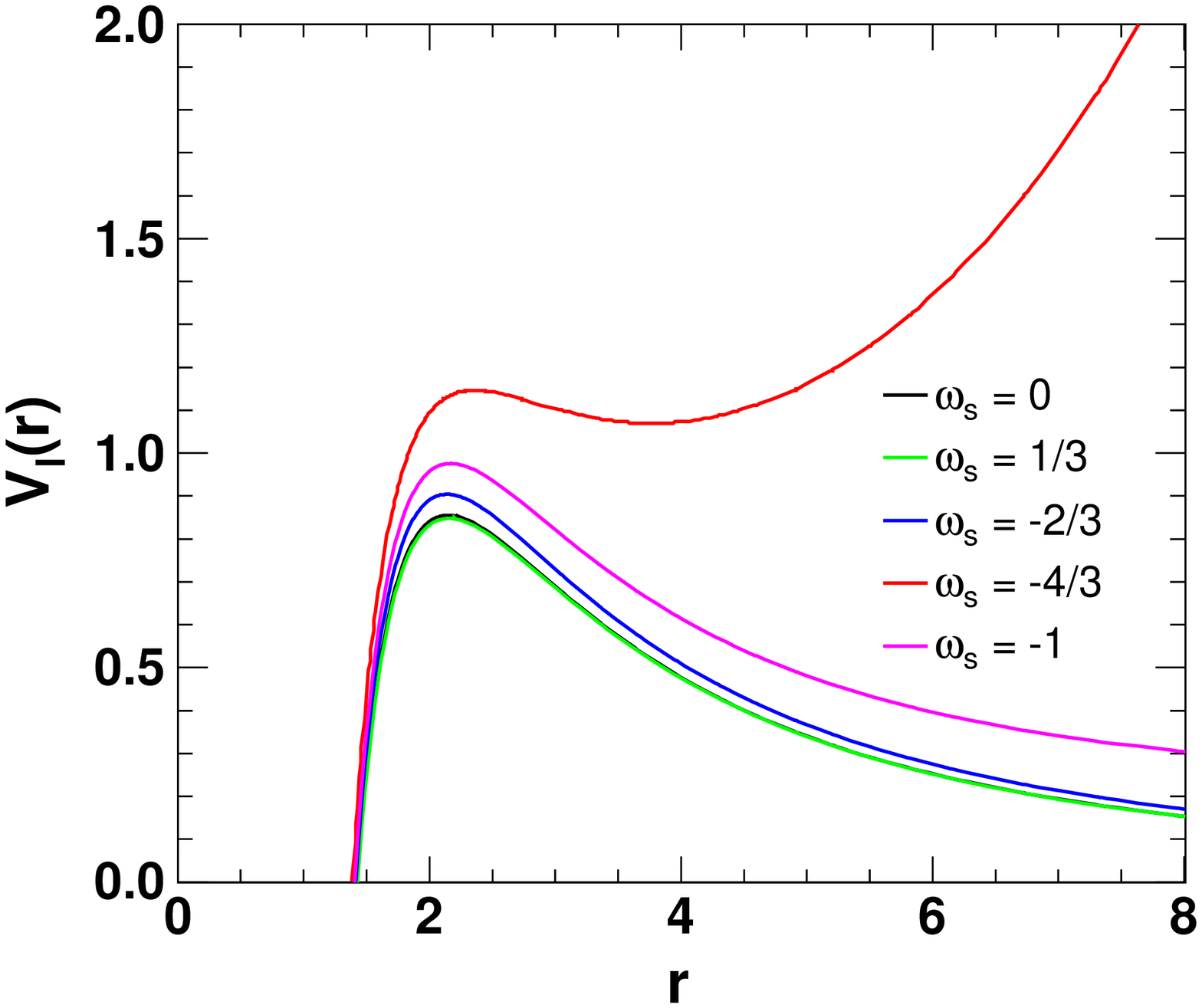}}
\vspace{-0.2cm}
\caption{Behaviour of the potential $V_l(r)$ w.\ r.\ t.\ $r$ for black holes 
defined by the metric \eqref{metric01} (left plot) and metric \eqref{metric03} 
(right plot) with $l=3$, $Q=0.5$, $\lambda=0.01$, $\kappa = 1$, $N_s = 0.01$ 
and $M = 0.8$ for different surrounding fields.}
\label{fig03}
\end{figure}

The potential defined in Eq.~\eqref{gen_pot} behaves differently for 
different surrounding fields of the black hole. To visualize this we have 
plotted the potential for different surrounding fields in Fig.~\ref{fig03} 
for both the black holes in Rastall gravity (defined by metrics 
\eqref{metric01} and \eqref{metric03}). It is seen that for the $r<2$ 
region, the behaviours of both the black holes are similar. The potential 
increases significantly with decrease in the value of $\omega_s$. For the 
phantom field the potential is having maximum value. For the quintessence 
field with $\omega_s = - 2/3$, the potential decreases further and finally 
for the radiation field, the potential is having minimum values.

\begin{figure}[htb]
   \centerline{
   \includegraphics[scale = 0.3]{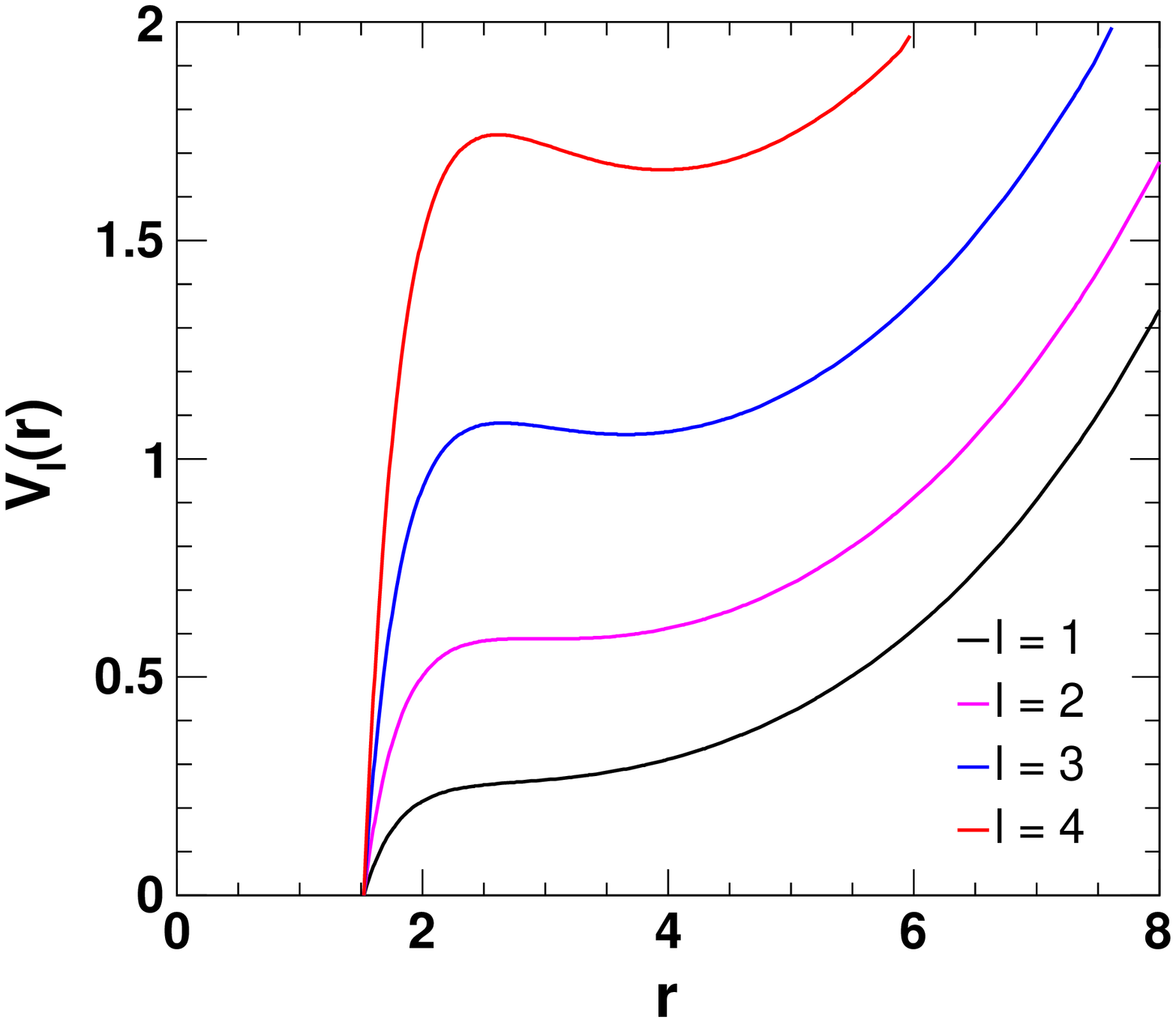}\hspace{0.5cm}
   \includegraphics[scale = 0.3]{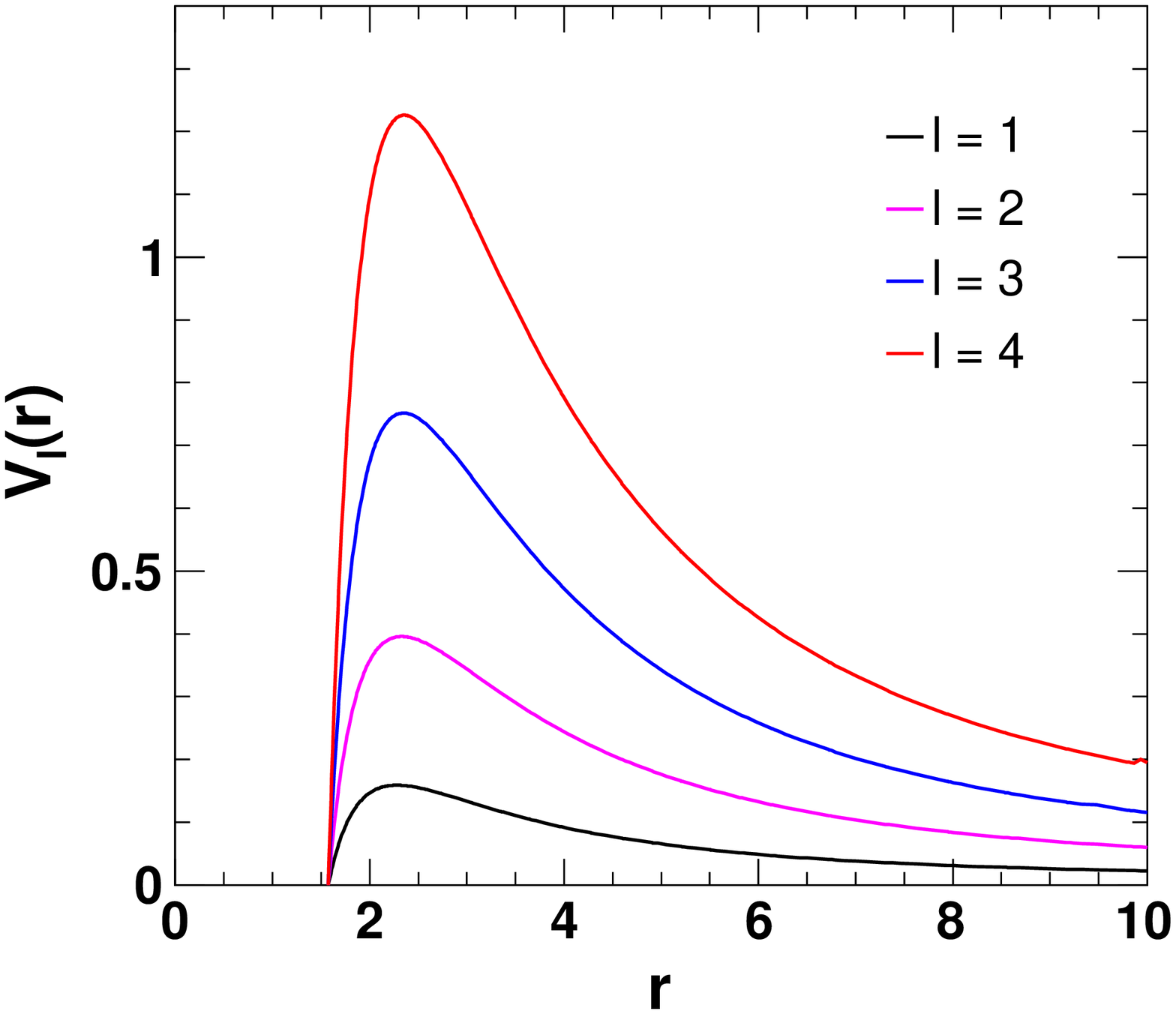}}
\vspace{-0.2cm}
\caption{Variation of the potential $V_l(r)$ w.\ r.\ t.\ $r$ for black holes 
defined by the metric \eqref{metric03} surrounded by phantom field with 
$Q=0.2$, $\lambda=0.01$, $\kappa = 1$, $M = 0.8$ and $N_s = 0.01$ (left plot) 
and $N_s = 0.0001$ (right plot).}
\label{fig04}
\end{figure}

\begin{figure}[htb]
\centerline{
   \includegraphics[scale = 0.3]{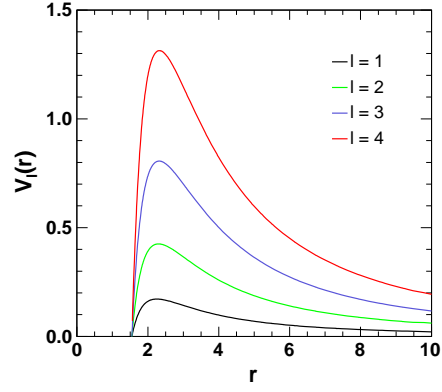}}
\vspace{-0.2cm}
\caption{Variation of the potential $V_l(r)$ w.\ r.\ t.\ $r$ for black holes 
defined by the metric \eqref{metric03} surrounded by quintessence field with 
$Q=0.2$, $\lambda=0.01$, $\kappa = 1$, $M = 0.8$ and $N_S = 0.01$.}
\label{fig05}
\end{figure}

The variation of the potential with $l$ is shown in Fig.s \ref{fig04} and 
\ref{fig05} for phantom field and quintessence field respectively for the 
black hole with non-linear electrodynamic source in Rastall gravity. The 
Fig.~\ref{fig04} shows that with higher values of the structural constant 
$N_s$ in phantom field, the potential increases further after crossing the 
first bump (see the left plot). This anomalous behaviour can be diminished by 
decreasing the structural constant to a comparatively smaller value as shown
in the right plot of this figure. Since all other fields behave similar to
the quintessence field as clear from the Fig.~\ref{fig03}, so the behaviour
of the potential with $l$ would be similar to Fig.~\ref{fig05} for all other 
fields as in the case of the quintessence field. 

\begin{figure}[htb]
\centerline{
   \includegraphics[scale = 0.3]{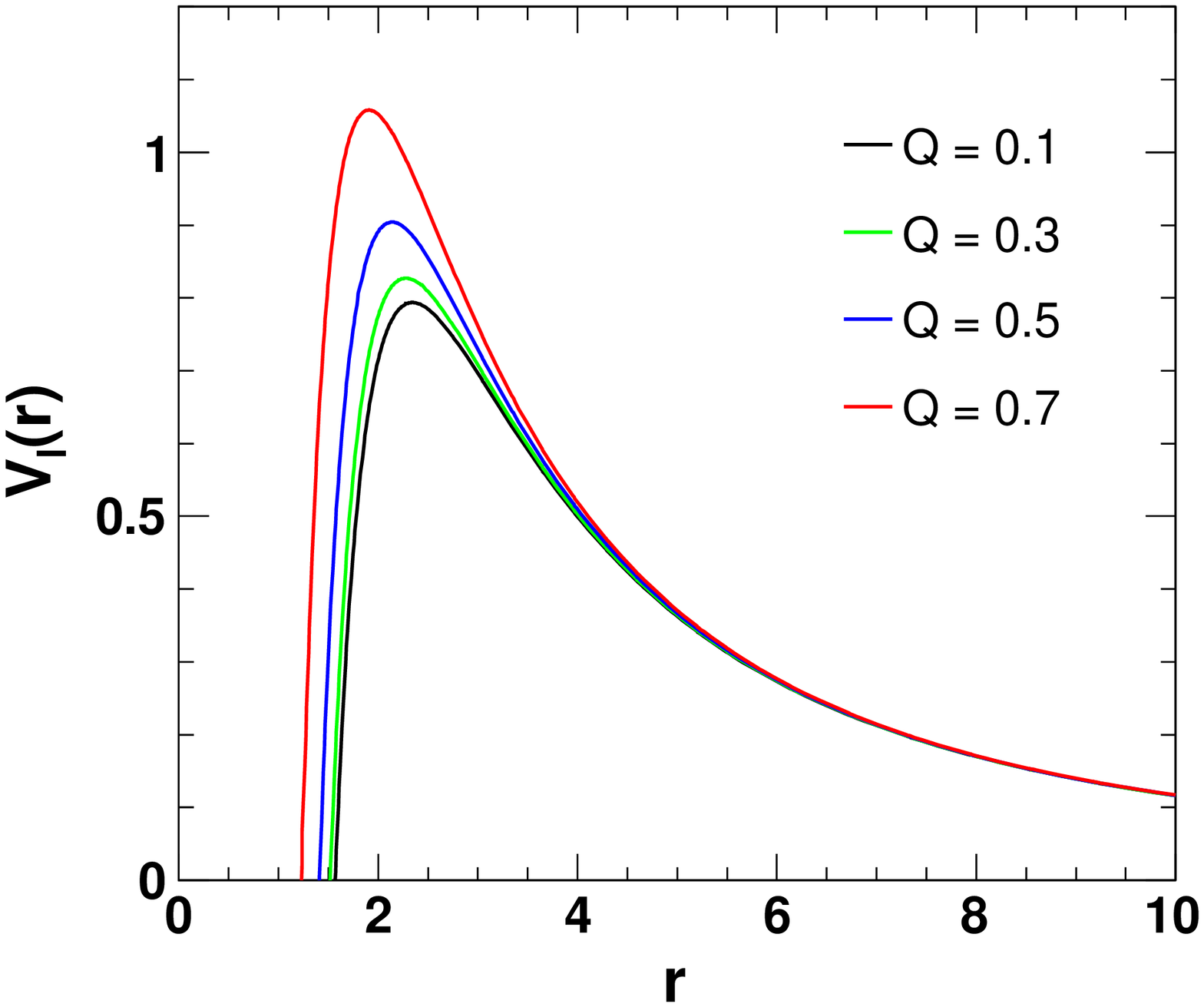}\hspace{0.5cm}
   \includegraphics[scale = 0.3]{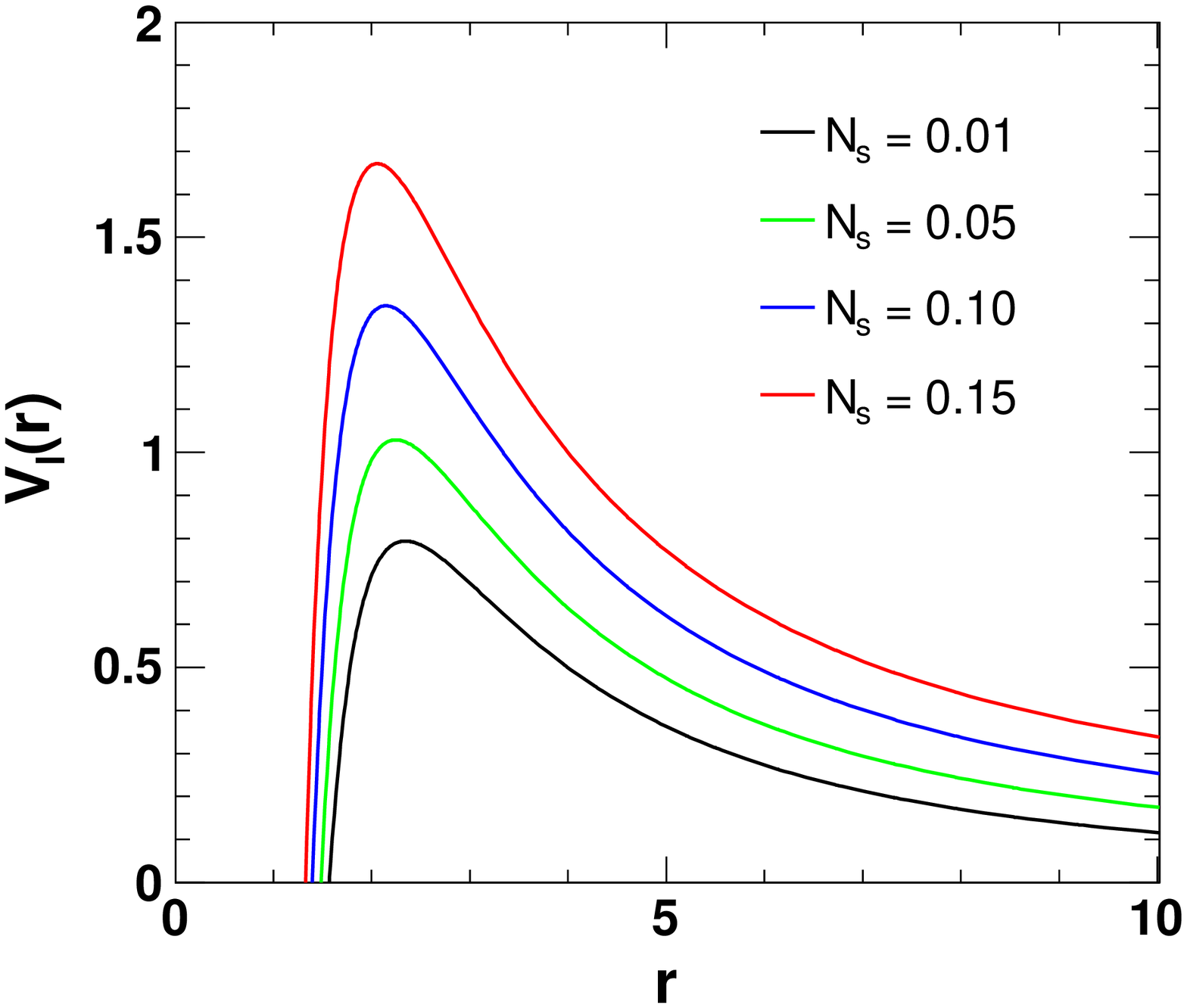}}
\vspace{-0.2cm}
\caption{Behaviour of the potential $V_l(r)$ w.\ r.\ t.\ $r$ for black holes 
defined by the metric \eqref{metric03} surrounded by quintessence field with 
$l=3$, $\lambda=0.01$, $\kappa = 1$ and $M = 0.8$. For the plot on left 
$N_s = 0.01$ and for the plot on right $ Q = 0.1$.}
\label{fig06}
\end{figure}

To see the impacts of charge $Q$ and the structural parameter $N_s$ 
over the potential, we have further plotted the potential $V_l$ versus $r$ 
for different values of $Q$ and $N_s$ (see Fig.~\ref{fig06}) for the 
quintessence field. The potential increases for increase in the values of 
$Q$ and $N_s$ and these results are in agreement with Fig.~\ref{fig02}.
These behaviours of $V_l$ will be similar for all other fields excluding the 
phantom field. However, for very small values of $N_s$ the phantom field 
potential will behave in a similar fashion.

To see the dependency of the quasinormal modes on the surrounding fields, we 
have studied the corresponding black holes explicitly as follows.

\subsection{Surrounded by cosmological constant field}
\begin{figure}[htb]
\centerline{
   \includegraphics[scale = 0.3]{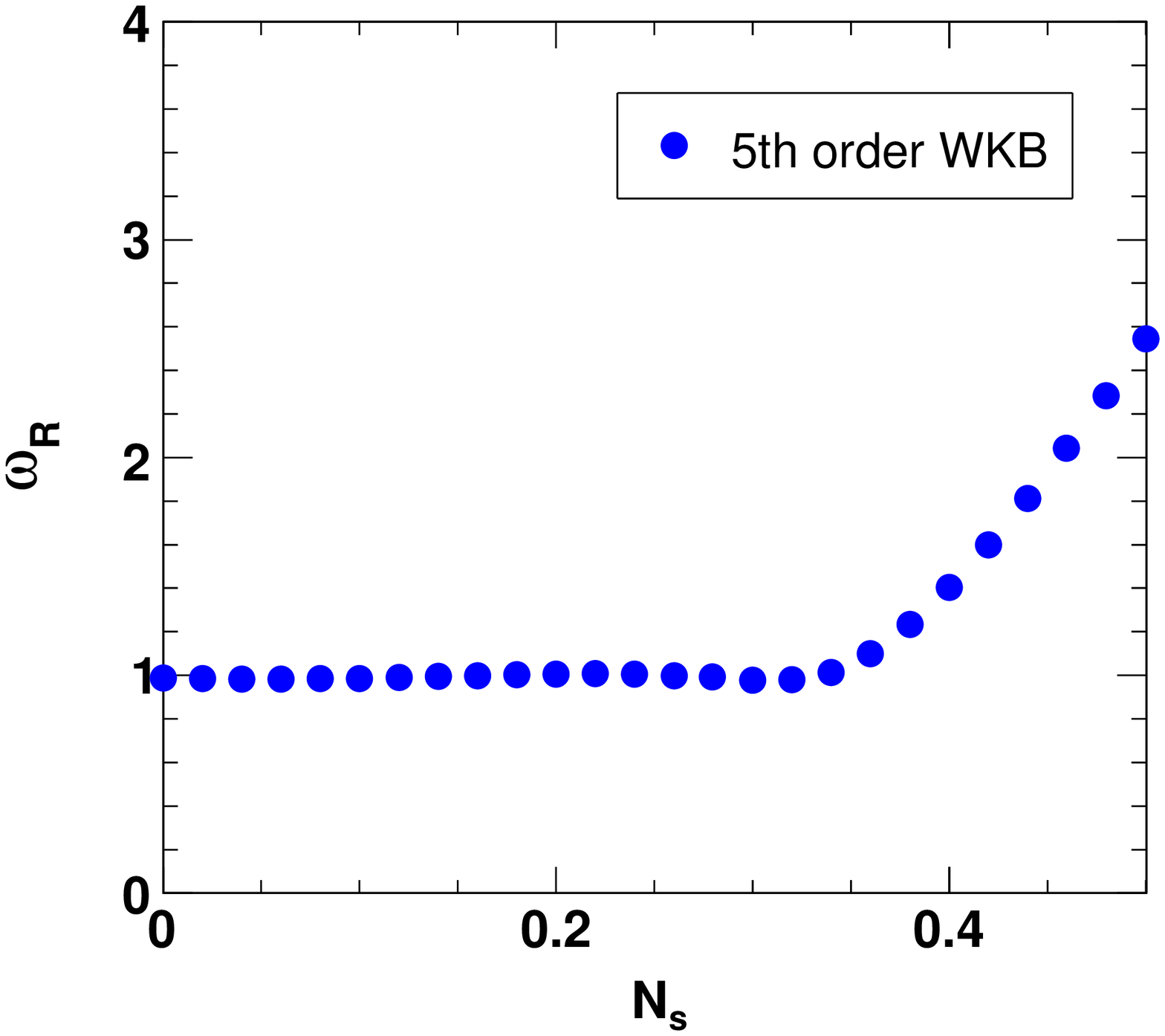}\hspace{0.5cm}
   \includegraphics[scale = 0.3]{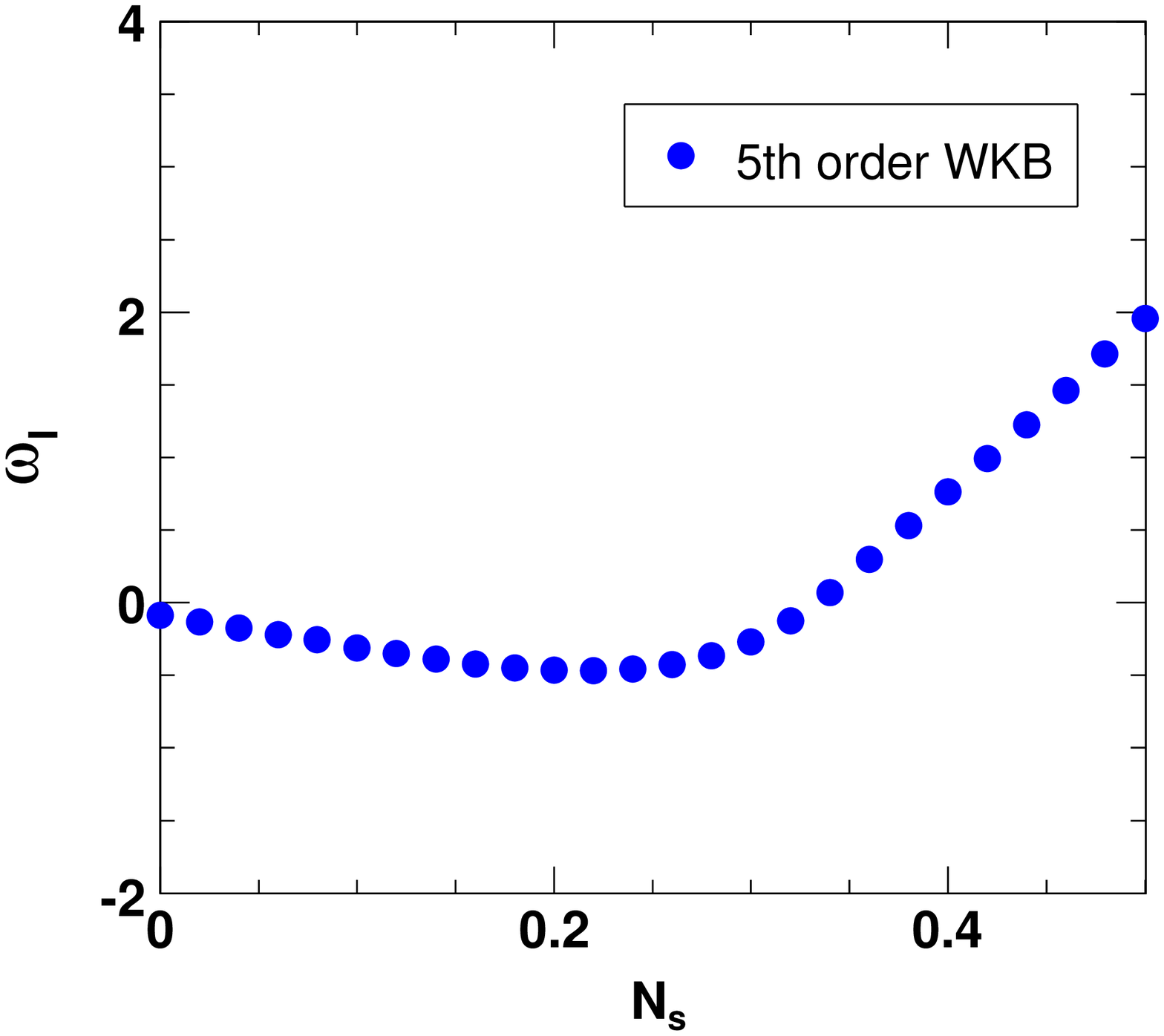}}
\vspace{-0.2cm}
\caption{Fundamental quasinormal mode frequencies in terms of $N_s$ for the 
black holes defined by the metric \eqref{Cosmological_metric03} surrounded by 
cosmological constant field with $l=4$, $Q = 0.2$ and $M = 1$ obtained by 
using the 5th order WKB approximation method.}
\label{figCNvary}
\end{figure}

\begin{figure}[htb]
\centerline{
   \includegraphics[scale = 0.3]{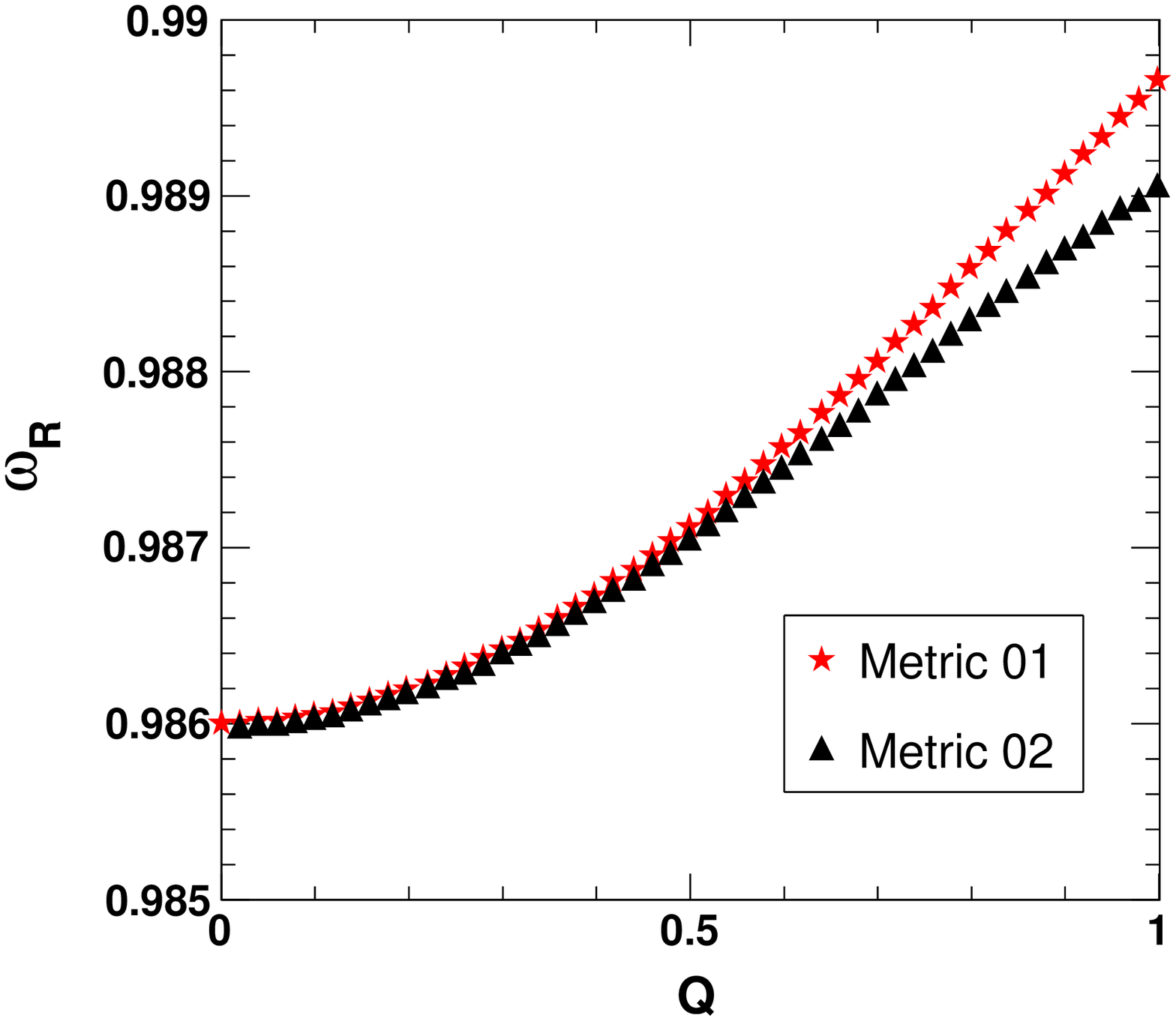}\hspace{0.5cm}
   \includegraphics[scale = 0.3]{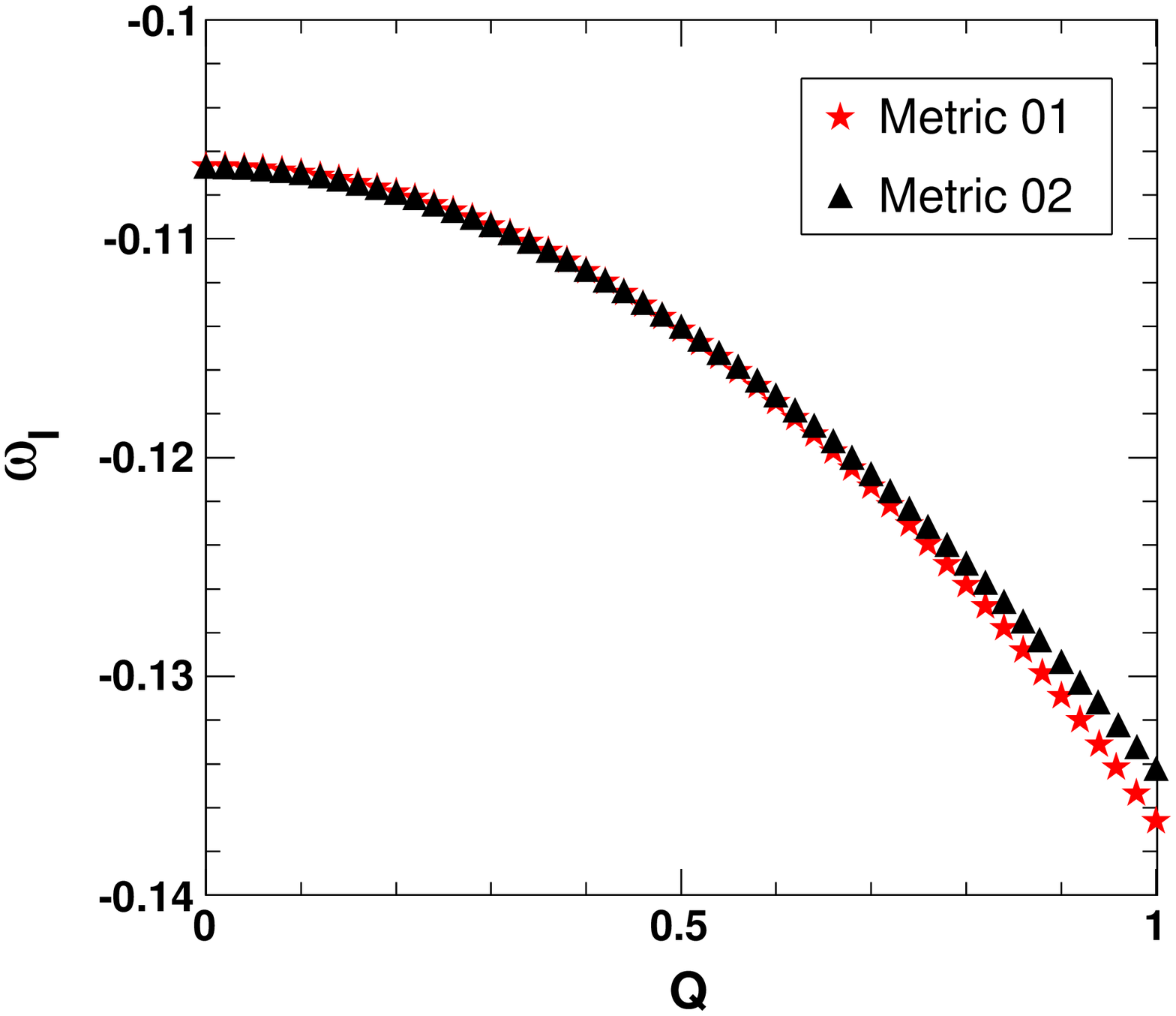}}
\vspace{-0.2cm}
\caption{Fundamental quasinormal mode frequencies in terms of $Q$ for the 
black holes defined by the metric \eqref{Cosmological_metric01} (Metric 01) and 
\eqref{Cosmological_metric03} (Metric 02) surrounded by cosmological constant 
field with $l=4$, $N_s = 0.01$ and $M = 1$ obtained by using 5th order WKB 
approximation method.}
\label{figCQvary}
\end{figure}

For the cosmological constant field, $\omega_s = -1$, the metrics 
\eqref{metric01} and \eqref{metric02} take the same form, which is  

\begin{equation}\label{Cosmological_metric01}
ds^2=-\left(1-\frac{2 M}{r}+\frac{Q^2}{r^2} + r^2 N_s \right)dt^2
+\frac{dr^2}{1-\frac{2 M}{r}+\frac{Q^2}{r^2} + r^2 N_s}
+r^2 d\Omega^2.
\end{equation}
This is identical to the metric obtained by Kiselev in GR \cite{Kiselev}. It 
implies that the black hole metric behaves identically in both Rastall gravity 
and GR in cosmological constant background. Hence there will be no changes 
in the quasinormal modes in this case in comparison to the case in GR. 
Similarly, for this field the metric \eqref{metric03} with non-linear 
electrodynamic sources takes the following form,
\begin{equation}\label{Cosmological_metric03}
ds^2=-\left(1-\frac{2m(r)}{r} + r^2 N_s\right)dt^2
+\frac{dr^2}{1-\frac{2m(r)}{r} + r^2 N_s}
+r^2 d\Omega^2,
\end{equation}
For this metric, the Ricci scalar given in Eq.~\eqref{ricci_m02} takes the 
form,
\begin{equation}
R = \frac{Q^4 e^{-\frac{Q^2}{2 M r}}}{2 M r^5}-12 N_s,
\end{equation}
which is regular everywhere. Similarly, the Ricci squared and Kretschmann 
scalar are also regular everywhere. Thus the metric 
\eqref{Cosmological_metric03} represents a regular black hole with non-linear 
electrodynamic source. In Table \ref{table01}, we have listed a few 
quasinormal mode frequencies for some selected parameters.
To have an idea of the error associated with the calculations, we have 
calculated the deviations of 6th order WKB approximation from 5th order WKB 
approximation for all the cases. This deviation term is denoted by 
$|\omega_5 - \omega_6|$, where $\omega_5$ represents quasinormal modes 
obtained by using the 5th order WKB approximation and $\omega_6$ represents 
quasinormal modes calculated by using the 6th order WKB approximation method. 
We have also calculated a quantity defined by \cite{Konoplya2019}, 
$$ \vartriangle_k = |\dfrac{\omega_{k+1} - \omega_{k-1}}{2}|,$$ where $k$ 
denotes the order of WKB approximation. This quantity gives a good measurement 
of the error order \cite{Konoplya2019}. In this study, we have used this 
expression to calculate the error $\vartriangle_5$ associated with the 5th 
order WKB approximation. It is seen from the
table that the quasinormal mode frequencies depend on the metric as well
as on the multipole number $l$. This observation is applicable
to quasinormal frequencies for all the cases calculated in this work. Here, in case of 6th order WKB approximation, the real
frequencies given by the metric \eqref{Cosmological_metric03} are slightly 
higher than that given by the metric \eqref{Cosmological_metric01} for
$l<3$. But for the $l=3$ the situation is appeared in the reverse way. 
In case of the 5th order WKB approximation, the real frequencies given 
by the metric \eqref{Cosmological_metric03} are slightly lower than that 
given by the metric \eqref{Cosmological_metric01} for all cases. The anomaly 
in the real frequencies in case of the 6th order WKB approximation can 
be explained in the following way. One can see the Table \ref{table01}, where 
we have calculated the corresponding deviations and errors associated with 
the calculation. It is clear from the table that the errors and deviations 
associated with $l=1$ and $l=2$ are comparatively high and in the case of 
$l=3$, they are lowest. Hence, the reverse pattern of real quasinormal 
frequencies calculated by the 6th order WKB approximation can be an 
implication of the errors. If it is true, then the real frequencies given by 
the metric \eqref{Cosmological_metric03} should be, in general, lower than 
that given by the metric \eqref{Cosmological_metric01}.
To understand the behaviour of variation of quasinormal modes with respect to 
the parameters $Q$ and $N_{s}$, we have plotted both the real and imaginary 
parts of frequencies $\omega_R$ and $\omega_I$ respectively with respect to 
these two parameters in Fig.s \ref{figCNvary} and \ref{figCQvary}. The 
variation of quasinormal modes with respect to $N_{s}$ is almost similar for 
both the metrics defined by \eqref{metric01} and \eqref{metric03} for small 
values of charge $Q$. However, in case of the variation with respect to 
charge $Q$ of the black holes, we notice significant variations of quasinormal 
modes for larger values of $Q$. It is seen that the magnitudes of both 
real and imaginary quasinormal modes are smaller for the regular black hole 
with non-linear charge distribution. This confirms the conclusion made on the 
results obtained in Table \ref{table01} i.e., the real quasinormal modes of 
the non-linearly charged black hole are expected to be lower.
The asymptotic behaviours are identical and it is due to the choice of the 
distribution function as mentioned earlier.

\begin{table}[h!]
\centering
\begin{tabular}{||c|c|c||}
\hline\hline
    & Metric \eqref{Cosmological_metric01} & Metric \eqref{Cosmological_metric03} \\ \hline
$l = 1$ (6th order WKB) & $0.336678 - 0.10427i$ & $0.338903 - 0.10357 i$   \\ 
$l = 1$ (5th order WKB)& $0.339734 - 0.103332i$ & $0.339723 - 0.10332 i$ \\ 
$|\omega_5 - \omega_6|$ & $0.00319671$  & $0.000857263$  \\
$\vartriangle_5$ & $0.00178137$ & $0.000904052$ \\
\hline
$l = 2$ (6th order WKB) & $0.553448 - 0.106688i$ & $0.553569 - 0.106663 i$  \\ 
$l = 2$ (5th order WKB)& $0.55353 - 0.106672i$ & $0.553519 - 0.106673 i$ \\ 
$|\omega_5 - \omega_6|$ &  $0.0000835464$ & $0.0000509902 $  \\
$\vartriangle_5$ & $0.0000425029$ & $0.0000269258 $ \\
\hline
$l = 3$ (6th order WKB) & $0.769435 - 0.107499 i$  & $0.76936 - 0.107509 i$ \\ 
$l = 3$ (5th order WKB)& $0.769444 - 0.107498i$ & $0.76943 - 0.107499 i$ \\
$|\omega_5 - \omega_6|$ &  $9.05539\times10^{-6}$ & $ 0.0000707107$  \\
$\vartriangle_5$ & $5.31507\times10^{-6}$ & $0.000035203 $ \\
\hline \hline
\end{tabular}
\caption{Quasinormal modes of black holes for $n=0$, $M=1$, $\omega_s = -1$ (cosmological constant field), 
$N_s = 0.01$ and $Q = 0.2$.}
\label{table01}
\end{table}

In this case, the geometric parameter defined in Eq.~\eqref{Ws} takes the 
following form \cite{Heydarzade2},
\begin{equation}
\mathcal{W}_{s} = -(1-4 \kappa \lambda).
\end{equation}
Hence, the weak energy condition \eqref{weak_energy} demands $N_s > 0$ for $ 0 \le \kappa \lambda < \dfrac{1}{4}$ and $N_s < 0$ for $\kappa \lambda > \dfrac{1}{4}$.

It is important to mention that both the black holes are independent of the 
Rastall parameter $\lambda$ and hence there will be no changes in the 
quasinormal modes for different values of this parameter.

\subsection{Surrounded by dust field}
For the dust field, $\omega_s = 0$ \cite{Kiselev, Vikman} and hence the metric 
\eqref{metric01} becomes,
\begin{equation} \label{Dust_metric01}
ds^2=-\left(1-\frac{2M}{r}
+\frac{Q^2}{r^2}+\frac{N_s}{r^{\frac{1 - 6\kappa\lambda}{1-3\kappa\lambda}}}\right)dt^2
+\frac{dr^2}{1-\frac{2M}{r}+\frac{Q^2}{r^2}
+\frac{N_s}{r^{\frac{1 - 6\kappa\lambda}{1-3\kappa\lambda}}}}
+r^2 d\Omega^2.
\end{equation}
This metric has a different form in comparison to the similar case in GR 
\cite{Kiselev, Heydarzade}. Thus we expect variation in the quasinormal modes 
for this solution. Similarly for the metric \eqref{metric03}, we have for dust 
field,
\begin{equation} \label{Dust_metric03}
ds^2=-\left(1-\frac{m(r)}{r}
+\frac{N_s}{r^{\frac{1 - 6\kappa\lambda}{1-3\kappa\lambda}}}\right)dt^2
+\frac{dr^2}{1-\frac{2m(r)}{r}
+\frac{N_s}{r^{\frac{1 - 6\kappa\lambda}{1-3\kappa\lambda}}}}
+r^2 d\Omega^2.
\end{equation} 
For the limit $\lambda \rightarrow 0$, the above metric can result black hole 
solutions similar to those found in \cite{Kiselev} with a mass distribution 
function similar to \cite{Balart}. One can easily recover the black hole 
solutions found in \cite{Balart} in the limit $N_s \rightarrow 0$. Another 
interesting point mentioned for the metric \eqref{Dust_metric01} in 
\cite{Heydarzade2} is that, in the limit  $\lambda \rightarrow 0$, the 
structural constant behaves more like a contribution to the mass of the black 
hole, where the modified mass term can be given as $(M-N_s)$. Similarly, in 
this limit, the  other black hole with non-linear electrodynamic source will 
also have a modified mass distribution function given by $(m(r)-N_s)$. In this 
case, the black hole will be a regular black hole for any value of the parameter 
$N_s$. Thus it is seen that the structural parameter or constant $N_s$ can 
mimic the behaviour of mass of the black hole in the limit 
$\lambda \rightarrow 0$.

Here, the geometric parameter $\mathcal{W}_s$ is (Eq.~\eqref{Ws})
\begin{equation}
\mathcal{W}_{s}=-\frac{\kappa\lambda\left(1-4\kappa\lambda\right)}{(1-3\kappa\lambda)^2}.
\end{equation}
The weak energy condition for this type of black holes (for both) are given by, 
$0\leq\kappa\lambda<\frac{1}{4}$ demands $N_s>0$,
and $\kappa\lambda<0~\cup~\kappa\lambda>\frac{1}{4}$, demands $N_s<0$
for the structural constant \cite{Heydarzade2}.

\begin{figure}[htb]
\centerline{
   \includegraphics[scale = 0.3]{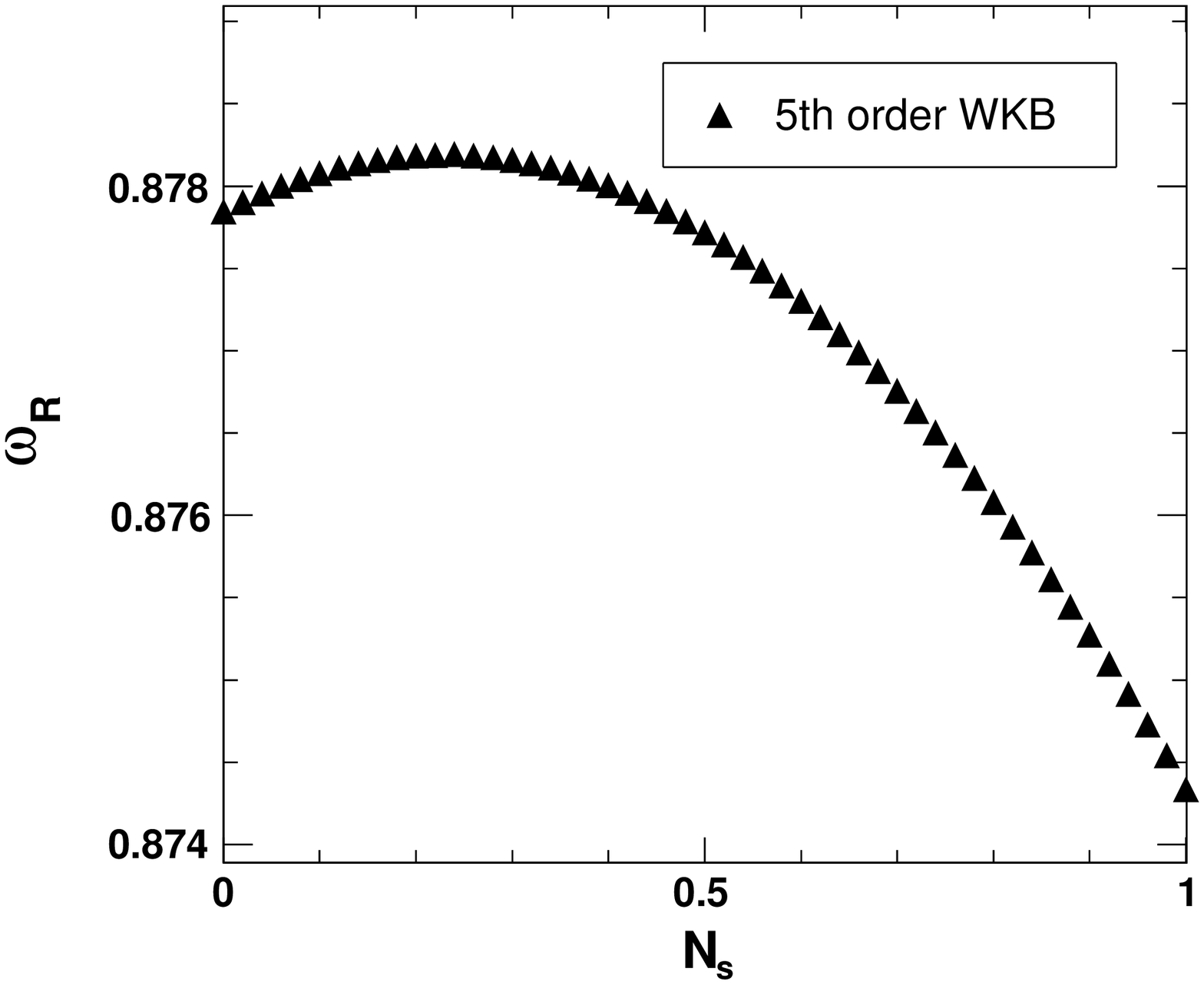}\hspace{0.5cm}
   \includegraphics[scale = 0.3]{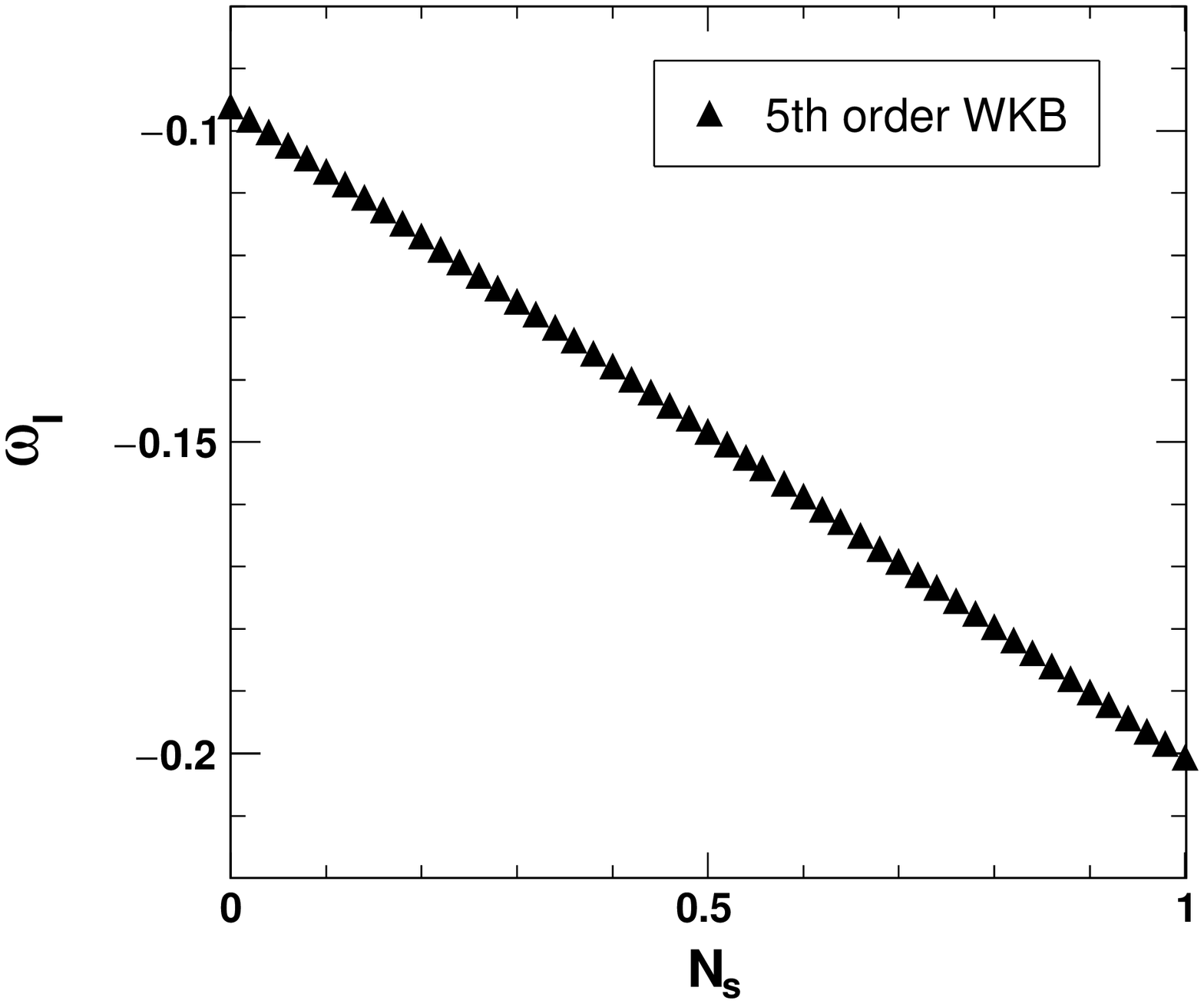}}
\vspace{-0.2cm}
\caption{Fundamental quasinormal mode frequencies w.\ r.\ t.\ $N_s$ for the 
black holes defined by metric \eqref{Dust_metric03} surrounded by dust field 
with $l=4$, $\lambda=0.01$, $\kappa = 1$, $Q = 0.2$ and $M = 1$ obtained by
using the 5th order WKB approximation method.}
\label{figDustNsvary}
\end{figure}

The quasinormal modes for both cases are shown in Table \ref{table02}. The 
metrics \eqref{Dust_metric01} and \eqref{Dust_metric03} give the different 
values of quasinormal modes in comparison to the case of GR. Further, a similar
observation can be made here in relation to $l$ as in the case of the 
cosmological constant field. The deviations of quasinormal modes obtained 
by using the 5th order and 6th order WKB approximations i.e., 
$|\omega_5-\omega_6|$ and error term associated with the 5th order WKB 
approximation $\vartriangle_5$ show that the errors are higher for smaller 
values of $l$ and for $l=3$, the errors are lowest. Hence, the quasinormal 
modes with higher $l$ value, more precisely higher $(l-n)$ values can have 
less errors in WKB approximation. Due to this reason, in the graphical 
analysis part, we have used $l=4$ and $n=0$. Again, the deviations of 
quasinormal modes obtained by using the 6th order WKB approximation of 
non-linearly charged black hole in Rastall gravity from the 
Reissner-Nordstr\"om black hole $\triangle_{NL-GR}$ are higher than that 
corresponding to linearly charged black hole in Rastall gravity from the 
Reissner-Nordstr\"om black hole $\triangle_{L-GR}$. However, in the case of 
5th order WKB approximation both $\triangle_{NL-GR}$ and $\triangle_{L-GR}$ 
are of similar order. From the following part, it can be seen that with higher 
charge $(Q)$ values, these deviations may increase.
\begin{table}[h!]
\centering
\begin{tabular}{||c|c|c|c|c|c||}
\hline \hline
  & Metric \eqref{Dust_metric01} & Metric \eqref{metric02} & Metric \eqref{Dust_metric03} & $\triangle_{NL-GR}$ & $\triangle_{L-GR}$ \\ \hline
$l = 1$ (6th order WKB) & $0.296996 - 0.0983033 i$ &  $0.296422 - 0.098453 i$ & $0.30379 - 0.0960863 i$ & $2.47764\%$  & $ 0.189919\%$\\ 
$l = 1$ (5th order WKB)& $0.296621 - 0.0984277i$ & $0.296569 - 0.0984042i$ & $0.296608 - 0.0984128i$ & $0.0127811\%$ & $ 0.0182622\%$\\
$|\omega_5-\omega_6|$& $0.000395095$ & $0.000154888$ & $0.00754942$ & &\\
$\vartriangle_5$& $0.000259325$ & $0.000185536$ & $0.00377913$ & & \\ \hline
$l = 2$ (6th order WKB) & $0.489542 - 0.0974789 i$ & $0.48941 - 0.0974632 i$ & $0.494892 - 0.0964322 i$ & $1.11781\%$  & $0.0266383\% $\\ 
$l = 2$ (5th order WKB)&$0.48951 - 0.0974851i$ & $0.489425 - 0.0974603i$& $0.4895 - 0.0974943i$ & $0.0165012\%$ & $ 0.0177431\%$\\
$|\omega_5-\omega_6|$& $0.0000325951$  & $0.0000152778$ & $0.00549561$ & & \\
$\vartriangle_5$& $0.0000286051$ & $0.000024683$ & $0.00274767$ & &\\ \hline
$l = 3$ (6th order WKB) & $0.683541 - 0.0972213i$  & $0.68341 - 0.0971983 i$ & $0.682585 - 0.0973599 i$ & $0.121787\%$ & $0.0192679\% $ \\ 
$l = 3$ (5th order WKB)& $0.683532 - 0.0972226i$ &$0.683413 - 0.0971979i$ & $0.683516 - 0.0972274i$ & $0.0155212\%$ & $0.0176066\% $ \\
$|\omega_5-\omega_6|$& $9.0934\times10^{-6}$  & $3.02655\times10^{-6}$ & $0.000940381$ & &\\
$\vartriangle_5$& $7.43303\times10^{-6}$ & $5.87133\times10^{-6}$  & $0.000470279$  & &\\ \hline \hline
\end{tabular}
\caption{Quasinormal modes of black holes for $n=0$, $M=1$, $\omega = 0$ (dust field),
$\kappa \lambda = 0.01$, $N_s = 0.01$ and $Q = 0.2$.}
\label{table02}
\end{table}
To see the variations of quasinormal mode frequencies with respect to the 
structural constant $N_s$ we have plotted the quasinormal frequencies versus 
$N_s$ in Fig.~\ref{figDustNsvary}, where the frequencies are calculated by 
using the 5th order WKB approximation method. Since the asymptotic behaviour of 
both the black holes are identical for smaller charges, we have shown the 
variations for the second black hole with non-linear electrodynamic source 
only. From Fig.~\ref{figDustNsvary} it is clear that the real quasinormal 
frequencies slightly increase for smaller region of $N_s$ (upto near 
$N_s = 0.2$) giving rise to a small bump and beyond $N_s = 0.5$ the frequencies
decrease sharply. However, in case of the imaginary quasinormal frequencies, 
no such bump is observed and the frequencies fall off linearly with increasing
values of $N_s$. It is important to mention that the magnitude of imaginary 
frequencies increase more rapidly in comparison to the real frequencies. To be 
precise, an increase in the parameter $N_s$ introduces more variations in the 
imaginary quasinormal frequencies.

\begin{figure}[htb]
\centerline{
   \includegraphics[scale = 0.3]{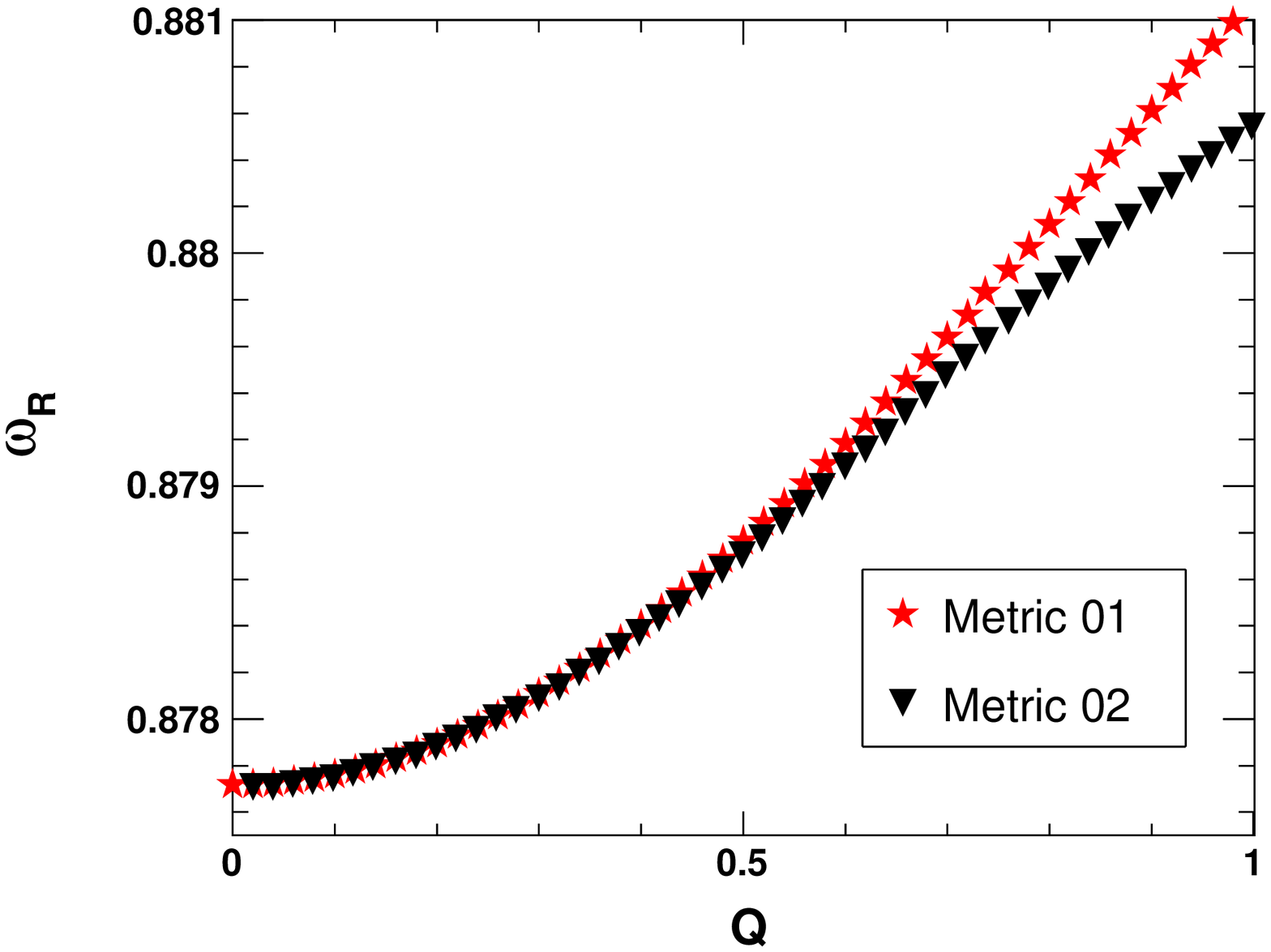}\hspace{0.5cm}
   \includegraphics[scale = 0.3]{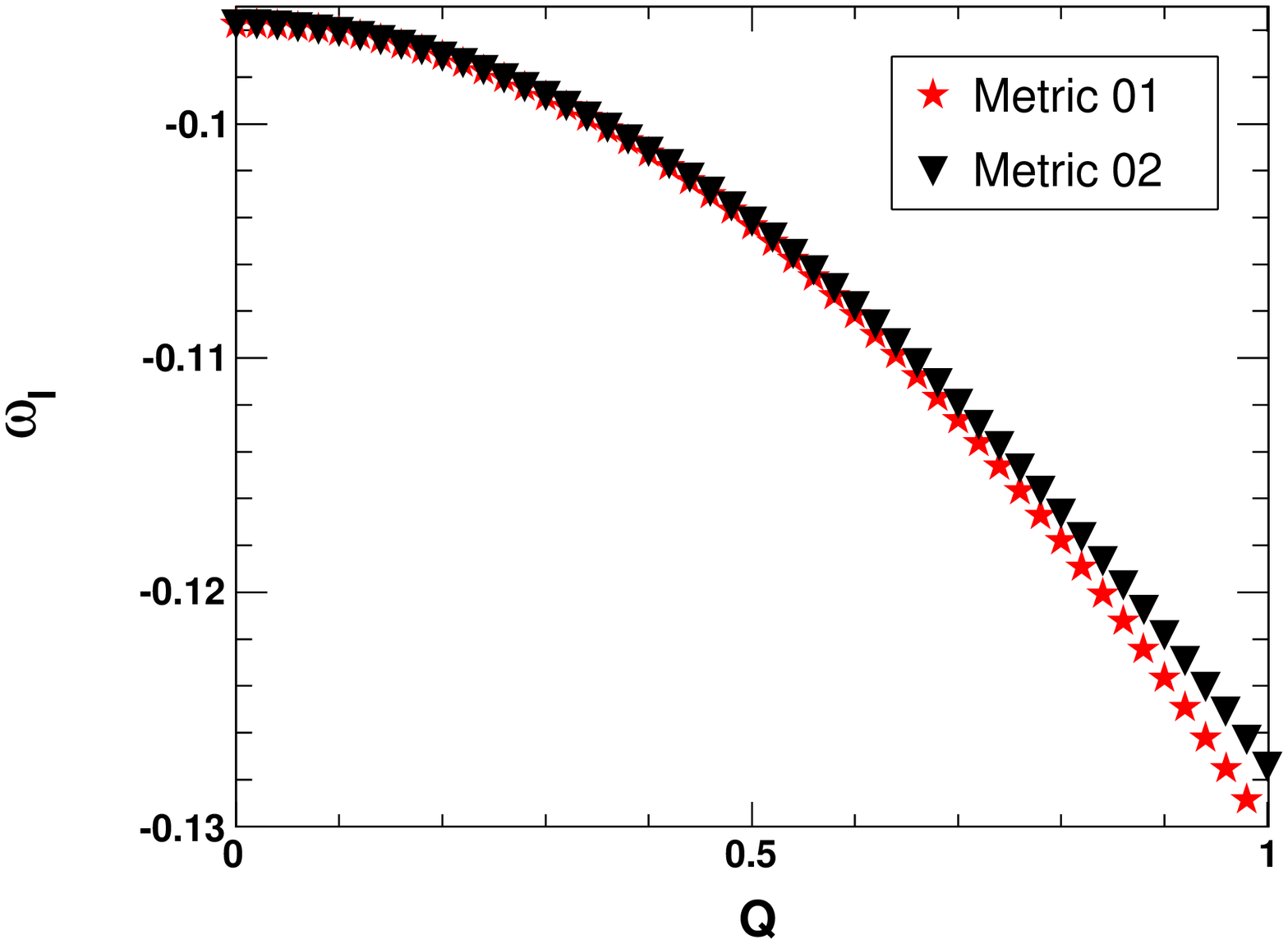}}
\vspace{-0.2cm}
\caption{Fundamental quasinormal mode frequencies w.\ r.\ t.\ $Q$ for the 
black holes defined by metric \eqref{Dust_metric01} (Metric 01) and 
\eqref{Dust_metric03} (Metric 02) surrounded by dust field with $l=4$, 
$\lambda=0.01$, $\kappa = 1$, $N_s = 0.01$ and $M = 1$ obtained by using the 
5th order WKB approximation method.}
\label{figDustQvary}
\end{figure}

Similar to the previous case, here also the quasinormal modes are dependent on 
the electrodynamic source of the black hole. The dependency of the frequencies 
on the charge $Q$ can be seen in Fig. \ref{figDustQvary}. The variations are 
quite similar to that of the previous case.

\begin{figure}[htb]
\centerline{
   \includegraphics[scale = 0.3]{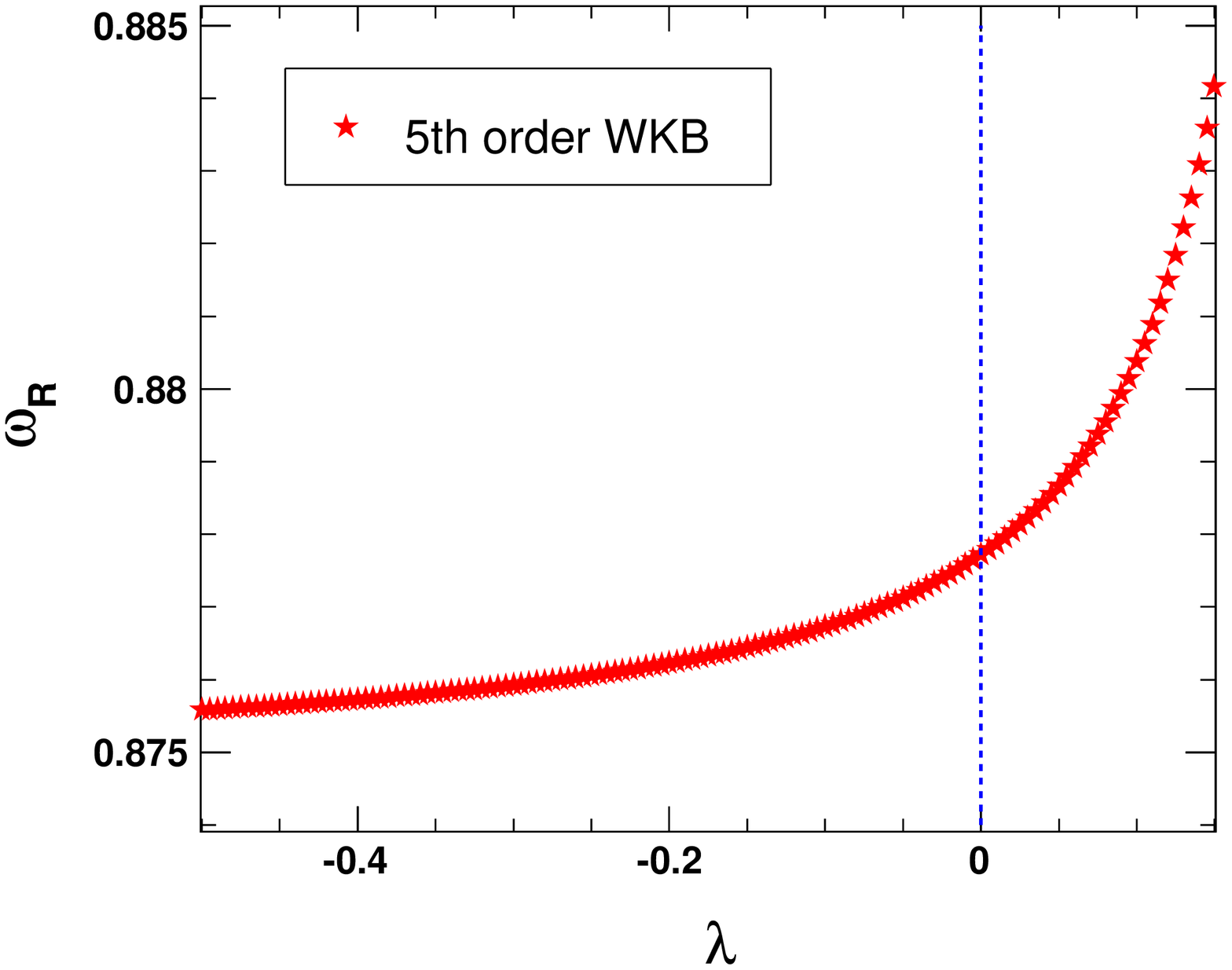}\hspace{0.5cm}
   \includegraphics[scale = 0.3]{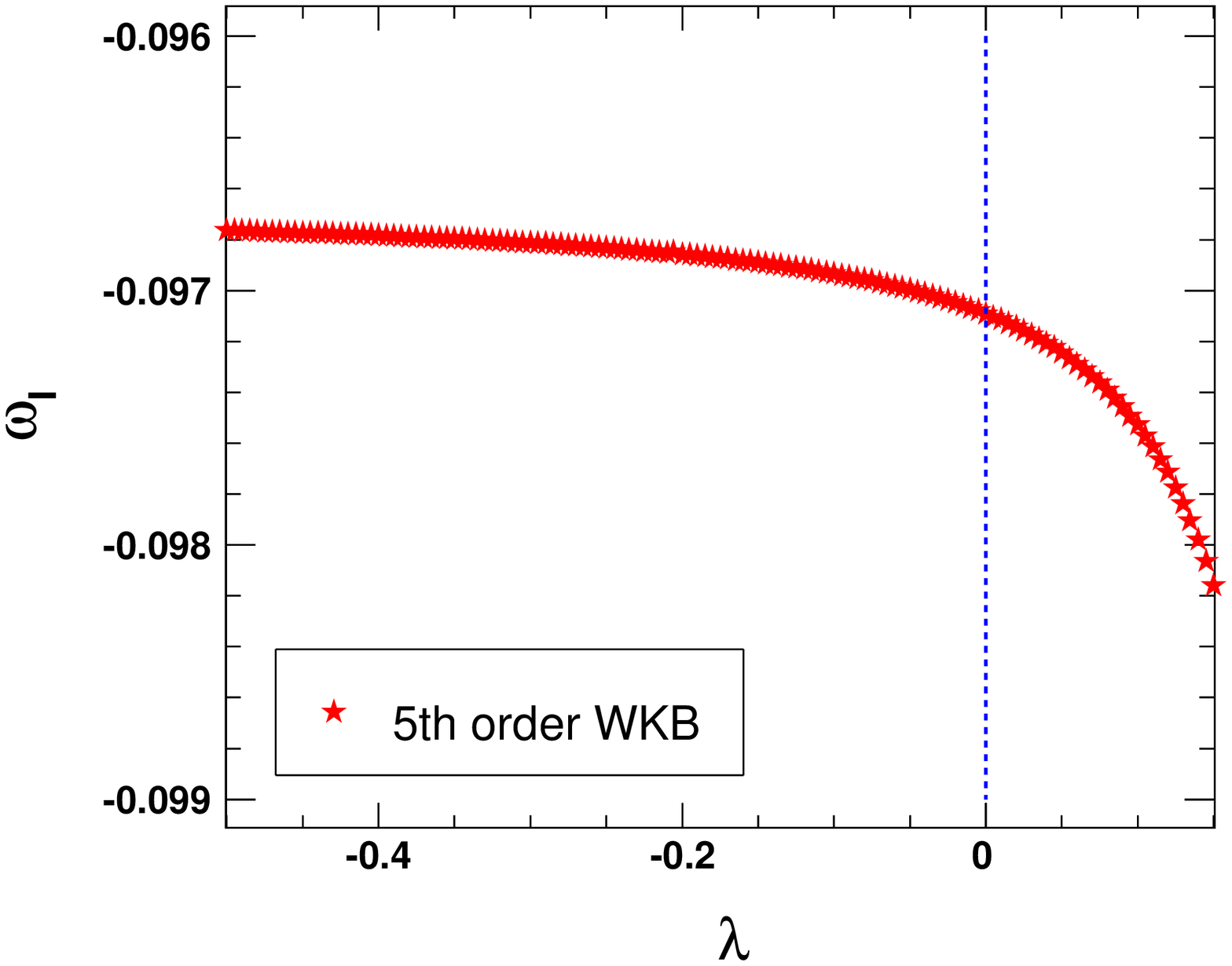}}
\vspace{-0.2cm}
\caption{Fundamental quasinormal mode frequencies w.\ r.\ t.\ the Rastall 
parameter $\lambda$ for the black holes defined by the metric 
\eqref{Dust_metric03} surrounded by dust field with $l=4$, $Q=0.2$, 
$\kappa = 1$, $N_s = 0.01$ and $M = 1$ obtained by using the 5th order WKB 
approximation method. The blue line shows the GR limit.}
\label{figDustlambdavary}
\end{figure}
We have checked the dependency of quasinormal mode frequencies with respect 
to the Rastall parameter $\lambda$, which is shown in 
Fig.~\ref{figDustlambdavary}. It seen that the real quasinormal mode 
frequencies increase gradually with the increasing value of $\lambda$ upto the
GR limit and after crossing this limit the frequencies increase rapidly with
the increasing positive value of $\lambda$. On the other hand 
the magnitude of imaginary
quasinormal mode frequencies increase with $\lambda$ following a similar 
pattern of the real frequencies before and after the GR limit of $\lambda$. 

\subsection{Surrounded by phantom field}
\begin{figure}[htb]
\centerline{
   \includegraphics[scale = 0.3]{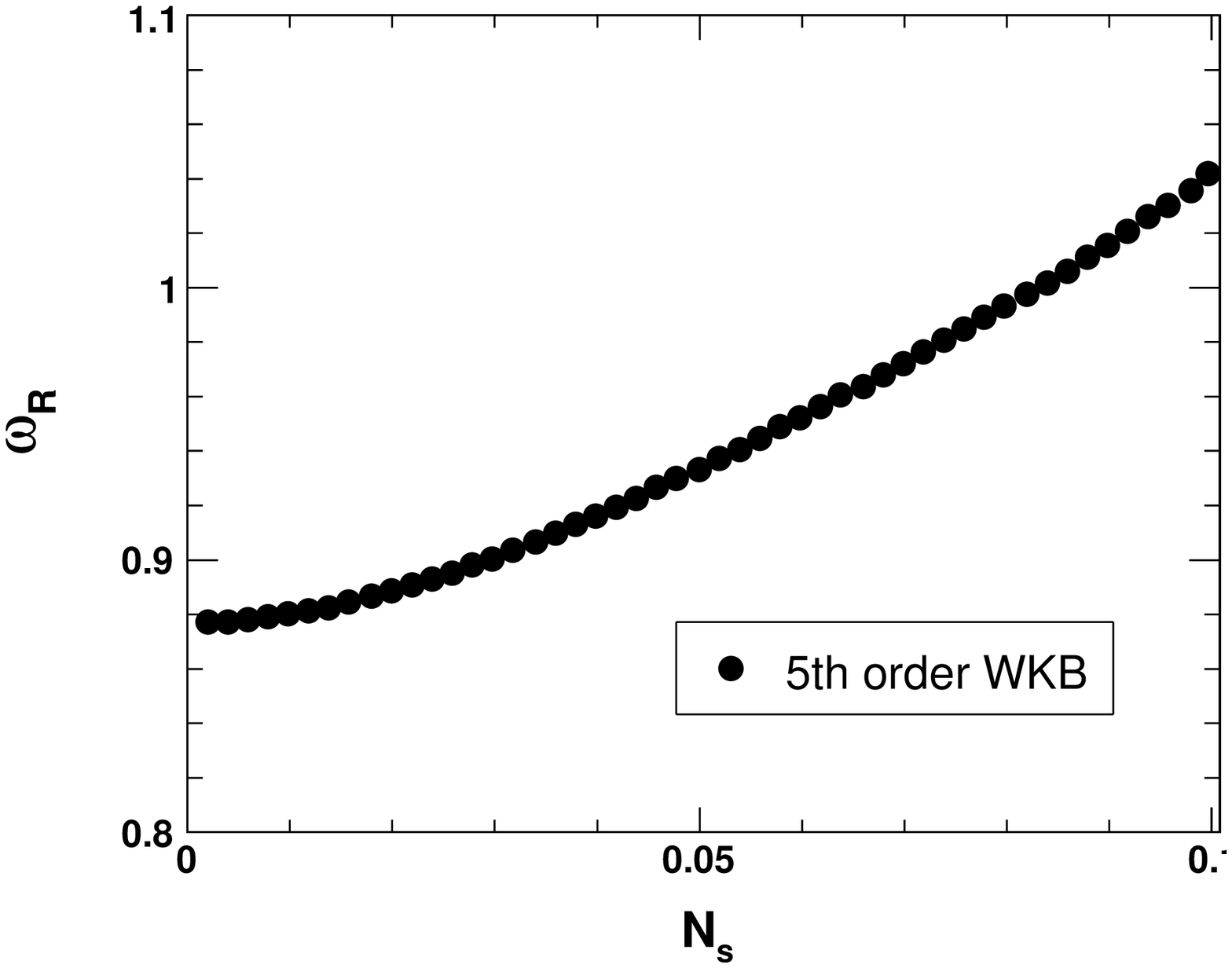}\hspace{0.5cm}
   \includegraphics[scale = 0.3]{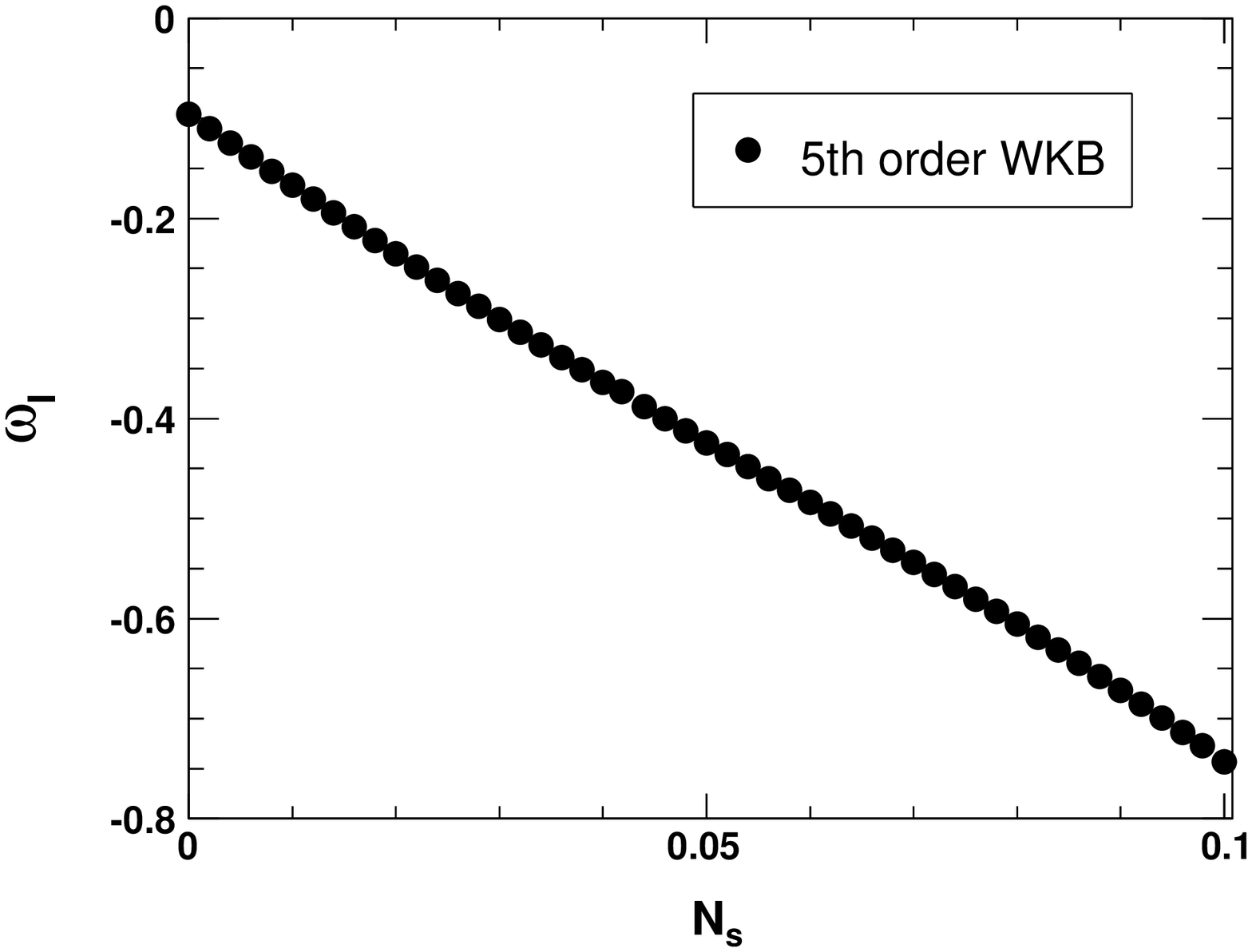}}
\vspace{-0.2cm}
\caption{Behaviour of fundamental quasinormal mode frequencies w.\ r.\ t.\ 
$N_s$ for the black holes defined by the metric \eqref{Phantom_metric03} 
surrounded by the phantom field with $l=4$, $\lambda=0.01$, $\kappa = 1$, 
$Q = 0.2$ and $M = 1$ obtained using the 5th order WKB approximation method.}
\label{figPhantomNsvary}
\end{figure}

\begin{figure}[htb]
\centerline{
   \includegraphics[scale = 0.3]{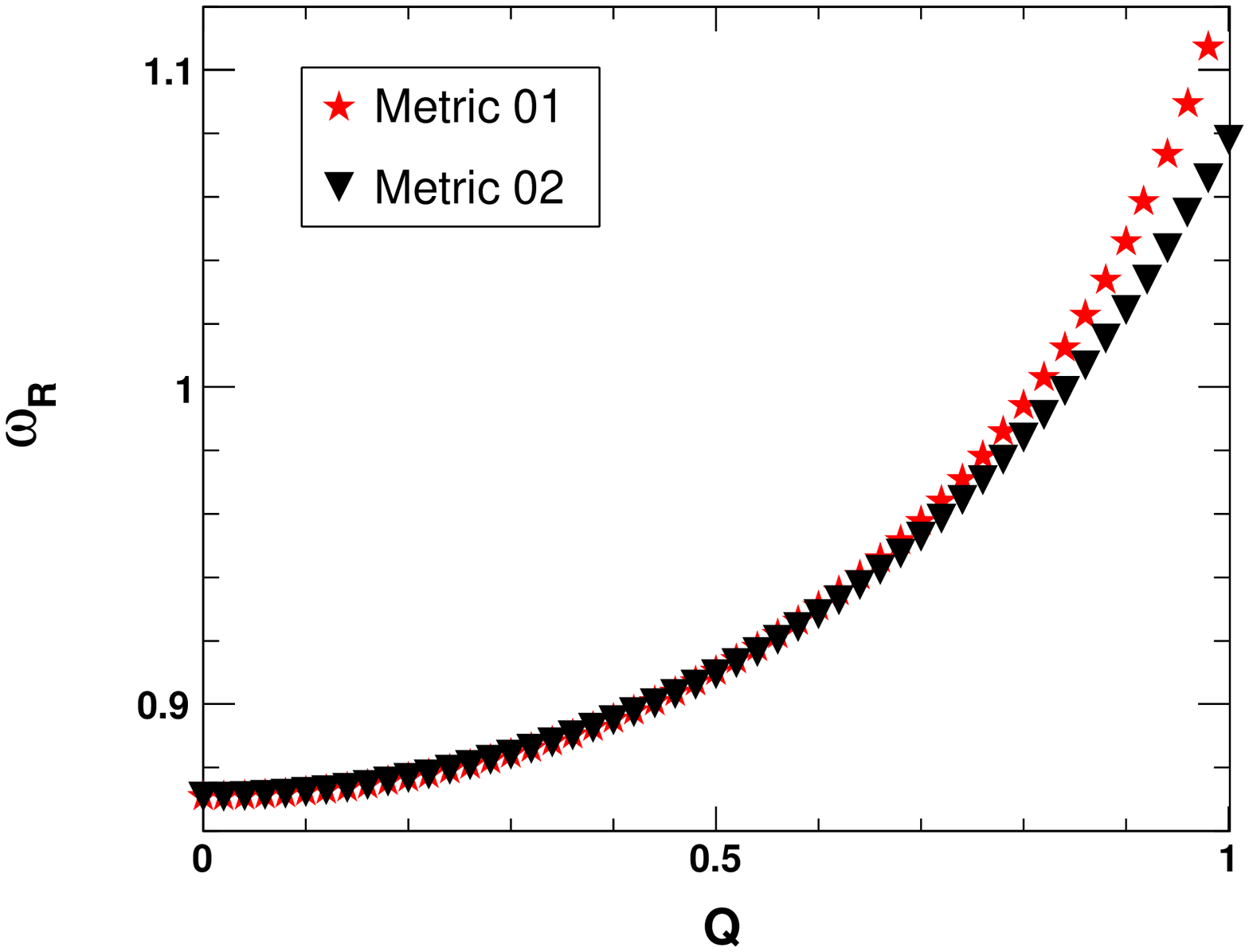}\hspace{0.5cm}
   \includegraphics[scale = 0.3]{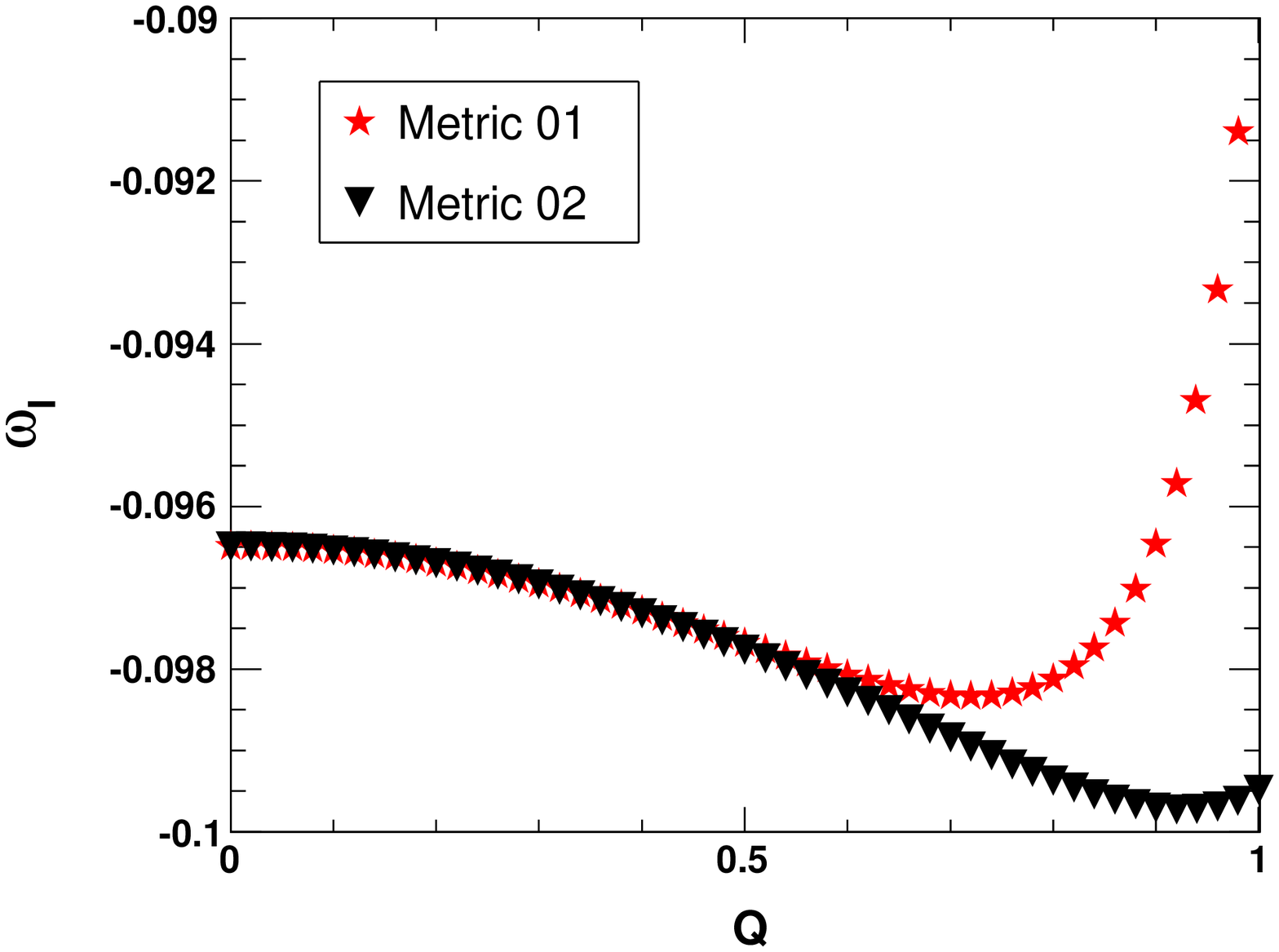}}
\vspace{-0.2cm}
\caption{Behaviour of fundamental quasinormal mode frequencies w.\ r.\ t.\ 
$Q$ for the black holes defined by the metric \eqref{Phantom_metric01} 
(Metric 01) and \eqref{Phantom_metric03} (Metric 02) surrounded by phantom 
field with $l=4$, $\lambda=0.01$, $\kappa = 1$, $N_s = 0.0001$ and $M = 1$ 
obtained by using the 5th order WKB approximation method.}
\label{figPhantomQvary}
\end{figure}

\begin{figure}[htb]
\centerline{
   \includegraphics[scale = 0.3]{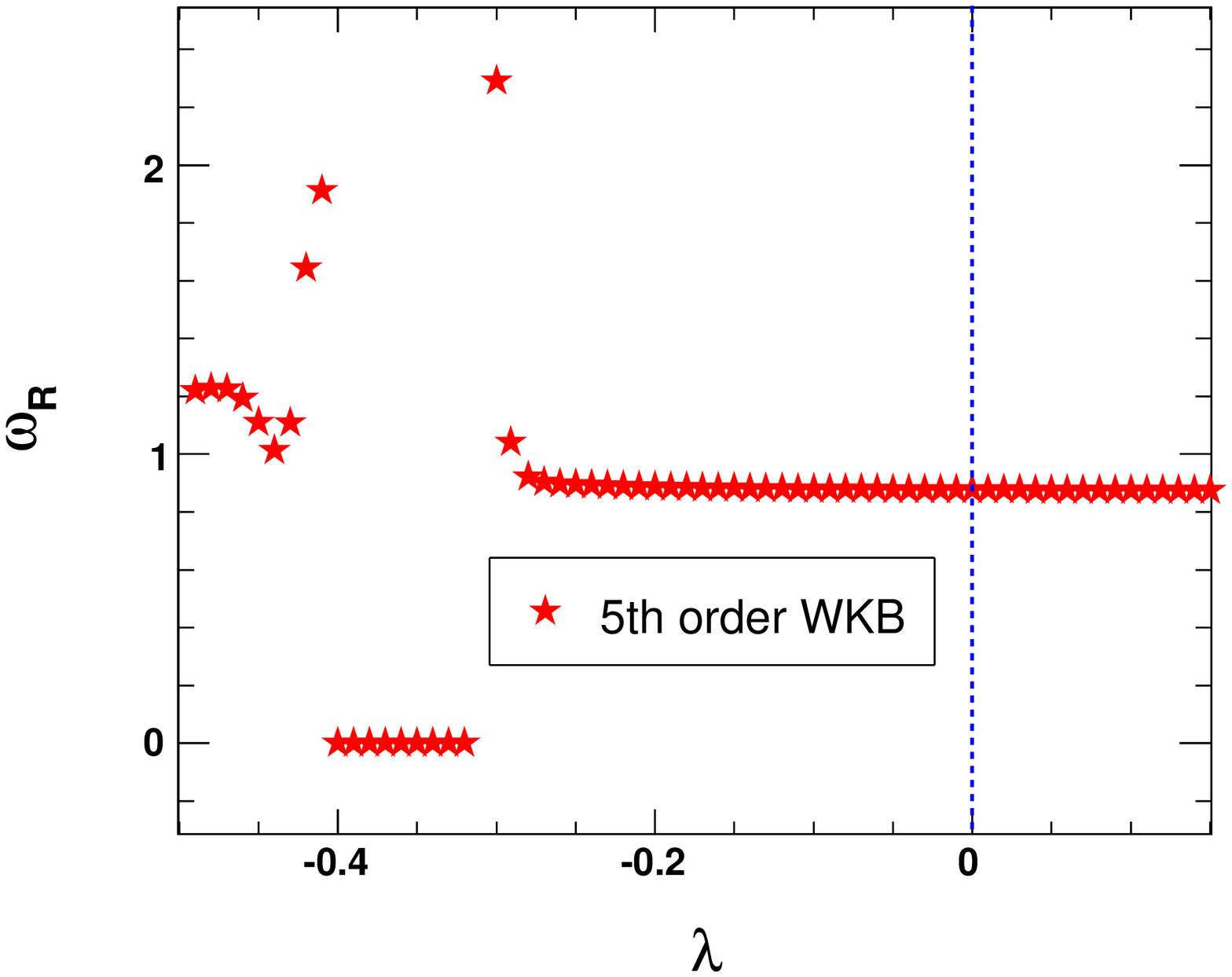}\hspace{0.5cm}
   \includegraphics[scale = 0.3]{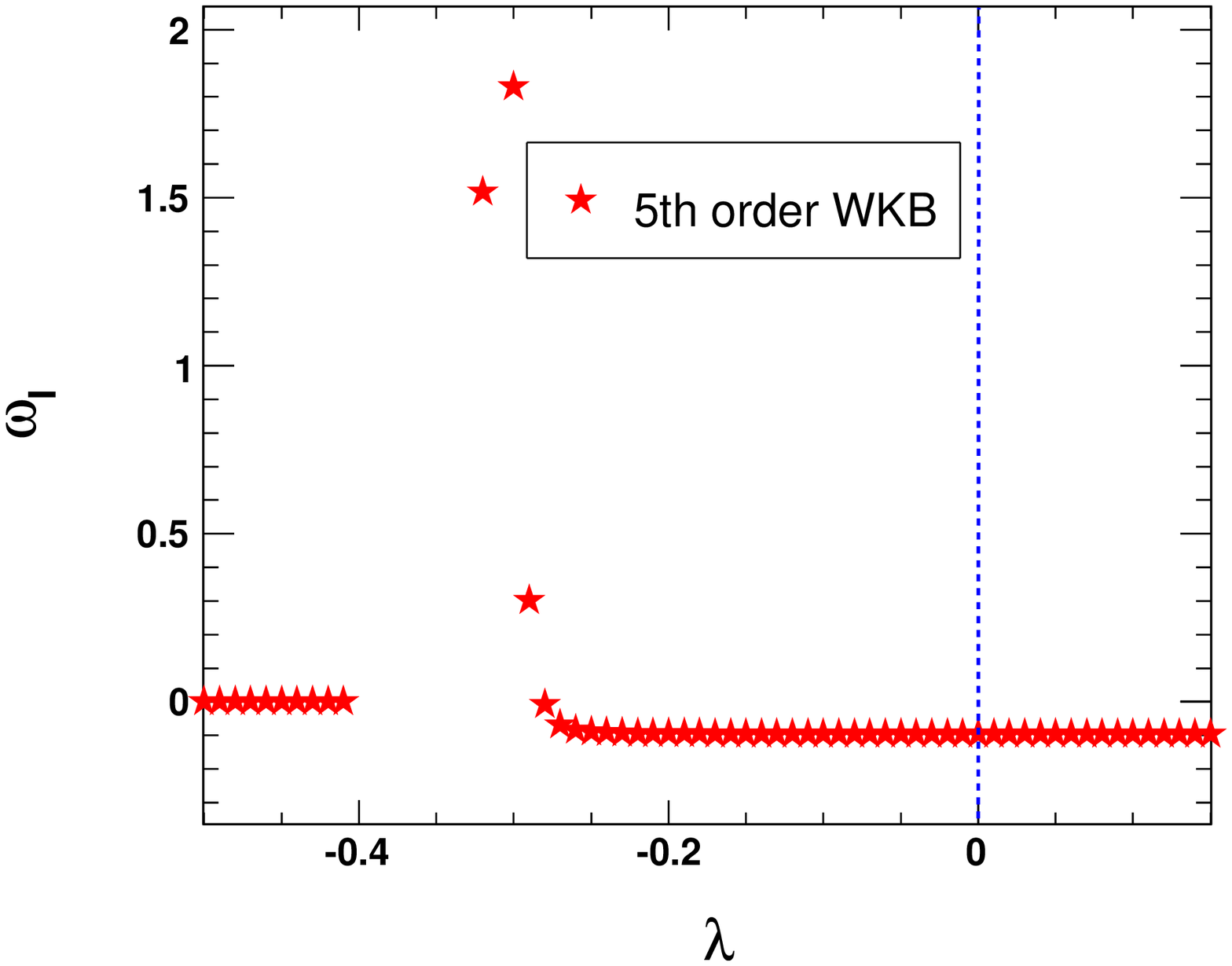}}
\vspace{-0.2cm}
\caption{Behaviour of fundamental quasinormal mode frequencies w.\ r.\ t.\ 
the Rastall parameter $\lambda$ for the black holes defined by the 
metric \eqref{Phantom_metric03} surrounded by the phantom field with 
$l=4$, $Q=0.2$, $\kappa = 1$, $N_s = 0.0001$ and $M = 1$ obtained by using 
the 5th order WKB approximation method. The blue line shows GR limit.}
\label{figPhantomlambdavary}
\end{figure}

In case of phantom field $\omega_s = -\dfrac{4}{3}$ \cite{Vikman}. Thus for 
this case the metric \eqref{metric01} becomes,
\begin{equation} \label{Phantom_metric01}
ds^2=-\left(1-\frac{2M}{r}
+\frac{Q^2}{r^2}+\frac{N_s}{r^{\frac{-3 + 2\kappa\lambda}{1+\kappa\lambda}}}\right)dt^2
+\frac{dr^2}{1-\frac{2M}{r}+\frac{Q^2}{r^2}
+\frac{N_s}{r^{\frac{-3 + 2\kappa\lambda}{1+\kappa\lambda}}}}
+r^2 d\Omega^2.
\end{equation}
Similarly metric \eqref{metric03} becomes,
\begin{equation} \label{Phantom_metric03}
ds^2=-\left(1-\frac{2m(r)}{r}
+\frac{N_s}{r^{\frac{-3 + 2\kappa\lambda}{1+\kappa\lambda}}}\right)dt^2
+\frac{dr^2}{1-\frac{2m(r)}{r}
+\frac{N_s}{r^{\frac{-3 + 2\kappa\lambda}{1+\kappa\lambda}}}}
+r^2 d\Omega^2.
\end{equation}
These solutions are different from the Kiselev black hole \cite{Kiselev}. Here 
the Rastall parameter, charge and the structural constant play significant 
role in the variations of quasinormal frequencies. To see these variations, we 
have plotted Fig.s \ref{figPhantomNsvary}, \ref{figPhantomQvary} and 
\ref{figPhantomlambdavary}. In Fig.~\ref{figPhantomNsvary} we see that an 
increase in the value of $N_s$ increases the real quasinormal frequencies. 
However, in case of the imaginary quasinormal frequencies, with increase in 
$N_s$, frequencies decrease more rapidly. Here $N_s$ acts as an damping factor 
increasing the damping rate. In this figure we have used the black hole 
defined by the metric \eqref{Phantom_metric03}. Again, for small values of 
charges, the quasinormal frequencies of black hole defined by the metric 
\eqref{Phantom_metric01} coincides with that of the other black hole with 
non-linear electrodynamic sources (see Fig.~\ref{figPhantomQvary}). We found 
that the black hole with non-linear electrodynamic source i.e.\ the black hole 
defined by the metric \eqref{Phantom_metric03}, can result a regular or 
non-singular black hole if $\kappa \lambda < 4$. The quasinormal frequencies 
of the black holes do not show any significant variation with respect to 
parameter $\lambda$. However, the frequencies show a discontinuous behaviour 
for a region with the value of $\lambda \le -\,0.3$ (see 
Fig.~\ref{figPhantomlambdavary}). This is due to the fact that the phantom 
field potential becomes non feasible for the black hole solution 
for such values of $\lambda$. In this both cases, the geometric parameter 
$\mathcal{W}_{s}$ from Eq.~\eqref{Ws} is obtained as
\begin{equation}
\mathcal{W}_{s} = -\dfrac{1}{3} \dfrac{(1-4 \kappa \lambda)(4 - \kappa \lambda)}{(1+\kappa \lambda)^2}.
\end{equation}
From this expression, weak energy condition can be obtained. The weak energy 
condition allows  $N_s>0$  for $0\leq\kappa\lambda<\frac{1}{4}~\cup ~ 
\kappa\lambda>4$ and $N_s<0$ for $\frac{1}{4}<\kappa\lambda<4$. The quasinormal modes for these black holes for $l=1$ to $3$ are shown in Table \ref{table03}.
In this case the metric \eqref{Phantom_metric03} gives the smaller values of 
quasinormal modes than that given by the metric \eqref{Phantom_metric01} for
all three values of $l$ and all these frequencies are different in comparison 
to the case of GR. $|\omega_5-\omega_6|$ and $\vartriangle_5$ also show 
a similar pattern as in the previous cases. For the non-linearly charged black 
hole in Rastall gravity, both these terms give higher values. This is due to 
the non-linear charge distribution function as mentioned earlier. The deviation
terms $\triangle_{NL-GR}$ and $\triangle_{L-GR}$ have equivalent order 
for $l=3$ in the case of both 5th order and 6th order WKB approximations. But, 
it is seen that for $l=1$ and $2$, in the case of 6th order WKB approximation, 
$\triangle_{NL-GR}$ is much higher than $\triangle_{L-GR}$. It can be due to 
the error associated with the approximation, because for the non-linear black 
hole with $l=1$ and $2$, both $|\omega_5-\omega_6|$ and $\vartriangle_5$ have 
higher values which can impact the results significantly.

\begin{table}[h!]
\centering
\begin{tabular}{||c|c|c|c|c|c||}
\hline \hline
 & Metric \eqref{Phantom_metric01} & Metric \eqref{metric02}               & Metric \eqref{Phantom_metric03} & $\triangle_{NL-GR}$ & $\triangle_{L-GR}$ \\ \hline
$l = 1$ (6th order WKB) & $0.296112 - 0.0978824 i$ &  $0.296164 - 0.0978612 i$ & $0.289416 - 0.10012 i$ & $2.28141\%$ & $0.0180036\%$  \\
$l = 1$ (5th order WKB)& $0.296264 - 0.0978322i$ & $0.296318 - 0.0978103i$ & $0.296249 - 0.0978105i$ & $0.0221124\%$ & $0.0186743\%$\\
$|\omega_5-\omega_6|$& $0.000160075$ & $0.000162194$ & $0.00721274$ & &\\
$\vartriangle_5$& $0.00018417$ & $0.000183679$ & $0.00361083$ & &\\ \hline
$l = 2$ (6th order WKB) & $0.488881 - 0.0970006 i$ & $0.488984 - 0.0969835 i$ & $0.487678 - 0.097239 i$ & $0.266948\%$ & $0.0209444\%$ \\ 
$l = 2$ (5th order WKB) & $0.488896 - 0.0969976i$ & $0.488999 - 0.0969805i$ & $0.488884 - 0.0969991i$ & $0.0233679\%$ & $0.0209438\%$\\
$|\omega_5-\omega_6|$& $0.0000152971$ & $0.0000152971$ & $0.00122963$ & &\\
$\vartriangle_5$& $0.0000245341$ & $0.0000244845$ & $0.000615288$ & &\\ \hline
$l = 3$ (6th order WKB) & $0.682603 - 0.0967835 i$  & $0.682748 - 0.0967698 i$ & $0.682546 - 0.0967902 i$ & $0.0294425\%$ & $0.0211212\%$\\ 
$l = 3$ (5th order WKB) & $0.682606 - 0.0967831i$ & $0.682751 - 0.0967695i$ & $0.68259 - 0.0967839i$ & $0.0234409\%$ & $0.0211197\%$\\
$|\omega_5-\omega_6|$& $3.02655\times10^{-6}$ & $3.01496\times10^{-6}$ & $0.0000444487$ & &\\
$\vartriangle_5$& $5.82151\times10^{-6}$ & $5.82151\times10^{-6}$ & $0.0000230827$ & &\\ \hline \hline
\end{tabular}
\caption{Quasinormal modes of black holes for $n=0$, $M=1$, $\omega = -\dfrac{4}{3}$ (phantom field), $\kappa \lambda = 0.01$, $N_s = 0.0001$ and $Q = 0.2$. }
\label{table03}
\end{table}

\subsection{Surrounded by quintessence field}
Using $\omega_s = - \dfrac{2}{3}$ for the quintessence field \cite{Kiselev}, 
the metric \eqref{metric01} for this surrounding field can be written as
\begin{equation} \label{Quin_metric01}
ds^2=-\left(1-\frac{2M}{r}
+\frac{Q^2}{r^2}+\frac{N_s}{r^{\frac{-1 - 2\kappa\lambda}{1-\kappa\lambda}}}\right)dt^2
+\frac{dr^2}{1-\frac{2M}{r}+\frac{Q^2}{r^2}
+\frac{N_s}{r^{\frac{-1 - 2\kappa\lambda}{1-\kappa\lambda}}}}
+r^2 d\Omega^2.
\end{equation}
Similarly, for this field the metric \eqref{metric03} becomes,
\begin{equation} \label{Quin_metric03}
ds^2=-\left(1-\frac{2m(r)}{r}
+\frac{N_s}{r^{\frac{-1 - 2\kappa\lambda}{1-\kappa\lambda}}}\right)dt^2
+\frac{dr^2}{1-\frac{2m(r)}{r}
+\frac{N_s}{r^{\frac{-1 - 2\kappa\lambda}{1-\kappa\lambda}}}}
+r^2 d\Omega^2.
\end{equation}
The geometric parameter in these both cases takes the form,
\begin{equation}
\mathcal{W}_{s}=
-\frac{\left(1-4\kappa\lambda\right)(2+\kappa\lambda)}{3(1-\kappa\lambda)^2}.
\end{equation}
Using this parameter we can see that the black holes can respect the weak 
energy condition for  $N_s>0$ with $0\leq\kappa\lambda<\frac{1}{4}$ 
and $N_s<0$ with $\kappa\lambda>\frac{1}{4}$ \cite{Heydarzade2}.

\begin{figure}[htb]
\centerline{
   \includegraphics[scale = 0.3]{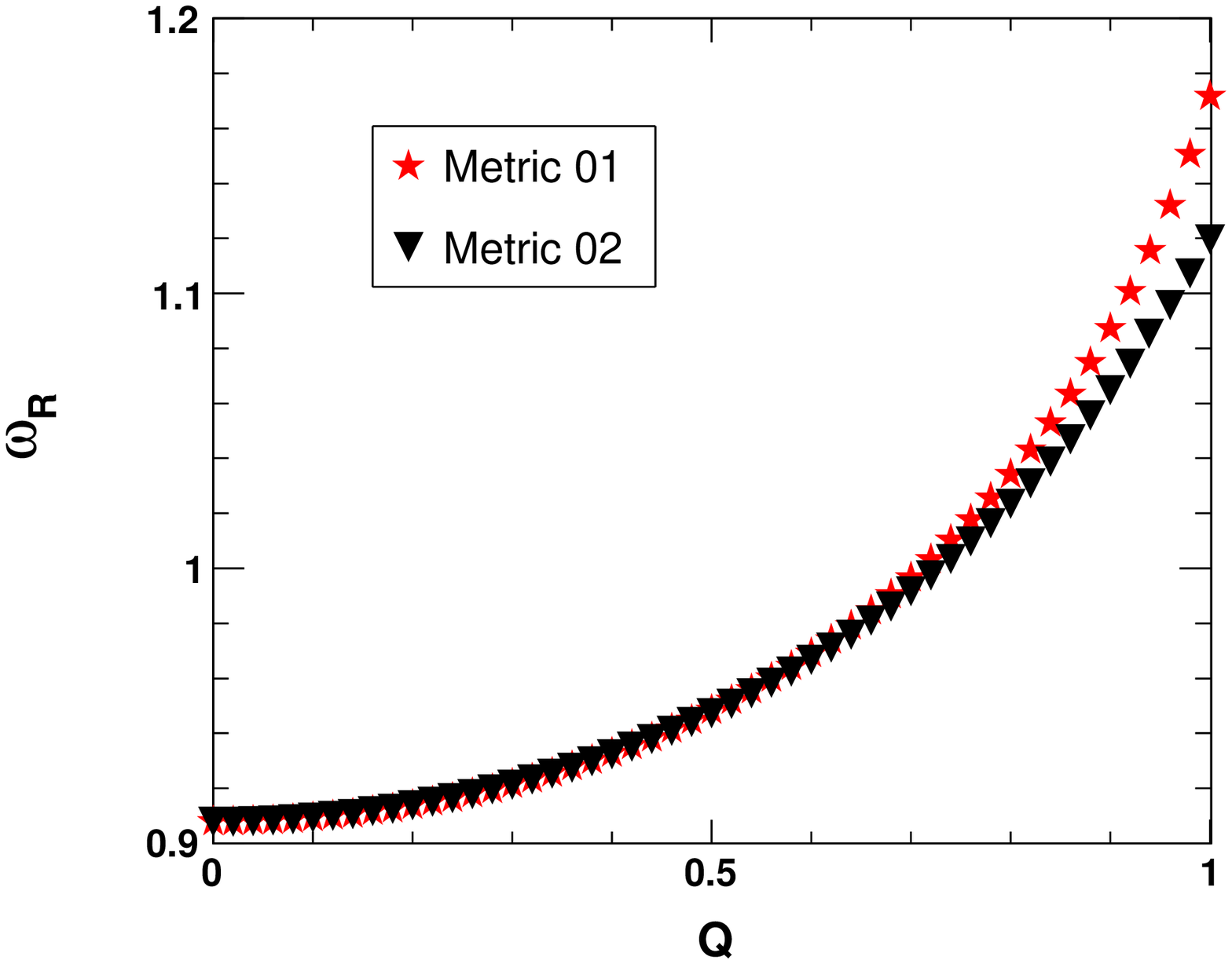}\hspace{0.5cm}
   \includegraphics[scale = 0.3]{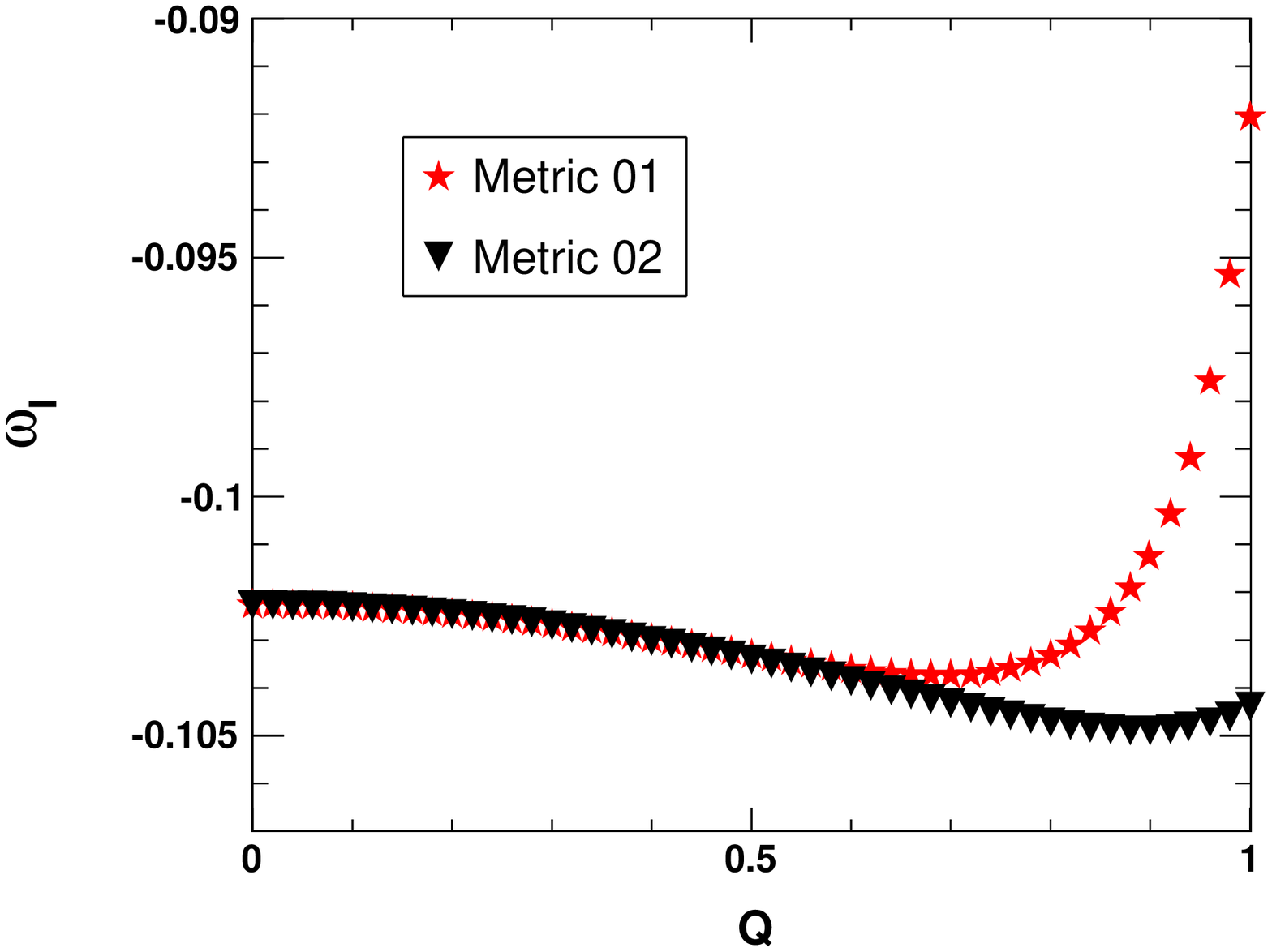}}
\vspace{-0.2cm}
\caption{Fundamental quasinormal mode frequencies w.\ r.\ t.\ $Q$ for the 
black holes defined by the metric \eqref{Quin_metric01} (Metric 01) and 
\eqref{Quin_metric03} (Metric 02) surrounded by the quintessence field with 
$l=4$, $\lambda=0.01$, $\kappa = 1$, $N_s = 0.01$ and $M = 1$ 
obtained by using the 5th order WKB approximation method.}
\label{figQuinQvary}
\end{figure}

Here also both the black holes differ from Kiselev type black holes 
\cite{Kiselev, Heydarzade2}. The second black hole with non-linear 
electrodynamic sources can result a regular or non-singular black hole 
if $\kappa \lambda <-2$. The variations of quasinormal modes with respect to 
charge for both the black holes are shown in Fig.~\ref{figQuinQvary}. For the 
real quasinormal frequencies, with increase in charge $Q$, the frequencies 
increase non-linearly. However, similar to the other cases, the black 
hole with non-linear electrodynamic source or the regular black hole shows 
comparatively slower variation with respect to the singular black hole. In 
case of the imaginary quasinormal frequencies, there is a small dip in the 
variation pattern and near $1$ the frequencies increase again. In this region 
also the regular black hole has smaller quasinormal frequencies. These 
behaviours are very similar to the case of phantom field. 
 
\begin{figure}[htb]
\centerline{
   \includegraphics[scale = 0.3]{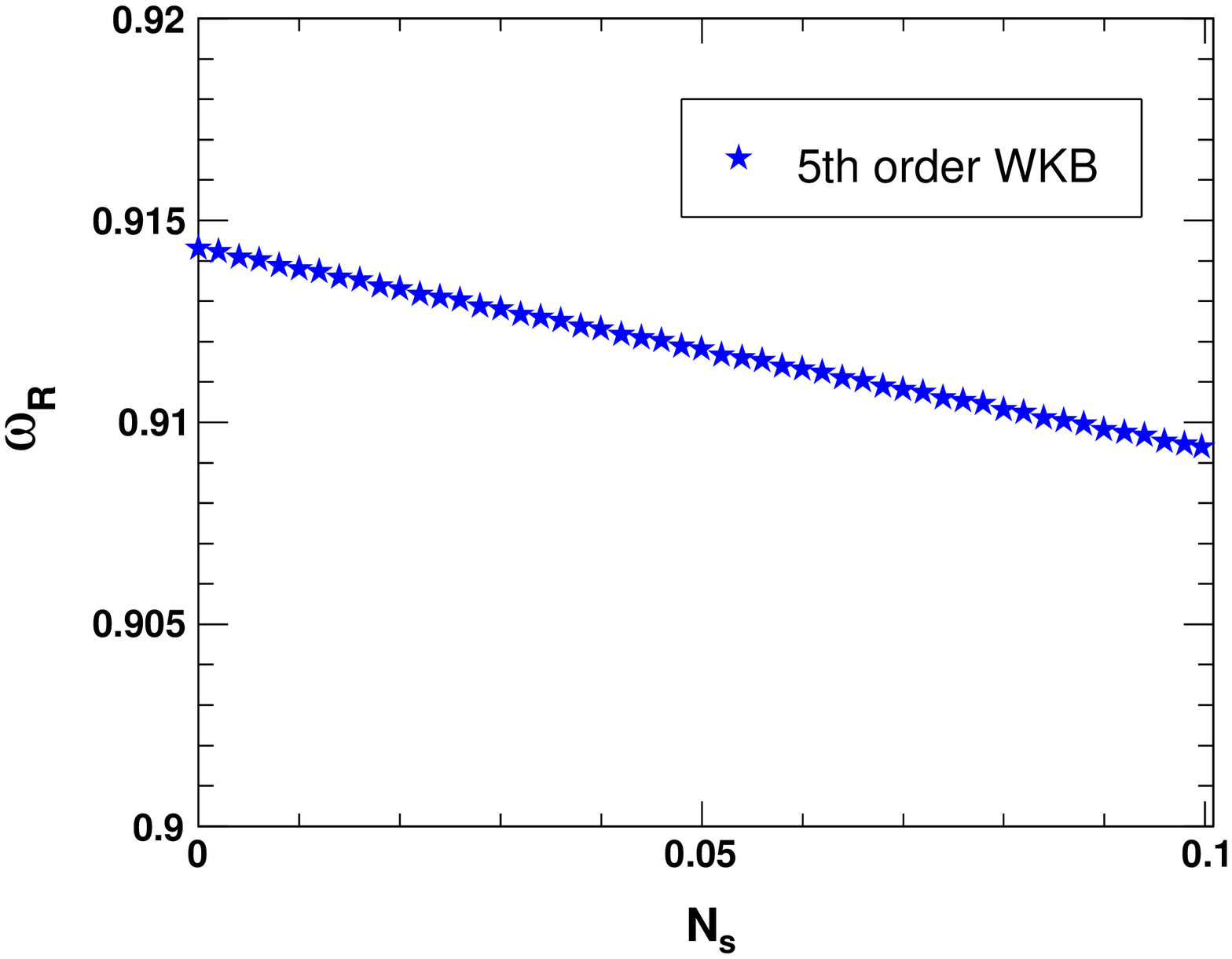}\hspace{0.5cm}
   \includegraphics[scale = 0.3]{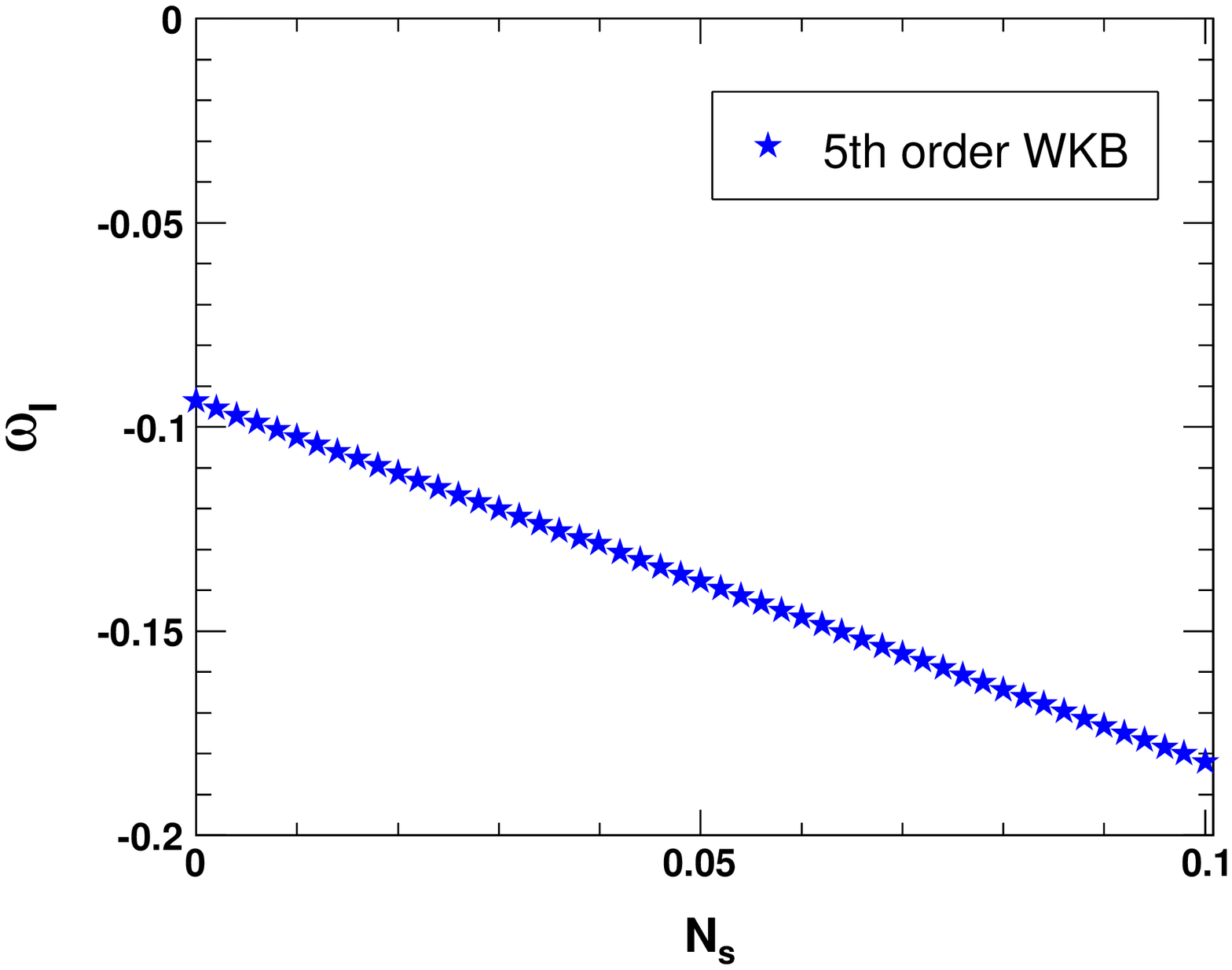}}
\vspace{-0.2cm}
\caption{Fundamental quasinormal mode frequencies w.\ r.\ t.\ $N_s$ for the
black holes defined by the metric \eqref{Quin_metric03} surrounded by
the quintessence field with $l=4$, $\lambda=0.01$, $\kappa = 1$, $Q = 0.2$
and $M = 1$ obtained by using the 5th order WKB approximation method.}
\label{figQuinNsvary}
\end{figure}

The behaviour of the quasinormal modes with the structural parameter $N_s$ for 
the regular black hole is shown in the Fig.~\ref{figQuinNsvary}. In both the 
cases increase in the parameter $N_s$ increases the damping rate. 
The increase in the magnitude of imaginary frequencies show that, with increase
in $N_s$, the typical damping time decreases.
The decrease in the frequencies are linear for both cases.
 
\begin{figure}[htb]
\centerline{
   \includegraphics[scale = 0.3]{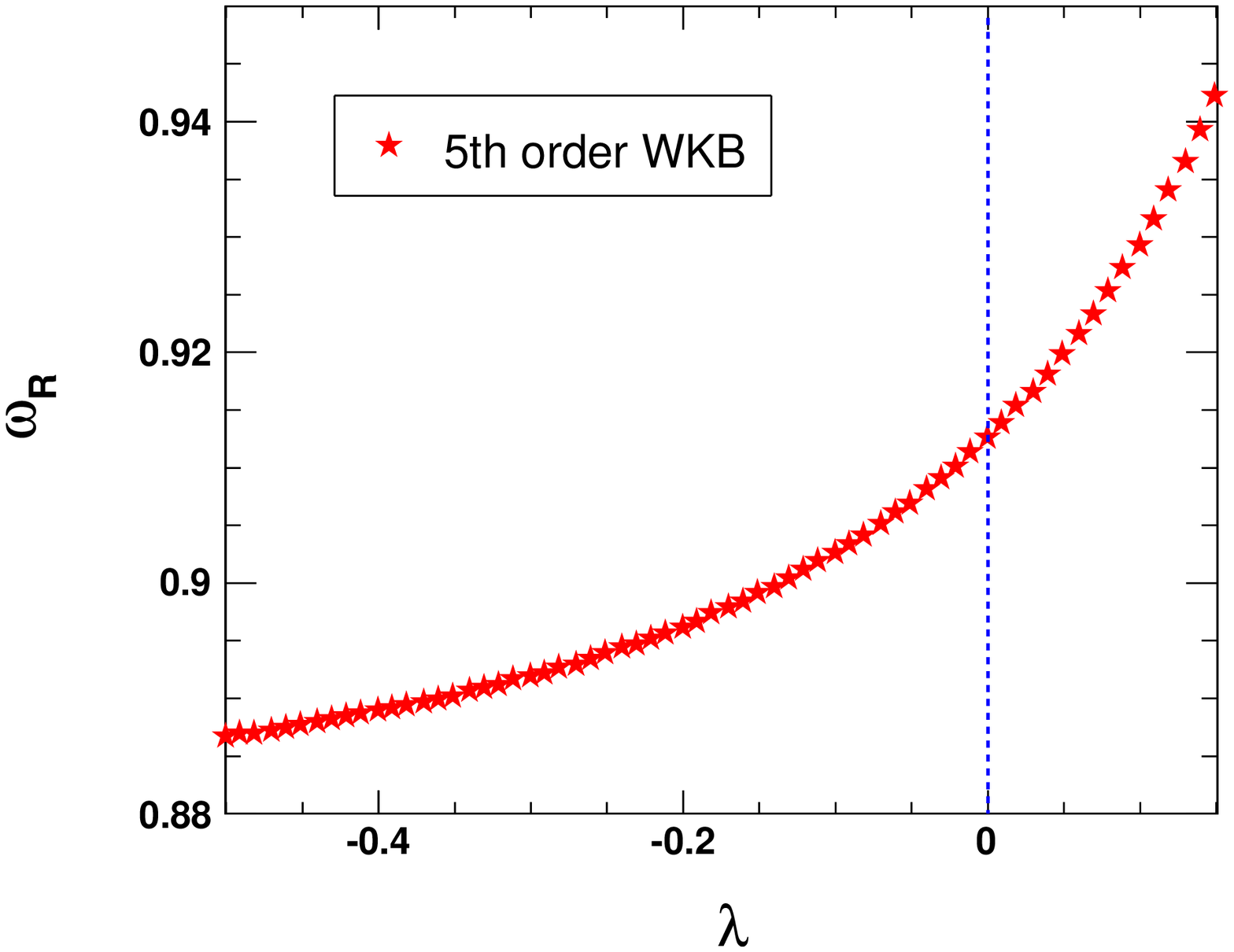}\hspace{0.5cm}
   \includegraphics[scale = 0.3]{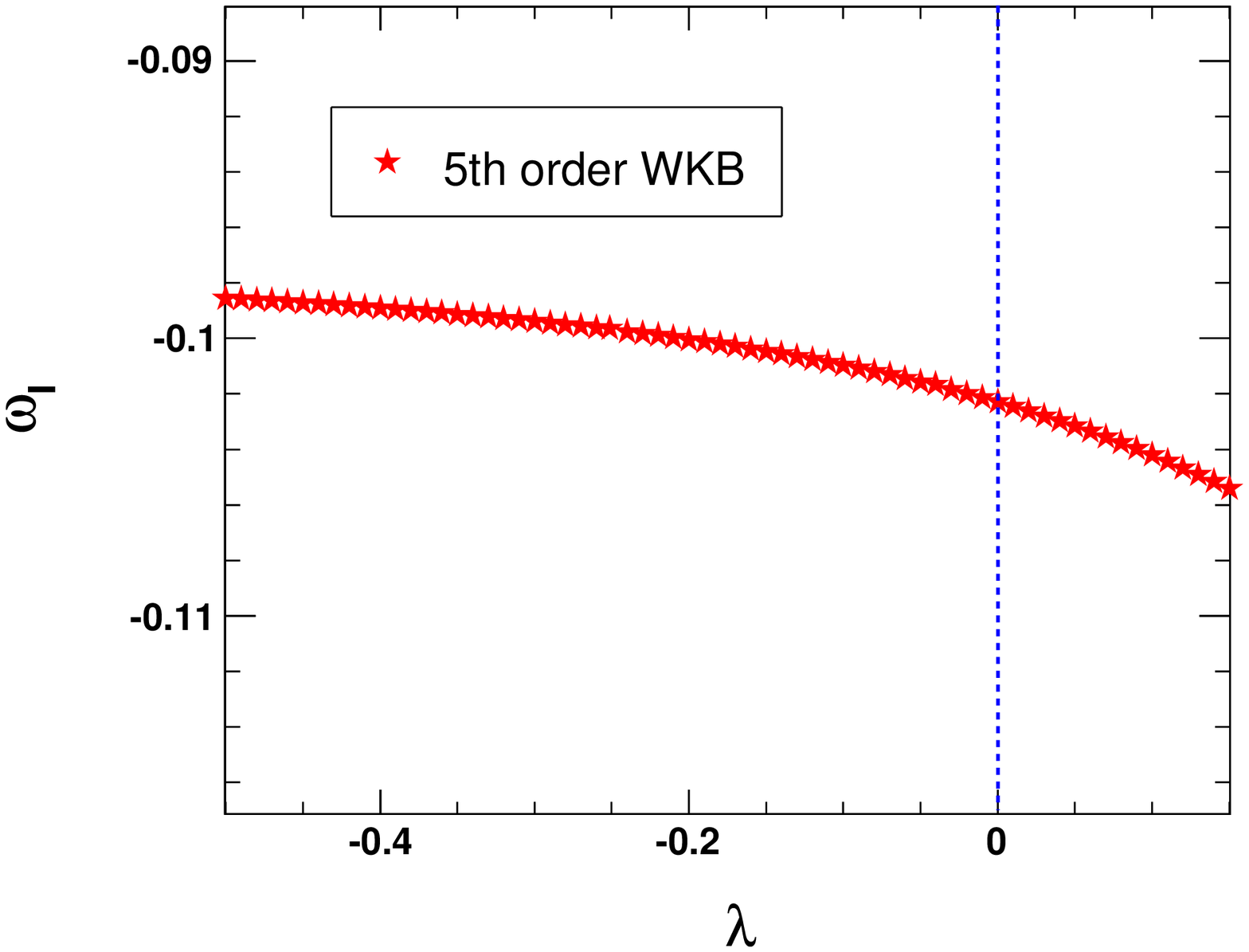}}
\vspace{-0.2cm}
\caption{Fundamental quasinormal mode frequencies w.\ r.\ t.\ $\lambda$ for 
the black holes defined by the metric \eqref{Quin_metric03} surrounded by the
quintessence field with $l=4$, $Q=0.2$, $\kappa = 1$, $N_s = 0.01$ and 
$M = 1$ obtained by using the 5th order WKB approximation method. The blue 
line shows GR limit.}
\label{figQuinlambdavary}
\end{figure}

The Rastall parameter $\lambda$ introduces a significant signature on the 
quasinormal modes of the regular black hole (see Fig.~\ref{figQuinlambdavary}). In case of the real quasinormal frequencies, an increase in $\lambda$ 
increases the frequencies non-linearly. Similarly, in case of the imaginary 
frequencies, the magnitude of the frequencies increase non-linearly but the 
variation is slower in comparison to the real frequencies. In both the plots, 
the blue line corresponds to the GR limit. The observational constraints on the 
Rastall parameter from \cite{Tang} shows that in Rastall gravity, the 
magnitude of quasinormal modes will increase in comparison to GR. Quasinormal 
modes for $l=1$ to $3$ are shown in Table \ref{table04}. For all three $l$
the quasinormal modes given by singular black hole are found to be larger than
that for the case of GR. The deviations $|\omega_5-\omega_6|$ and error 
term $\vartriangle_5$ are comparatively smaller for the linearly charged black 
hole in Rastall gravity and GR. However, for the non-linearly charged black 
hole in Rastall gravity, they are higher as expected. It is due to the 
non-linear charge distribution function present in the non-linearly charged 
black hole. In general, the deviation of quasinormal modes of non-linearly 
charged black hole from the Reissner-Nordstr\"om black hole i.e., 
$\triangle_{NL-GR}$ is higher than that of linearly charged black hole 
in Rastall gravity from the Reissner-Nordstr\"om black hole 
$(\triangle_{L-GR})$. It is again due to the effect of the non-linear charge 
distribution function. The Fig.~\ref{figQuinQvary} suggests that 
$\triangle_{NL-GR}$ increases with increase in the value of charge $Q$ of 
the non-linearly charged black hole.
 
\begin{table}[h!]
\centering
\begin{tabular}{||c|c|c|c|c|c||}
\hline \hline
 & Metric \eqref{Quin_metric01}  & Metric \eqref{metric02}               & Metric \eqref{Quin_metric03} & $\triangle_{NL-GR}$ & $\triangle_{L-GR}$ \\ \hline
$l = 1$ (6th order WKB) & $0.295123 - 0.0979993 i$ &  $0.295069 - 0.0980137 i$ & $0.274429 - 0.105352 i$ & $7.04541\%$ & $0.0179746\%$  \\
$l = 1$ (5th order WKB)&$0.295221 - 0.0979667i$ & $0.295216 - 0.097965 i$ & $0.295204 - 0.0979378i$& $0.00955789\%$ & $0.00169785\%$\\
$|\omega_5-\omega_6|$& $0.00010328$ & $0.000154857$ & $0.0220584$ & & \\
$\vartriangle_5$& $0.000175935$ & $0.00018489$ & $0.0110335$ &  & \\
\hline
$l = 2$ (6th order WKB) & $0.487181 - 0.097029 i$ & $0.487159 - 0.0970304 i$ & $0.485045 - 0.0974635 i$ &$0.434425\%$ & $0.00443794\%$ \\ 
$l = 2$ (5th order WKB)&$0.487181 - 0.0970289 i$ & $0.487173 - 0.0970274i$& $0.487171 - 0.097038i$ &$0.00217156\%$ & $0.00163856\%$ \\
$|\omega_5-\omega_6|$& $1.0\times10^{-7}$ & $0.0000143178$ & $0.00216816$& & \\
$\vartriangle_5$& $0.000023387$ & $0.0000245775$ & $0.00108475$& & \\
\hline
$l = 3$ (6th order WKB) & $0.680272 - 0.0967685 i$  & $0.680259 - 0.0967672 i$ & $0.680935 - 0.0966769 i$ & $0.0992574\%$& $0.00190143\%$\\ 
$l = 3$ (5th order WKB)&$0.680272 - 0.0967685i$ & $0.680261 - 0.0967668i$ & $0.680257 - 0.0967733i$& $0.00111077\%$ & $0.00161992\%$ \\
$|\omega_5-\omega_6|$& $0.$ & $2.03961\times10^{-6}$ &$0.000684819$ & & \\
$\vartriangle_5$& $5.73781\times10^{-6}$ & $5.82151\times10^{-6}$ & $0.000342457$ & & \\
\hline \hline
\end{tabular}
\caption{Quasinormal modes of black holes for $n=0$, $M=1$, $\omega = -\dfrac{2}{3}$ (quintessence field), $\kappa \lambda = 0.01$, $N_s = 0.0001$ and $Q = 0.2$.}
\label{table04}
\end{table} 
 
\begin{figure}[htb]
\centerline{
   \includegraphics[scale = 0.3]{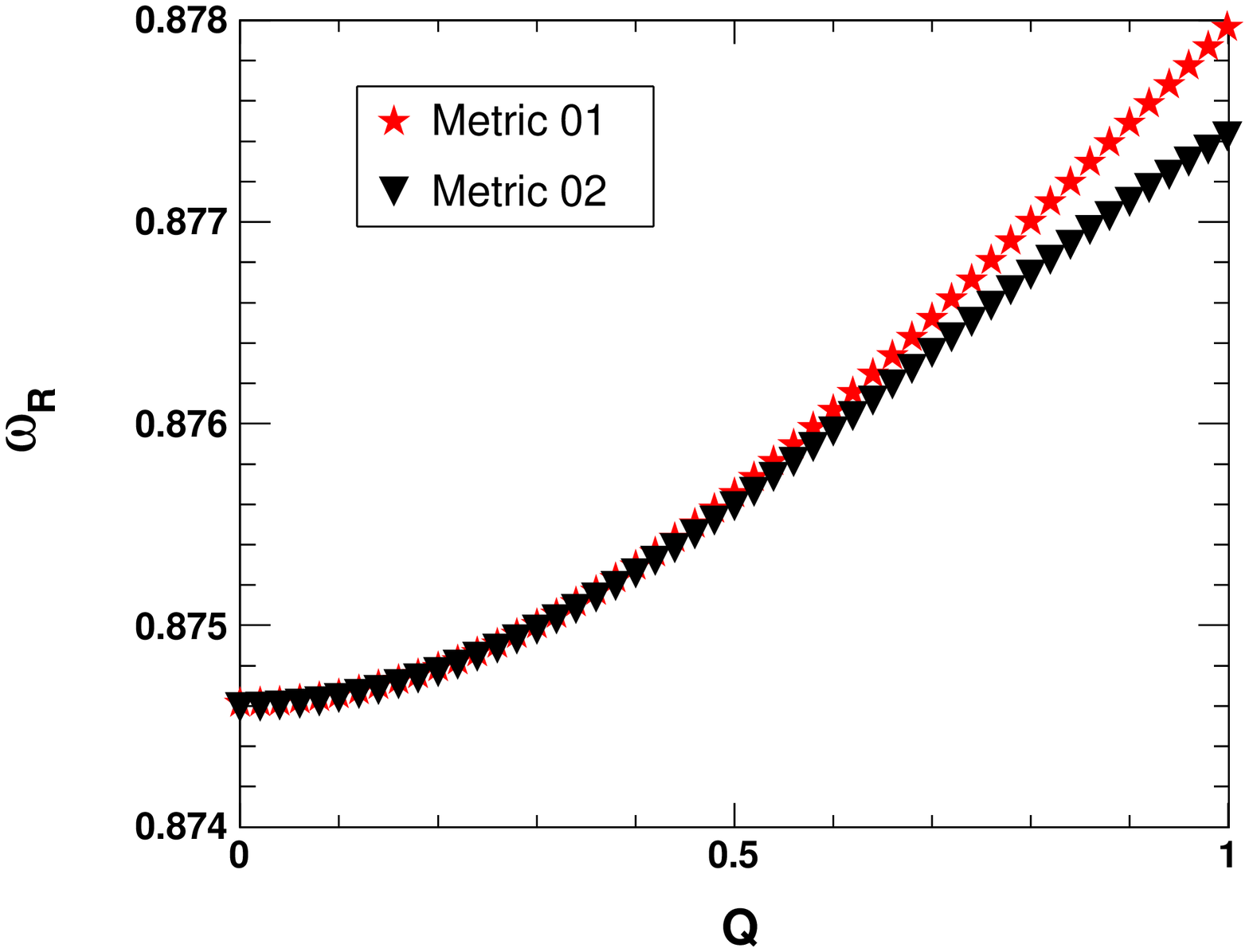}\hspace{0.5cm}
   \includegraphics[scale = 0.3]{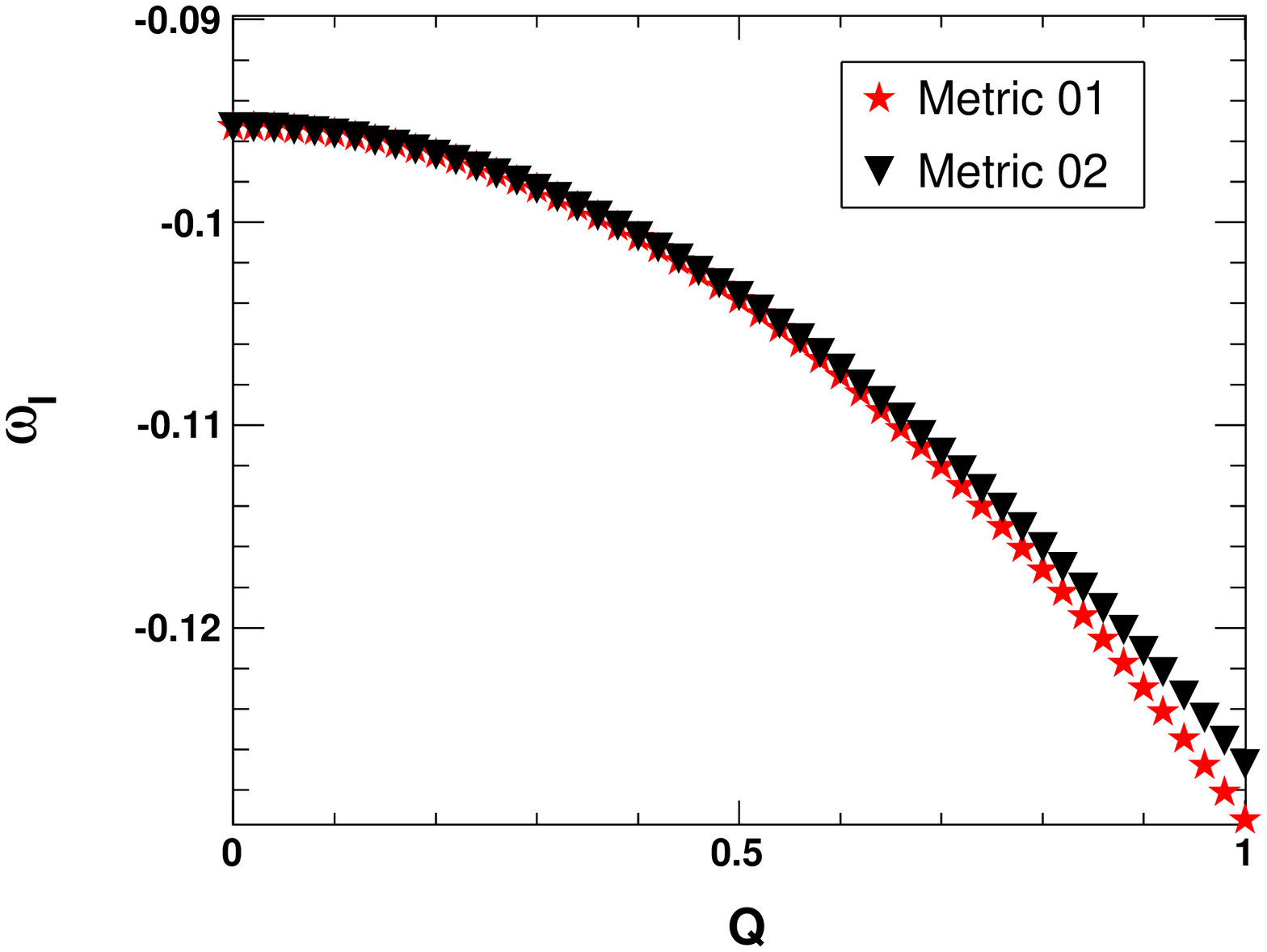}}
\vspace{-0.2cm}
\caption{Fundamental quasinormal mode frequencies w.\ r.\ t.\ $Q$ for the 
black holes defined by the metric \eqref{Rad_metric01} (Metric 01) and 
\eqref{Rad_metric03} (Metric 02) surrounded by the radiation field with 
$l=4$, $N_s = 0.01$ and $M = 1$ obtained by using 5th order WKB approximation 
method.}
\label{figRadQvary}
\end{figure}

\begin{figure}[htb]
\centerline{
   \includegraphics[scale = 0.3]{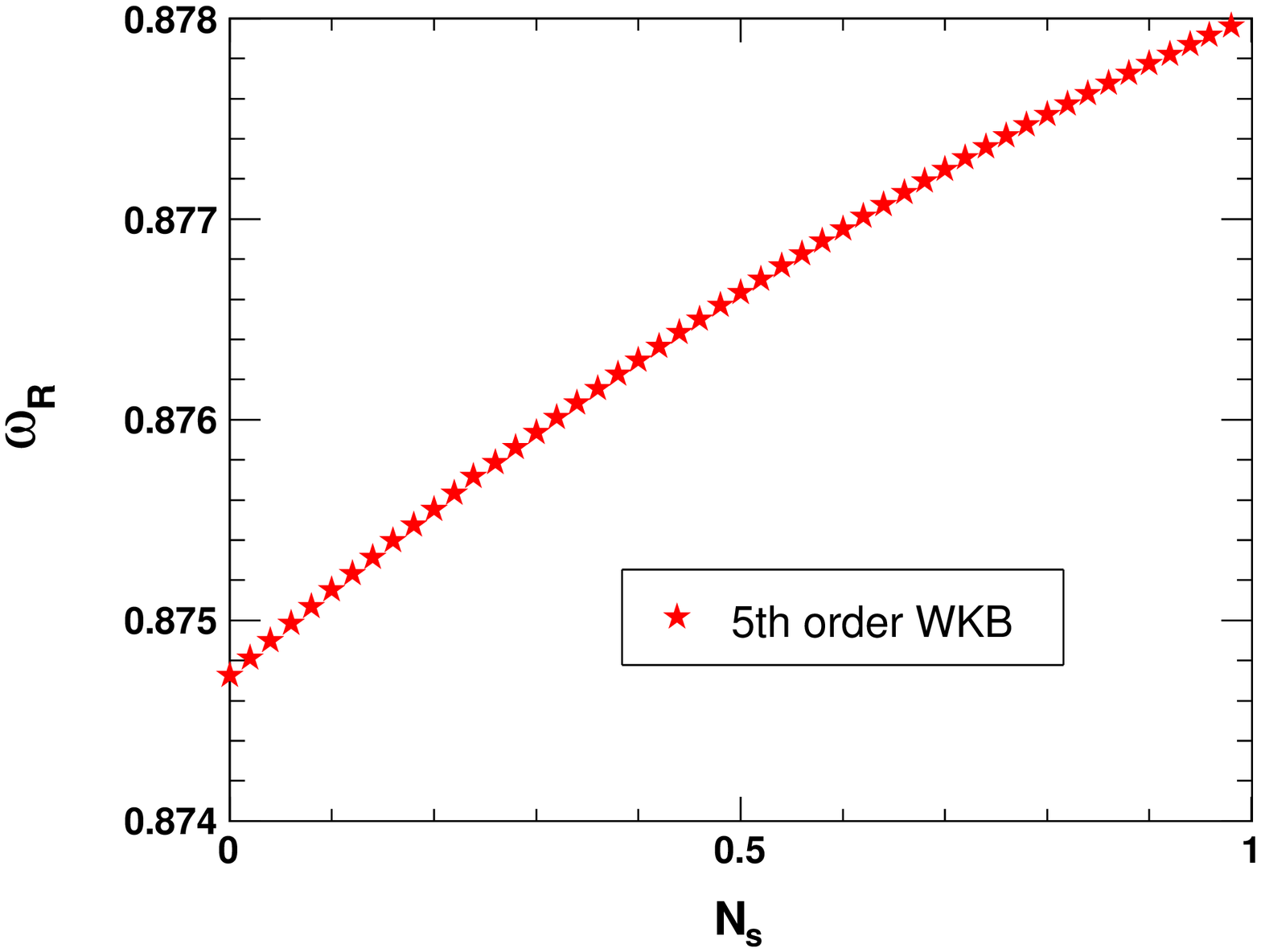}\hspace{0.5cm}
   \includegraphics[scale = 0.3]{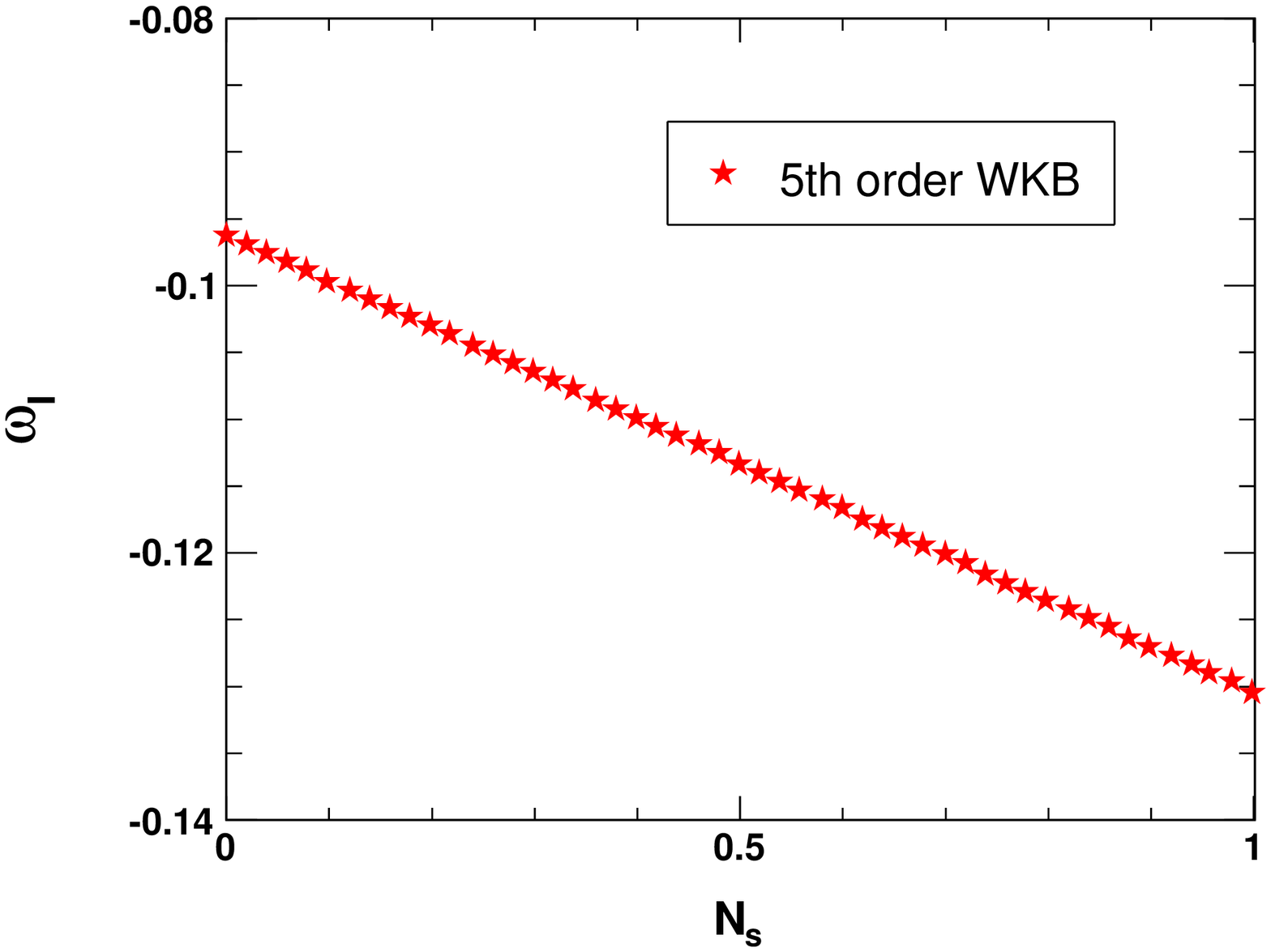}}
\vspace{-0.2cm}
\caption{Fundamental quasinormal mode frequencies w.\ r.\ t.\ $N_s$ for the 
black holes defined by the metric \eqref{Rad_metric03} surrounded by the 
radiation field with $l=4$, $Q = 0.2$ and $M = 1$ obtained by using the 
5th order WKB approximation method.}
\label{figRadNsvary}
\end{figure}

\begin{figure}[htb]
\centerline{
   \includegraphics[scale = 0.3]{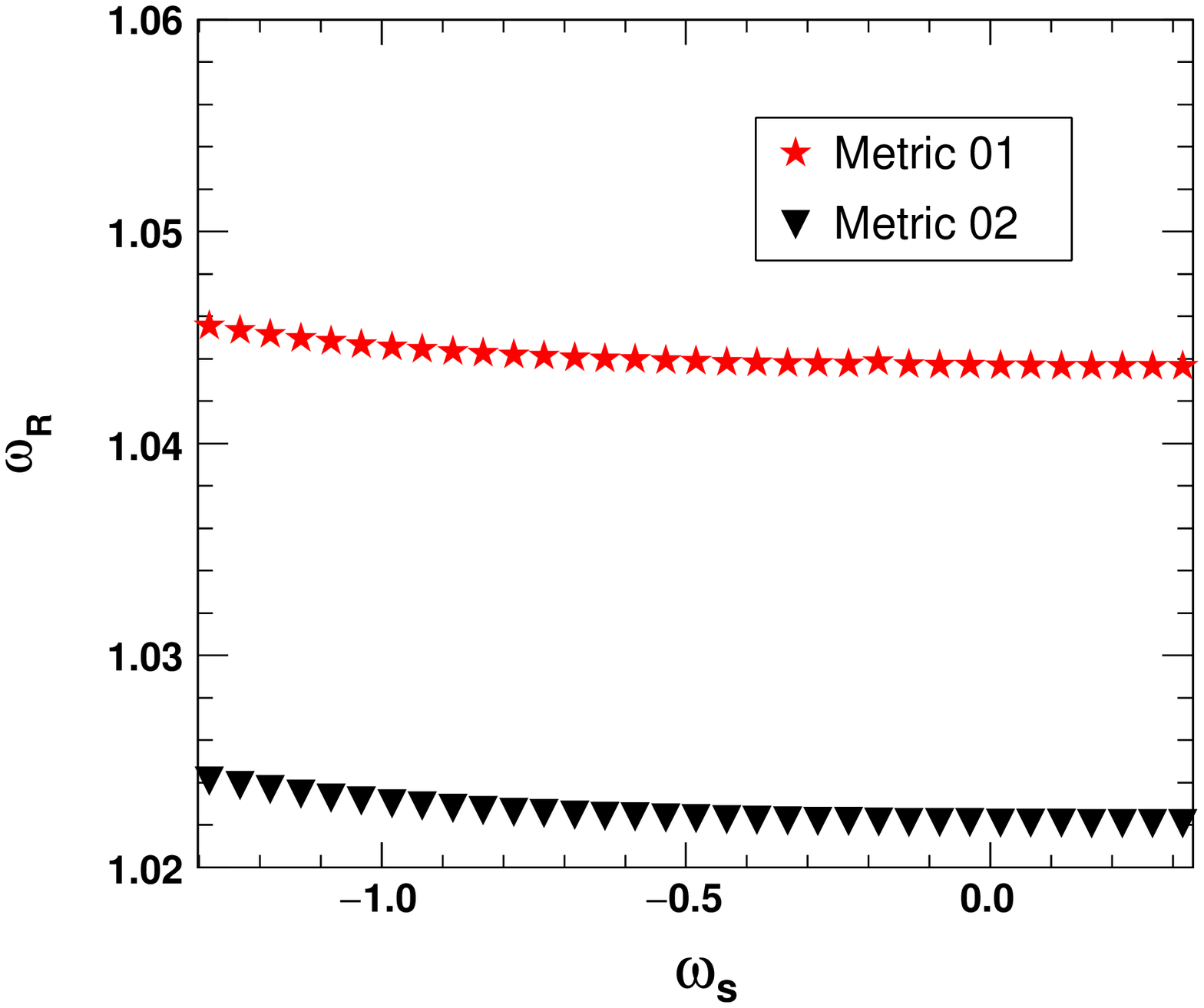}\hspace{0.5cm}
   \includegraphics[scale = 0.3]{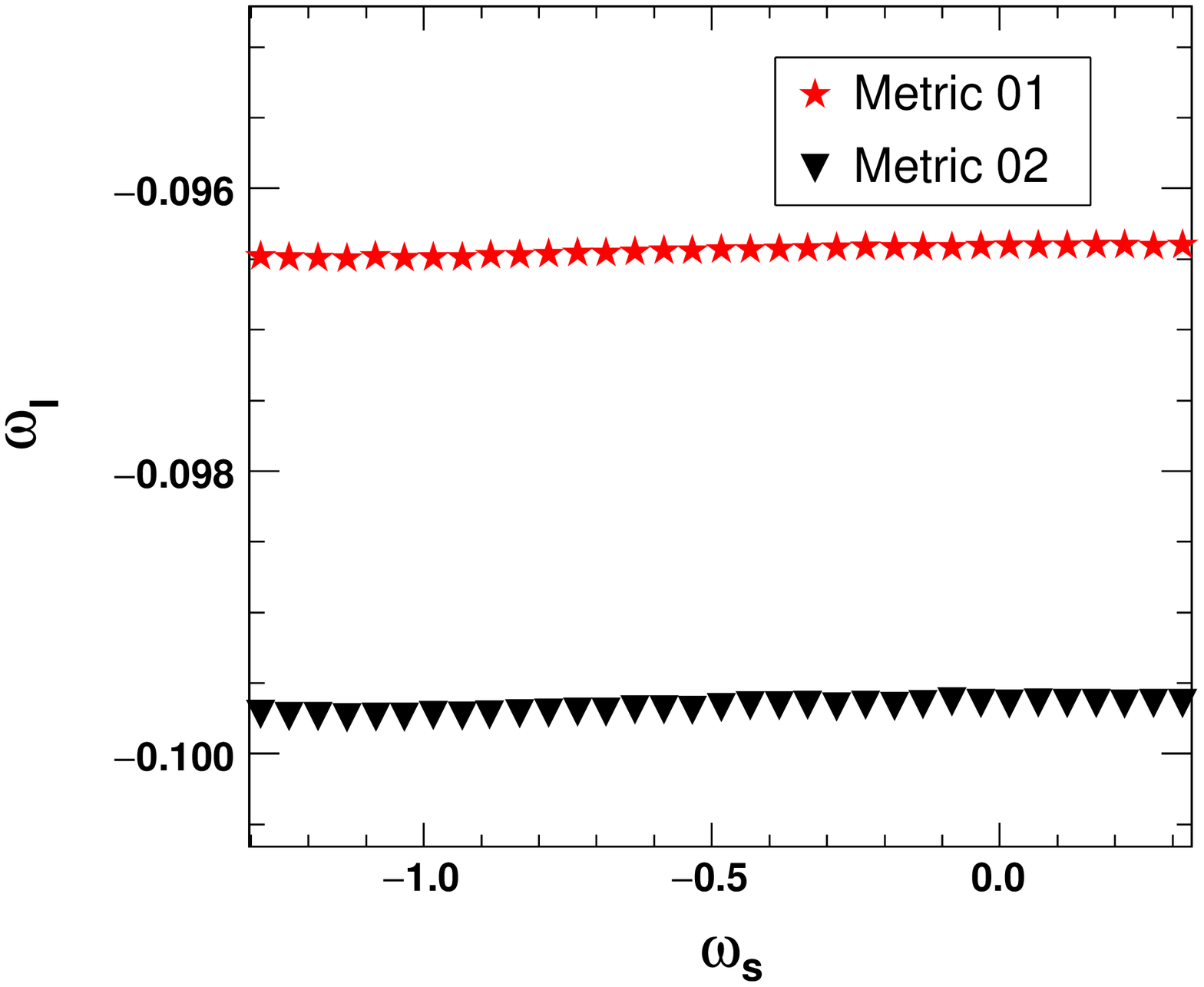}}
\vspace{-0.2cm}
\caption{Variation of fundamental quasinormal mode frequencies w.\ r.\ t.\ 
$\omega_s$ for the black holes defined by the metric \eqref{metric01} 
(Metric 01) and \eqref{metric03} (Metric 02) with $l=4$, $\lambda=0.01$, 
$Q = 0.9$, $\kappa = 1$, $N_s = 0.0001$ and $M = 1$ obtained by using the 
5th order WKB approximation method.}
\label{figomegavary}
\end{figure}

\subsection{Surrounded by radiation field}
We can use the relativistic matter state parameter $\omega_s = \dfrac{1}{3}$ 
in order to denote any electrically charged black hole surrounded by static 
spherically symmetric electric field \cite{Kiselev}.
Hence, for the radiation field the metric \eqref{metric01} becomes,
\begin{equation} \label{Rad_metric01}
ds^2=-\left(1-\frac{2M}{r}
+\frac{Q^2}{r^2}+\frac{N_s}{r^2}\right)dt^2
+\frac{dr^2}{1-\frac{2M}{r}+\frac{Q^2}{r^2}
+\frac{N_s}{r^2}}
+r^2 d\Omega^2.
\end{equation}
This metric is same as the metric in the GR limit for this field (see the 
metric \eqref{metric02}). That is the radiation field surrounded singular 
black holes are same for GR and Rastall gravity.  
For the radiation field the Metric \eqref{metric03} becomes,
\begin{equation} \label{Rad_metric03}
ds^2=-\left(1-\frac{2m(r)}{r}
+\frac{N_s}{r^2}\right)dt^2
+\frac{dr^2}{1-\frac{2m(r)}{r}
+\frac{N_s}{r^2}}
+r^2 d\Omega^2.
\end{equation}
From the above metric it is clear that the black hole with non-linear 
electrodynamic source can not result a regular black hole unless the 
structural parameter $N_s = 0$. In other words, with non-vanishing structural 
parameter both the metrics results singular black holes. The geometric 
parameter for both the black holes is
\begin{equation}
\mathcal{W}_{s}=\frac{1}{3}.
\end{equation}
Thus to respect the weak energy condition, one needs  $N_s\le0$. The 
metric \eqref{Rad_metric01} clearly shows that the charge $Q$ and the parameter 
$N_s$ terms can be combined to have $\frac{Q^2+N_s}{r^2}$ i.e., $N_s$ 
contributes to the square of charge in the metric in the case of the radiation 
field. This implies that for the radiation field, $N_s$ can behave as the 
square of some arbitrary charge. But $N_s<0$ predicts an imaginary charge 
behaviour and that may not be a physical situation to consider. However, in 
Fig.s \ref{figRadQvary}, \ref{figRadNsvary} and \ref{figomegavary}, 
we have considered positive $N_s$ values to compare the results with the 
previous cases although such values of $N_s$ do not respect the weak energy 
condition. We suggest that the case $N_s=0$ will be the most feasible case for 
a physically acceptable black hole as it is the only value that has no issues 
with the weak energy condition and imaginary charge problem.

In this surrounding field case, the black hole solutions are independent of 
the Rastall parameter $\lambda$ i.e., the black holes represented by 
the metrics in the radiation field are in fact the GR black holes. The 
variations of quasinormal modes with respect to charge $Q$ and the structural 
parameter $N_s$ are shown in Fig.~\ref{figRadQvary} and \ref{figRadNsvary} 
respectively. The variations of the frequencies with respect to the charge 
$Q$ is almost similar to the previous cases. However, in case of the imaginary 
frequencies, the bump is not present near $Q =1$. The magnitude of quasinormal 
frequencies increase with increase in the value of $N_s$. The quasinormal 
modes for $l=1$ to $3$ are shown in Table \ref{table05}. Here the quasinormal 
modes calculated using 6th order WKB approximation show the similarpattern as 
that in case of cosmological constant field.

\begin{table}[h!]
\centering
\begin{tabular}{||c|c|c||}
\hline \hline
 & Metric \eqref{Rad_metric01} & Metric \eqref{Rad_metric03} \\ \hline
$l=1$ (6th order WKB)& $0.294925 - 0.0979592 i$ & $0.320666 - 0.0901262 i$   \\
$l = 1$ (5th order WKB)& $0.295071 - 0.0979107i$ & $0.295074 - 0.0979431i$  \\
$|\omega_5-\omega_6|$& $0.000153845$ & $0.0267592$ \\
$\vartriangle_5$& $0.000184587$ & $0.0133854$\\
 \hline
$l=2$ (6th order WKB)& $0.486938 - 0.0969744 i$ & $0.491358 - 0.0960924 i$  \\ 
$l = 2$ (5th order WKB)& $0.486953 - 0.0969714i$& $0.486939 - 0.0969644i$  \\
$|\omega_5-\omega_6|$&$0.0000152971$ & $0.00450421$  \\
$\vartriangle_5$& $0.0000245277$ & $0.00225202$\\
\hline
$l=3$(6th order WKB) & $0.679958 - 0.0967107 i$ & $0.679922 - 0.0967148  i$ \\
$l = 3$ (5th order WKB)& $0.67996 - 0.0967104i$ & $0.679944 - 0.0967117i$  \\
$|\omega_5-\omega_6|$& $2.02237\times10^{-6}$ & $0.0000222173$ \\
$\vartriangle_5$& $5.82151\times10^{-6}$ & $0.0000126752$ \\
 \hline \hline
\end{tabular}
\caption{Quasinormal modes of black holes for $n=0$, $M=1$, $\omega = 
\dfrac{1}{3}$ (radiation field), $N_s = 0.0001$ and $Q = 0.2$.}
\label{table05}
\end{table} 

In the previous parts, the Tables \ref{table01}, \ref{table02}, 
\ref{table03}, \ref{table04} and \ref{table05} show the quasinormal modes 
for $l=1, 2$ and $3$ obtained by using both 5th order and 6th order WKB 
approximation method with errors. We observe some anomalies and reverse 
patterns in some tables corresponding to different surrounding fields. 
Moreover, from the variations of quasinormal modes w.r.t. charge $Q$, one can 
easily see that for the higher multipole $(l=4)$ values and towards higher 
values of charge $Q$, the non-linearly charged black hole always has smaller 
real frequencies. However, this pattern is also slightly affected by the 
surrounding field. But the reverse pattern observed in some cases can be due 
to the error in WKB approximation method. It is because the WKB approximation 
does not allow to calculate quasinormal modes corresponding to $n\ge l$ with 
the satisfactory accuracy \cite{Konoplya2019} and again for smaller $l$ values
the error of the approximation method is likely to be high. Consequently, in 
all the cases, it is observed that the errors and deviations corresponding to 
$l=1$ and $2$ are higher than $l=3$. Moreover, higher orders of WKB approximations may not always result in less error \cite{Konoplya2019}. This can be a reason for higher fluctuations or anomalous patterns of quasinormal modes calculated using 6th order WKB approximation. Apart from these, in general, the 
deviations and errors are higher for the non-linear black hole. This can be 
due to the non-linear charge distribution function. Thus, as mentioned earlier,
these reverse patterns for smaller $l$ values in some tables corresponding to 
different surrounding fields are most likely due to the errors associated with 
the approximation method. 

Another observation obtained from all the cases is that the imaginary part of 
the quasinormal modes are more affected by the parameter $N_s$. From 
Eq.\ \eqref{rho} along with the condition \eqref{weak_energy}, one can see 
that the increase in the parameter $N_s$ increases the energy density and it 
eventually increases the damping. As a reason of which we can see a direct 
dependence of the imaginary quasinormal modes with the structural parameter 
$N_s$. However, for the nature of a field both $\rho_s(r)$ and $f(r)$ get 
affected and the quasinormal modes also vary accordingly.

Finally, to have a comparative picture, we have studied the variation of the 
quasinormal frequencies for the linearly charged black hole and the black hole with 
non-linear electrodynamic source in Rastall gravity with respect to the 
surrounding field in Fig.~\ref{figomegavary}. The magnitude of both the 
frequencies decrease slowly with increase in the value of $\omega_s$ from 
phantom field to radiation field. Thus increasing the value of $\omega_s$ 
introduces a damping effect on the quasinormal frequencies. The effect of 
damping is small for the phantom field and maximum for the radiation field.

We have also considered the constrained values of the Rastall parameter 
$\lambda$ from Ref.~\cite{Tang} and calculated the fundamental quasinormal 
frequencies for $l=2$ for the black holes defined by the metrics 
\eqref{metric01} (charged black holes in Rastall gravity surrounded by a 
field), \eqref{metric02} (Reissner-Nordstr\"om black hole) and 
\eqref{metric03} (black hole in Rastall gravity surrounded by a field with 
non-linear electrodynamic sources) in Table \ref{experimental_comparison}.
This table will help to check the viability of Rastall gravity from the 
future observational data of quasinormal modes of black holes. In this
context it needs to mention that in the Ref.\ \cite{Tang}, from the 
constraining of the  Rastall parameter, authors found that Rastall gravity is
equivalent to GR. In contrast, we would like to point out that the quasinormal 
modes of black holes for the same set of constraint values of the Rastall 
parameter are different from the quasinormal modes in GR. This is clear from 
the fact that different $\kappa \lambda$ constraint values give different 
quasinormal modes as seen from Table  \ref{experimental_comparison}. It 
implies that Rastall gravity, as mentioned earlier, is not completely 
equivalent to GR although some results may show similar behaviour.

\section{Conclusion} \label{conclusion}

At weak field regime, Rastall gravity is 
identical to GR. Hence, it shares the same numbers of polarization modes of 
GWs with that of GR. But in the strong field regime, the theory deviates 
significantly from the GR. We have obtained charged black hole solutions 
surrounded by different fields in Rastall gravity. To obtain a regular black 
hole solution, we have further modified the metric by introducing a 
non-linear distribution function. However, we have seen that, for non-zero 
structural parameter $N_s$, the solutions can give regular black holes only 
for some selected surrounding fields. Although, the quasinormal modes of 
black hole with non-linear electrodynamic source mimic with that from the 
other one in terms of linear charge distribution in asymptotic regime, we have 
seen a significant difference between these two black holes for the large 
values of charge $Q$. Such a deviation can be clearly observed in 
Fig.~\ref{figomegavary} where $Q = 0.9$ is used. In all five cases of the 
surrounding fields, the real quasinormal frequencies increase with respect to 
charge of the black hole. However, for the black hole with non-linear charge 
distribution, the real quasinormal frequencies increase slowly. 
This work has shown that regular black holes can be obtained by introducing a 
non-linear electrodynamic source in the black hole metric of Rastall gravity 
for some selected surrounding fields with non-zero structural parameter $N_s$.

We have also shown the dependency of the quasinormal modes with respect to the 
Rastall parameter $\lambda$. Since, Rastall gravity shows deviation from GR 
only in presence of matter or non-zero curvature, a choice of surrounding 
field can be expected to have significant influences over the black hole 
metric in this theory. For cosmological constant field $(\omega_s = -1)$ and 
radiation field $(\omega_s = 1/3)$, the black hole metrics are independent of 
the Rastall parameter. Thus, the quasinormal modes for the black holes 
surrounded by these fields are independent of the Rastall parameter. The prime 
changes in the quasinormal modes are reflected by the structural parameter 
$N_s$ and the charge $Q$ of the black holes only. For the other three fields, 
the quasinormal modes depend highly on the Rastall parameter. However, the 
choice of field imposes changes differently in the quasinormal modes as seen 
from this study. It is seen that for dust field and quintessence field, the 
magnitudes of real and imaginary quasinormal modes increase with increase 
in the Rastall parameter and the variation in the quasinormal modes is more in 
the case of dust field than that of quintessence and phantom fields.

Finally, in this work we have mainly considered two types of charged black 
holes in Rastall gravity defined by \eqref{metric01} and \eqref{metric03}. 
Similar studies are done for uncharged black holes in Rastall gravity in 
Ref.s \cite{Graca, Liang}. In another similar work the authors considered 
a black hole in Rastall gravity surrounded by a cloud of strings \cite{Cai}. 
In this study, they observed decrease of the quasinormal frequency with 
increase in the Rastall parameter. But in our case, the results are distinctly 
opposite for dust field and quintessence field. Scalar quasinormal modes of 
non-linear charged black holes in Rastall gravity are studied recently in 
\cite{Hu}, in which they have showed a similar variation of quasinormal modes 
with respect to charge. However, in our work, we have considered a non-linear 
distribution function to modify the metric of a charged black hole surrounded 
by different fields in Rastall gravity. The quasinormal modes of such a
black hole in Rastall gravity have not been studied till now to the best of 
our knowledge. So we think that this study will contribute towards our 
knowledge and understanding of black holes in Rastall gravity. Apart from 
this, the present study may contribute to the ongoing studies in the GW 
astronomy. After detection of the GW events, the possibility of detection 
of quasinormal modes has increased. In the GW events, the ringdown phase has 
a very weak signal-to-noise ratio which makes it difficult to extract 
significant results from it. In a recent study, a method has been introduced 
to detect the weak quasinormal modes from such GW events once design 
sensitivity of ground based detectors is achieved \cite{Costa, Martynov}. In 
another study, it is shown that second generation GWs 
detectors such as Advanced LIGO (aLIGO) \cite{Aasi}, Advanced Virgo (AdV) 
\cite{Acernese} and KAGRA \cite{Aso} can play a very promising role in the 
detection of quasinormal modes \cite{Nakamura}. In the near future, with 
sufficient experimental results on quasinormal modes, such studies will help 
to check the viability of Rastall gravity and to discover more about the 
black hole spacetime. This work can be further extended by generalizing the 
Rastall gravity with the various aspects of spacetime geometry. For 
example, the modified covariant conservation condition 
$\nabla_\nu T^{\mu\nu} =\lambda \nabla^\mu R$ can be further extended by 
modifying the geometry part. This modification can be of two types: one in 
which usual conservation of the energy-momentum tensor can be recovered at 
weak field limit and the other in which conservation of the energy-momentum 
tensor violates in both the strong and weak field limits. It will be 
interesting to see the impacts of such modifications on the quasinormal modes 
of black holes. 
\begin{table*}
\centering

   \begin{scriptsize}{
  \begin{tabular}{llllll}
  \hline \hline
 Galaxy  & $\kappa \lambda$  & $\chi^{2}/dof$   & Non-linearly charged Rastall black hole  & Reissner-Nordstr\"om black hole  & Linearly charged Rastall black hole  \\
 (1)  & (2) & (3) & (4) & (5) & (6)  \\
  \hline
    F563-1 &  0.054    &  0.879 &  0.491994  - i0.096021\; 0.487031 - i0.096999 \;\;  & 0.483776 - i0.096784 \; 0.483753 - i0.096789\;\;  & 0.487037 - i0.096998 \; 0.487042 - i0.096996\;\; \\
    F568-3 &  0.154   &  0.866 &  0.483888  - i0.097674 \; 0.487135  - i0.097023\;\;  & 0.483822  - i0.096821\; 0.483858  - i0.096814\;\;   & 0.487132  - i0.097025 \; 0.487147  - i0.097022\;\; \\
    F583-1 &  0.15      &  0.836 &  0.486820  - i0.097075\; 0.487127  - i0.097014 \;\;  & 0.483838  - i0.096815 \; 0.483851  - i0.096812\;\;   & 0.487151  - i0.097018 \; 0.487139  - i0.097020 \;\; \\
    F571-8 &  0.143       &  0.882 &  0.487064  - i0.097024 \; 0.487115  - i0.097014 \;\; & 0.483822  - i0.096813 \; 0.483839  - i0.096810 \;\; & 0.487084  - i0.097026 \; 0.487127  - i0.097017 \;\;   \\
    F579-v1 &  0.048      &  0.13  &  0.489137  - i0.096563 \; 0.487024  - i0.096982  \;\;& 0.483737  - i0.096790 \; 0.483750  - i0.096788 \;\;  & 0.486974  - i0.097009 \; 0.487039  - i0.096996 \;\; \\
    F583-4 &  0.141        &  0.211  &  0.484901  - i0.097459 \; 0.487113  - i0.097016  \;\; & 0.483837  - i0.096808\; 0.483836  - i0.096809 \;\; & 0.487063  - i0.097029 \; 0.487124  - i0.097017 \;\; \\
    F730-v1 &  0.095    &  0.53  &  0.489195  - i0.096583 \; 0.487059  - i0.097007 \;\; & 0.483788  - i0.096795 \; 0.483782  - i0.096796 \;\;  & 0.487035  - i0.097010 \; 0.487070  - i0.097003 \;\;  \\
    U5750  &  0.146     &  0.82  &  0.484517  - i0.097542 \; 0.487121  - i0.097021 \;\; & 0.483836  - i0.096813 \; 0.483844  - i0.096811 \;\;  & 0.487144  - i0.097016 \; 0.487132  - i0.097018 \;\; \\
    U11454 &  0.118    &  0.826  &  0.486740  - i0.097090\; 0.487084  - i0.097021  \;\; & 0.483776  - i0.096807 \; 0.483805  - i0.096802 \;\;  & 0.487074  - i0.097013 \; 0.487093  - i0.097009 \;\;  \\
    U11616 &  0.122        &  0.808  &  0.487093  - i0.097002 \; 0.487085  - i0.097003  \;\; & 0.483810  - i0.096803 \;0.483809  - i0.096803 \;\;  & 0.487044  - i0.097021 \; 0.487098  - i0.097011\;\;  \\
    U11648 &  0.132      &  0.356  &  0.490119  - i0.096418\; 0.487100  - i0.097015  \;\; & 0.483798  - i0.096811 \;0.483822  - i0.096806 \;\;  & 0.487086  - i0.097019 \; 0.487111  - i0.097014\;\; \\
    U11819 &  0.147      &  0.958  &  0.488775  - i0.096693 \; 0.487123  - i0.097021  \;\; & 0.483852  - i0.096810 \; 0.483846  - i0.096811 \;\;  & 0.487111  - i0.097024 \; 0.487134  - i0.097019\;\;  \\
    ESO0140040 & 0.083     &  0.757  &  0.489062  - i0.096604\; 0.487049  - i0.097003  \;\; & 0.483741  - i0.096800 \; 0.483772  - i0.096793  \;\; & 0.487108  - i0.096992 \; 0.487061  - i0.097002\;\; \\
    ESO2060140 &  0.1        &  0.812  &  0.488801  - i0.096670 \; 0.487065  - i0.097015  \;\; & 0.483772  - i0.096800 \; 0.483786  - i0.096797 \;\; & 0.487016  - i0.097017 \; 0.487075  - i0.097005\;\; \\
    ESO3020120 &  0.135     &  0.397 &  0.484058  - i0.097626 \; 0.487104  - i0.097016  \;\; & 0.483833  - i0.096806 \; 0.483827  - i0.096807 \;\; & 0.487093  - i0.097019 \; 0.487115  - i0.097015\;\; \\
    ESO4250180 &  0.121     &  1.675  &  0.487036  - i0.097013 \; 0.487084  - i0.097003  \;\; & 0.483772  - i0.096810 \; 0.483808  - i0.096802 \;\; & 0.487136  - i0.097002 \; 0.487097  - i0.097010\;\; \\
   \hline \\
  \end{tabular}
 \caption{Quasinormal modes of black holes for the fundamental quadrupolar mode $(n=0, l=2)$ with $N_s = 0.0001$ and 
$Q = 0.2$ with the Rastall parameter $\lambda$ obtained from best fitting 
results from the rotation curves of 16 LSB spiral galaxies \cite{Tang}. 
Column (1), (2) and (3) are the name of galaxy, the fitting values of Rastall 
parameter $\kappa\lambda$ and the $\chi^{2}$ values, respectively. These 
values are taken from \cite{Tang} and are calculated by using 
$\omega = -1/3$ i.e. for perfect fluid dark matter (PFDM) or quintessence 
matter. Column (4) shows the quasinormal modes for the black hole in Rastall 
gravity with non-linear electrodynamic source (Eq.\ \ref{metric03}), column 
(5) shows the quasinormal modes for Reissner-Nordstr\"om black hole 
(Eq.\ \ref{metric02}) and column (6) shows the quasinormal modes for charged 
singular black hole in Rastall gravity (Eq.\ \ref{metric01}). In columns 
(4), (5) and (6) values of first subcolumn are calculated using the 6th order 
WKB approximation method and second subcolumn are calculated using the 5th 
order WKB approximation method}
 \label{experimental_comparison}  
  }
  \end{scriptsize}
\end{table*}


\bibliographystyle{apsrev}

\begin{thebibliography}{99}

\bibitem{will2014}
C. M. Will, \textit{The Confrontation between General Relativity and Experiment}, Living Rev. Relativ. {\bf 17}, 4 (2014) [arXiv:gr-qc/0510072].
\bibitem{hulse1975}
R. A. Hulse and J. H. Taylor, \textit{Discovery of a Pulsar in a Binary System}, The Astrophysical Journal Letters {\bf 195}, L51 (1975).
\bibitem{damour1992}
T. Damour and J. H. Taylor, \textit{Strong-Field Tests of Relativistic Gravity and Binary Pulsars}, Phys. Rev. D {\bf 45}, 1840 (1992).

\bibitem{stelle}
K. S. Stelle, \textit{Renormalization of Higher-Derivative Quantum Gravity}, Phys. Rev. D {\bf 16}, 953 (1977).
\bibitem{Riess1998}
A. G. Riess et al., \textit{Observational Evidence from Supernovae for an Accelerating Universe and a Cosmological Constant}, The Astronomical Journal {\bf 116}, 1009 (1998).
\bibitem{Perlmutter1999}
S. Perlmutter et al., \textit{Measurements of $\Omega$ and $\Lambda$ from 42 High-Redshift Supernovae}, ApJ {\bf 517}, 565 (1999).
\bibitem{Bahcall1999}
N. A. Bahcall, \textit{The Cosmic Triangle: Revealing the State of the Universe}, Science {\bf 284}, 1481 (1999).
\bibitem{Bull}
P. Bull et al., \textit{Beyond $\Lambda$CDM: Problems, Solutions, and the Road Ahead}, Physics of the Dark Universe {\bf 12}, 56 (2016).

\bibitem{Liang_2017}
D. Liang, Y. Gong, S. Hou and Y. Liu, \textit{Polarizations of Gravitational Waves in $f(R)$ Gravity}, Phys. Rev. D {\bf 95}, 104034 (2017) [arXiv:1701.05998].
\bibitem{gogoi1}
D. J. Gogoi and U. D. Goswami, \textit{A New f(R) Gravity Model and Properties of Gravitational Waves in It}, Eur. Phys. J. C {\bf 80}, 1101 (2020) [arXiv:2006.04011].
\bibitem{gogoi2}
D. J. Gogoi and U. D. Goswami, \textit{Gravitational Waves in $\mathbf {f(R)}$ Gravity Power Law Model}, Indian J. Phys. (2021) [arXiv:1901.11277].
\bibitem{Rastall72}
P. Rastall, \textit{Generalization of the Einstein Theory}, Phys. Rev. D {\bf 6}, 3357 (1972).

\bibitem{Saleem2021}
R. Saleem and Shahnila, \textit{Cosmological Evolution via Interacting/Non-Interacting Holographic Dark Energy Model for Curved FLRW Space–Time in Rastall Gravity}, Physics of the Dark Universe {\bf 32}, 100808 (2021).
\bibitem{Bronnikov2021}
K. A. Bronnikov, V. A. G. Barcellos, L. P. de Carvalho, and J. C. Fabris, \textit{The Simplest Wormhole in Rastall and K-Essence Theories}, Eur. Phys. J. C {\bf 81}, 395 (2021).
\bibitem{Ghosh2021}
S. Ghosh, S. Dey, A. Das, A. Chanda, and B. C. Paul, \textit{Study of Gravastars in Rastall Gravity}, J. Cosmol. Astropart. Phys. {\bf 07}, 004 (2021).
\bibitem{Sakti2021}
M. F. A. R. Sakti and F. P. Zen, \textit{CFT Duals on Rotating Charged Black Holes Surrounded by Quintessence}, Physics of the Dark Universe {\bf 31}, 100778 (2021).
\bibitem{Maurya2020}
S. K. Maurya and F. Tello-Ortiz, \textit{Decoupling Gravitational Sources by MGD Approach in Rastall Gravity}, Physics of the Dark Universe {\bf 29}, 100577 (2020).

\bibitem{Visser}
M. Visser, \textit{Rastall Gravity Is Equivalent to Einstein Gravity}, Physics Letters B 782, {\bf 83} (2018) [arXiv:1711.11500].

\bibitem{Darabi}
F. Darabi, H. Moradpour, I. Licata, Y. Heydarzade, and C. Corda, \textit{Einstein and Rastall Theories of Gravitation in Comparison}, Eur. Phys. J. C {\bf78},  25 (2018).
\bibitem{Moraes}
W. A. G. De Moraes and A. F. Santos, \textit{Lagrangian Formalism for Rastall Theory of Gravity and G\"{o}del-Type Universe}, Gen. Relativ. Gravit. {\bf 51}, 167 (2019).

\bibitem{Shabani}
H. Shabani and A. Hadi Ziaie, \textit{A Connection between Rastall-Type and f(R, T) Gravities}, EPL {\bf 129}, 20004 (2020).

\bibitem{Rawaf01}A. S. Al-Rawaf and M. O. Taha, \textit{Cosmology of General Relativity without Energy-Momentum Conservation}, Gen. Relat. Gravit. {\bf 28}, 935 (1996).

\bibitem{Rawaf02}A. S. Al-Rawaf and M. O. Taha, \textit{A Resolution of the Cosmological Age Puzzle}, Physics Letters B {\bf 366}, 69 (1996).

\bibitem{Fabris}J. C. Fabris, R. Kerner, and J. Tossa, \textit{Perturbative analysis of generalized Einstein theories}, Int. J. Mod. Phys. D {\bf 09}, 111 (2000).

\bibitem{Rahman}A.-M. M. Abdel-Rahman and M. H. A. Hashim, \textit{Gravitational Lensing in A Model With Non-Interacting Matter and Vacuum Energies}, Astrophys. Space Sci. {\bf 298}, 519 (2005).

\bibitem{Rahman2}A.-M. M. Abdel-Rahman, \textit{Gravitational Lensing Effects in a Modified General Relativity Model}, Astrophys. Space Sci. {\bf 278}, 385 (2001).

\bibitem{Moradpour}H. Moradpour, N. Sadeghnezhad, and S. H. Hendi, \textit{Traversable Asymptotically Flat Wormholes in Rastall Gravity}, Can. J. Phys. {\bf 95}, 1257 (2017).

\bibitem{Kokkotas}K. D. Kokkotas and B. G. Schmidt, \textit{Quasi-Normal Modes of Stars and Black Holes}, Living Rev. Relativ. {\bf 2}, 2 (1999).

\bibitem{Vishveshwara}
C. V. Vishveshwara, \textit{Stability of the Schwarzschild Metric}, Phys. Rev. D {\bf 1}, 2870 (1970).
\bibitem{Press}
W. H. Press, \textit{Long Wave Trains of Gravitational Waves from a Vibrating Black Hole}, ApJ {\bf 170}, L105 (1971).
\bibitem{Chandrasekhar}
S. Chandrasekhar and S. Detweiler, \textit{The Quasi-Normal Modes of the Schwarzschild Black Hole}, Proc. R. Soc. Lond. A {\bf 344}, 441 (1975).

\bibitem{Berti2015}E. Berti et al.,\textit{Testing General Relativity with Present and Future Astrophysical Observations}, Class. Quantum Grav. {\bf 32}, 243001 (2015).

\bibitem{Dreyer2004}O. Dreyer et al., \textit{Black-Hole Spectroscopy: Testing General Relativity through Gravitational-Wave Observations}, Class. Quantum Grav. {\bf 21}, 787 (2004).


\bibitem{Oliveira}
A. M. Oliveira, H. E. S. Velten, J. C. Fabris and L. Casarini, \textit{Neutron Stars in Rastall Gravity}, Phys. Rev. D {\bf 92}, 044020 (2015) [arXiv:1506.00567].

\bibitem{Heydarzade}
Y. Heydarzade, H. Moradpour, and F. Darabi, \textit{Black Hole Solutions in Rastall Theory}, Can. J. Phys. {\bf 95}, 1253 (2017) [arXiv:1610.03881].
\bibitem{Heydarzade2}
Y. Heydarzade and F. Darabi, \textit{Black Hole Solutions Surrounded by Perfect Fluid in Rastall Theory}, Physics Letters B {\bf 771}, 365 (2017) [arXiv:1702.07766].

\bibitem{Chen}
S. Chen, J. Jing, \textit{Quasinormal modes of a black hole surrounded by quintessence} Class. Quantum Gravity {\bf 22}, 4651 (2005) [arXiv:gr-qc/0511085]. 

\bibitem{Zhang}
Y. Zhang, Y.X. Gui, \textit{Quasinormal modes of gravitational perturbation around a Schwarzschild black hole surrounded by quintessence}, Class. Quantum Gravity {\bf 23}, 6141 (2006) [arXiv:gr-qc/0612009]. 

\bibitem{Zhang2}
Y. Zhang, Y.X. Gui, F. Li, \textit{Quasinormal modes of a Schwarzschild black hole surrounded by free static spherically symmetric quintessence: electromagnetic perturbations}, Gen. Relativ. Gravit. {\bf 39}, 1003 (2007) [arXiv:gr-qc/0612010]. 

\bibitem{Ma}
C. Ma, Y. Gui, W. Wang, F. Wang, \textit{Massive scalar field quasinormal modes of a Schwarzschild black hole surrounded by quintessence}, Cent. Eur. J. Phys. {\bf 6}, 194 (2008) [arXiv:gr-qc/0611146]. 

\bibitem{Zhang3}
Y. Zhang, Y.X. Gui, F. Yu, \textit{Dirac quasinormal modes of a Schwarzschild black hole surrounded by free static spherically symmetric quintessence}, Chin. Phys. Lett. {\bf 26}, 030401 (2009) [arXiv:0710.5064]. 

\bibitem{Graca}
J. P. M. Gra\c{c}a and I. P. Lobo, \textit{Scalar QNMs for Higher Dimensional Black Holes Surrounded by Quintessence in Rastall Gravity}, Eur. Phys. J. C {\bf 78}, 101 (2018) [arXiv:1711.08714].

\bibitem{Liang}
J. Liang, \textit{Quasinormal Modes of the Schwarzschild Black Hole Surrounded by the Quintessence Field in Rastall Gravity}, Commun. Theor. Phys. {\bf 70}, 695 (2018).
\bibitem{Hu}
Y. Hu, C.-Y. Shao, Y.-J. Tan, C.-G. Shao, K. Lin, and W.-L. Qian, \textit{Scalar Quasinormal Modes of Nonlinear Charged Black Holes in Rastall Gravity}, EPL {\bf 128}, 50006 (2020).
\bibitem{Xu}
Z. Xu, X. Hou, X. Gong, and J. Wang, \textit{Kerr–Newman-AdS Black Hole Surrounded by Perfect Fluid Matter in Rastall Gravity}, Eur. Phys. J. C {\bf 78}, 513 (2018).
\bibitem{Lin}
K. Lin and W.-L. Qian, \textit{Neutral Regular Black Hole Solution in Generalized Rastall Gravity}, Chinese Phys. C {\bf 43}, 083106 (2019) [arXiv:1812.10100].

\bibitem{Balart}
L. Balart and E. C. Vagenas, \textit{Regular Black Holes with a Nonlinear Electrodynamics Source}, Phys. Rev. D {\bf 90}, 124045 (2014) [arXiv:1408.0306].

\bibitem{Tang}
M. Tang, Z. Xu, and J. Wang, \textit{Observational Constraints on Rastall Gravity from Rotation Curves of Low Surface Brightness Galaxies}, Chinese Phys. C {\bf 44}, 085104 (2020) [arXiv:1903.01034].

\bibitem{Kiselev}V. V. Kiselev, \textit{Quintessence and Black Holes, Class. Quantum. Grav} {\bf 20}, 1187 (2003) [arXiv:gr-qc/0210040].
\bibitem{Ferrari}
V. Ferrari and L. Gualtieri, \textit{Quasi-Normal Modes and Gravitational Wave Astronomy}, Gen. Relativ. Gravit. {\bf 40}, 945 (2008) [arXiv:0709.0657].

\bibitem{Carter1971}B. Carter, \textit{Axisymmetric Black Hole Has Only Two Degrees of Freedom}, Phys. Rev. Lett. {\bf 26}, 331 (1971).

\bibitem{Israel1967}W. Israel, \textit{Event Horizons in Static Vacuum Space-Times}, Phys. Rev. {\bf 164}, 1776 (1967).

\bibitem{Gurlebeck}N. G\"{u}rlebeck, \textit{No-Hair Theorem for Black Holes in Astrophysical Environments}, Phys. Rev. Lett. {\bf 114}, 151102 (2015).

\bibitem{Bustillo}J. C. Bustillo, P. D. Lasky, and E. Thrane, \textit{Black-Hole Spectroscopy, the No-Hair Theorem, and GW150914: Kerr versus Occam}, Phys. Rev. D {\bf 103}, 024041 (2021).

\bibitem{Kono2003}
R. A. Konoplya, \textit{Quasinormal Behavior of the D -Dimensional Schwarzschild Black Hole and the Higher Order WKB Approach}, Phys. Rev. D {\bf 68}, 024018 (2003).

\bibitem{Konoplya2019}
R. A. Konoplya, A. Zhidenko, and A. F. Zinhailo, \textit{Higher Order WKB Formula for Quasinormal Modes and Grey-Body Factors: Recipes for Quick and Accurate Calculations}, Class. Quantum Grav. {\bf 36}, 155002 (2019).

\bibitem{Vikman} A. Vikman, \textit{Can Dark Energy Evolve to the Phantom?}, Phys. Rev. D {\bf 71}, 023515 (2005).



\bibitem{Cai}
X.-C. Cai and Y.-G. Miao, \textit{Quasinormal Modes and Spectroscopy of a Schwarzschild Black Hole Surrounded by a Cloud of Strings in Rastall Gravity}, Phys. Rev. D {\bf 101}, 104023 (2020) [arXiv:1911.09832].

\bibitem{Costa}C. F. Da Silva Costa, S. Tiwari, S. Klimenko, and F. Salemi, \textit{Detection of (2,2) Quasinormal Mode from a Population of Black Holes with a Constructive Summation Method}, Phys. Rev. D {\bf 98}, 024052 (2018).

\bibitem{Martynov} D. V. Martynov et al., \textit{Sensitivity of the Advanced LIGO Detectors at the Begining of Gravitational Wave Astronomy}, Phys. Rev. D {\bf 93}, 112004 (2016).



\bibitem{Aasi}J. Aasi et al. (LIGO Scientific Collaboration), \textit{Advanced
LIGO}, Class. Quant. Grav. {\bf 32}, 074001 (2015).

\bibitem{Acernese} F. Acernese et al. (VIRGO Collaboration), \textit{Advanced Virgo:
A second-generation interferometric gravitational wave
detector}, Class. Quant. Grav. {\bf 32}, 024001 (2015).

\bibitem{Aso} Y. Aso, Y. Michimura, K. Somiya, M. Ando, O. Miyakawa,
T. Sekiguchi, D. Tatsumi, and H. Yamamoto (KAGRA Collaboration), \textit{Interferometer design of the KAGRA gravitational wave detector}, Phys. Rev. D {\bf 88}, 043007 (2013).

\bibitem{Nakamura}T. Nakamura, H. Nakano, and T. Tanaka, \textit{Detecting Quasinormal Modes of Binary Black Hole Mergers with Second-Generation Gravitational-Wave Detectors}, Phys. Rev. D {\bf 93}, 044048 (2016).

\end{thebibliography}
\end{document}